\documentclass[journal,a4paper,10pt]{IEEEtran}
\onecolumn
\usepackage{amsmath,amssymb,amsthm}
\usepackage{float}
\usepackage{amsfonts}
\usepackage{epstopdf}
\usepackage{float}
\usepackage{cleveref}
\usepackage[export]{adjustbox}
\usepackage{lipsum} 
\usepackage{amsmath,amssymb,amsthm,cite}
\usepackage{graphicx}
\usepackage{caption}
\usepackage{subcaption}
\usepackage{kantlipsum} 
\usepackage{mwe} 
\usepackage{comment}
\DeclareCaptionLabelSeparator{periodspace}{.\quad}
\usepackage{psfrag}
\usepackage{dblfloatfix}
\usepackage{mathtools}
\usepackage{algpseudocode}
\usepackage{algorithm}
\usepackage{slashbox}
\usepackage{fancybox}
\usepackage{tikz}
\usepackage{tikzscale}
\usepackage{forest}
\usepackage[acronym]{glossaries}
\makeglossaries
\usetikzlibrary{shapes.geometric,backgrounds}
 \usetikzlibrary{decorations.pathreplacing}
 \usetikzlibrary{trees}
\usepackage{cleveref}
\usetikzlibrary{positioning}
\DeclareMathOperator*{\argmax}{\arg\max}

\allowdisplaybreaks
\newtheorem{lemma}{Lemma}
\newtheorem{theorem}{Theorem}

\newtheorem{proposition}{Proposition}
\newtheorem{remark}{Remark}
\newtheorem{claim}{Claim}
\theoremstyle{definition}

\newenvironment{condition}[1]
{\begin{trivlist}
\item[\hskip \labelsep {\bfseries Condition $\left( #1 \right)$:}]}{\end{trivlist}}
\usetikzlibrary{arrows,calc,shapes,decorations.pathreplacing}
\crefname{figure}{Fig.}{Figs.}
\colorlet{light blue}{blue!40}
\usepackage{nomencl}
\makenomenclature
\newacronym{nb}{NB}{Negative Binomial}
\newacronym{arq}{ARQ}{Automatic Repeat reQuest}
\newacronym{fec}{FEC}{Forward Error Correction}
\newacronym{evt}{EVT}{Extreme Value Theory}
\newacronym{oap}{OAP}{Over-the-Air Programming}
\newacronym{wsn}{WSN}{Wireless Sensor Networks}
\newacronym{dis}{DIS}{Distributed Interactive Simulation}
\newacronym{nsiv}{NSIV}{Non-Stationary Integer Valued}
\usepackage{amsthm}
\usepackage{latexsym}
\usepackage{amssymb}
\usepackage{fancyhdr}
\usepackage{amsmath}
\usepackage{color}
\usepackage{cite}
\usepackage{setspace}
\usepackage{epsfig, tabularx, multicol}
\usepackage[final]{pdfpages}
\usepackage{fancyhdr}

\theoremstyle{plain}
\begin{document}
\title{Asymptotic Analysis for Reliable Data Dissemination in Shared loss Multicast Trees}
\author{\IEEEauthorblockN{Ron Yadgar\IEEEauthorrefmark{1},
Asaf Cohen\IEEEauthorrefmark{2}, and
Omer Gurewitz\IEEEauthorrefmark{3}}\\
\IEEEauthorblockA{Department of Communication Systems Engineering
\\Ben-Gurion University of the Negev\\
Email: 
\IEEEauthorrefmark{1}ronyad@post.bgu.ac.il,
\IEEEauthorrefmark{2}coasaf@bgu.ac.il,
\IEEEauthorrefmark{3}gurewitz@bgu.ac.il
}}
\maketitle
\thispagestyle{plain}
\pagestyle{plain}
\begin{abstract}
The completion time for the dissemination (or alternatively, aggregation) of information from all nodes in a network plays a critical role in the design and analysis of communication systems, especially in real time applications for which delay is critical.
In this work, we analyse the completion time of data dissemination in a shared loss (i.e., unreliable links) multicast tree, at the limit of large number of nodes.
Specifically, analytic expressions for upper and lower bounds on the expected completion time are provided,
and, in particular, it is shown that both these bounds scale as $\alpha \log n$.
For example, on a full binary tree with $n$ end users, and packet loss probability of $0.1$, we bound the expected completion time for disseminating one packet from below by $1.41 \log_2 n+o \left( \log n \right)$ and from above by $1.78 \log_2 n+o \left( \log n \right)$.

Clearly, the completion time is determined by the last end user who receives the message, that is,  a maximum over all arrival times.
Hence, Extreme Value Theory (EVT) is an appropriate tool to explore this problem.
However, since arrival times are correlated, non-stationary, and furthermore, time slots are discrete, a thorough study of EVT for Non-Stationary Integer Valued (NSIV) sequences is required.
To the best of our knowledge, such processes were not studied before in the framework of EVT.
Consequently, we derive the asymptotic distribution of the maxima of NSIV sequences satisfying certain conditions, and give EVT results which are applicable also beyond the scope of this work.
These result are then used to derive tight bounds on the completion time.
Finally, the results are validated by extensive simulations and numerical analysis.
\end{abstract}

\section{Introduction}\label{intro}
Reliable data dissemination (one to many) and/or data collection (many to one), is highly important for a large variety of purposes.
For example, \acrfull{oap} is an essential utility which enables data distribution such as application downloads, software updates, configuration settings, etc., by one central entity to many devices (e.g., Deluge \cite{N82hui2004dynamic} , MNP \cite{N83MNPkulkarni2005mnp} ,and MOAP \cite{N85stathopoulos2003remote}).
OAP is becoming increasingly important with the expansion of \acrfull{wsn} and the Internet of Things, where the networks consist of \emph{hundreds or thousands} of nodes.
In addition, data collection, and specifically aggregating data packets from thousands of sensor nodes to a sink in large WSN's, has numerous applications, e.g., habitat monitoring \cite{N86mainwaring2002wireless}, microclimatic monitoring of a coastal redwood \cite{N87tolle2005macroscope} and even event detection of an active volcano \cite{N84werner2006fidelity}. 

Setting up a direct link between the source to all its intended receivers for data dissemination or from each source to the sink for data collection is usually infeasible, impractical or highly resource-wasteful.
For example, in WSN, setting up direct connectivity between the sink and all the sensors is usually impossible since sensors are distributed over a vast area while the sink transmission power is limited (i.e., the sensors have unlimited transmission power).
Even when possible, such a set up may result in enormous energy waste due to the proportionality of the required radio transmission power to the squared distance, or in poor 
 utilization, due to poor spatial reuse.
Accordingly, typical networks will rely on multi hop routing. Furthermore, since setting up an autonomous multi-hop path (unicast) between the sink to all devices is also impractical and results in poor resource utilization, especially on networks where the redundancy and contention can dramatically impair performance \cite{N91intanagonwiwat2000directed}, multicast tree-based data dissemination or collection is the common adopted solution both for data dissemination and data collection, respectively \cite{N94zhuang2001bayeux,N95szewczyk2004habitat}.

Throughout this study, we focus on one to many reliable data dissemination (downstream traffic), in which a source has information it needs to distribute to all other users in the network.
Nonetheless, all of our results also apply to data collection (ConvergeCast), namely, aggregating data from all nodes to the root of the tree, for a setup in which each intermediate node waits until all data coming from its children has been received successfully, before combining it into a single packet, and forwarding it to its parent.

Multi-cast transmission is usually done in two phases: The setup stage, which establishes and maintains the multicast tree, determines the routing tree to all nodes, usually under some constraints \cite{N90heinzelman1999adaptive}, and the continuous stage, which broadcasts the information to all nodes in the tree.
In this study, we focus on analyzing the second stage, and in particular, we investigate the \emph{completion time}, which we define as the amount of time required between the start of a block transmission initiated at the source (or the end users), and its completion, i.e., successful decoding at all the receivers (or sink) for data dissemination  (data collection). 

There are several studies in the literature which analyze tree-based reliable multicast.
For example,  \cite{N62bhagwat1994effect} found a recursive formula for the distribution of the required number of single packet transmissions, until all nodes receive it.
Specifically, \cite{N62bhagwat1994effect} assumed that the source which is the multicast tree root, retransmits the packet until it is received correctly by all receivers.
Note that under this scheme, denoted as End-to-End ARQ mechanism, intermediate nodes only forward each packet they receive from their parents to their children, and do not initiate a retransmission of the packet.
Extension to bulk data distribution utilizing \acrfull{fec} codes, and specifically to $C(n; k)$ erasure codes, in which receiving any $k$ out of $n$ packets is sufficient to decode the entire group is given in  \cite{N88mosko2000analysis}.
The benefit of granting each intermediate node the responsibility for reliable transmission of packets to its children, compared to end-to-end error control techniques was studied in \cite{N61ghaderi2007network}.
Specifically, \cite{N61ghaderi2007network} extend the model suggested in  \cite{N88mosko2000analysis} to capture also Link-by-Link ARQ and Link-by-Link FEC mechanisms, for which every node of the multicast tree is responsible for reliable delivery of packets it has received from its parent to its children, utilizing repeated retransmissions or rateless coding, respectively.
Note the inherent difference between the end-to-end and the link-by-link error controls; for the former, finding the number of transmissions from the source is equivalent to finding the completion (delay) time, whereas for the latter this is clearly not the case.
Delay is an important performance measure in many applications, for example in multimedia application such as distributed games, and/or \acrfull{dis}, emergency response, vehicular communication and even financial information. 


In this study, we consider the problem of Data Dissemination in shared loss multicast tree with \emph{link by link error control}, where the information is propagated from the source (root) to all nodes in the tree according to the following procedure: Each node, starting from the root, retransmits the packet it has received from its predecessor until it is correctly received by \emph{all of its successors}.
The process terminates when all the nodes in the tree receive the message.
We assume that the channel between each parent and its children has packet loss probability $p$, independent of each other.
Accordingly, the distribution of the number of transmissions, or time slots, needed for each child to receive the packet successfully follows a geometric distribution with parameter $p$.

Clearly, the completion time which is defined as the number of time slots it takes until all nodes in the tree receive the message/s is determined by the time the message arrives at end users of the tree, and, in particular, \emph{by the leaf that takes the longest time to receive} the messages.
Since the completion time is a random variable, we explore the distribution, and in particular, the expected number of time slots that it takes until the last leaf of the tree has received the messages.
Mathematically, denote by $T_i$ the arrival time, that is, the time it takes leaf $i$ to receive the message; we are interested in finding the distribution of the maximum of these random variables, denoted by $M_n$, or formally $M_n = \max (T_1, \ldots , T_n)$. 
We focus on large networks, comprising a large number of nodes (large $n$).
Note that $M_n$ can be defined recursively, $M_n = \max_{1 \leq i \leq K} \left(W_i+M_{\frac{n}{K}}^{(i)} \right)$, where $K$ indicate the number of children each node, $M_{\frac{n}{K}}^{(i)}$ being the completion time of the subtree which belongs to the child $i$, and $W_i$ is the time to forward the message to the child $i$.
Even though exploring $M_n$ for low values may be possible, for large values of $n$, this becomes impractical.
As a result, our analysis will focus on the \emph{scaling laws and asymptotic behaviour of $M_n$}.
We will, however show via simulations that our analytical results hold even for relatively small $n$, in the order of present-day networks.
Another aspect that asymptotic analysis allows is examining the impact of increasing/decreasing system parameters, such as the number of nodes or the packet loss probability $p$ on the completion time.
Such analysis can be useful, for instance in determining the amount of sensors in a single tree or the power to invest in a transmission (which affects $p$), under the constraint of the completion time.
Similarly, the structure of the tree (e.g., the number of offsprings) can also affect the completion time.

The main tool that we will utilize is \acrfull{evt}, which is a branch of statistics that deals with the asymptotic distribution of the maxima of random variables.
Even though EVT has been widely explored in the literature (e.g., \cite{N28leadbetter1983extremes,lawsOfSmallNumber,Bookde2007extreme,BookResnick2007extreme}), two main challenges need to be addressed in the context of this work: First, classic EVT considers continuous random variables, while we are dealing with number of time slots, hence \emph{discrete random variables}.
The second issue which is not less challenging, is that the random variables we are dealing with are \emph{correlated}, i.e., are not \emph{i.i.d.}, and in particular, are not independent, which is due to the structure of the tree.
Namely, for each two random variables, $T_i$ and $T_j$, there might be some \emph{mutual channels} where the message must go through, hence, times $T_i$ and $T_j$ might be correlated. Although there are several studies on this topic, the vast majority of them copes with stationary processes (for example \cite{N9,N10} for continuous and \cite{N30mccormick1992asymptotic,N53hall1996maximum} for integer valued).
In our case, one can see that the process $(T_1, \ldots , T_n)$ is \emph{non-stationary}.
Regarding the first issue, we adopt the technique initially suggested by Anderson \cite{N31anderson1970extreme}, to bound the CDF of the discrete variables by that of two continuous random variables, and apply the rules of EVT on both.
For the second issue, inspired by previous work,  we formulate sufficient conditions in order to asymptotically bound the distribution of the maximum of \acrfull{nsiv} sequences.

Note that the techniques used in this paper to analyse the dissemination time over \emph{trees} can be used to extend known results for simpler topologies, e.g., \cite{N99stojanovic2009data,N19,N100swapna2013throughput}, which considered the problem of disseminating $\kappa$ messages to $n$ users, in an unreliable \emph{single hop broadcast} network.
Specifically, we find the asymptotic distribution of the completion time for the cases of $\kappa = M \log_2 n$, for any constant $M$ as well as for $\kappa = o \left( \log n \right)$ under certain conditions, extending the known results which were valid only for constant $\kappa$ and $\kappa=\omega \left( n \right)$.
\subsection*{Main contribution}
Our contributions in this work are thus as follows.
\begin{itemize}
\item We analyse the extremal properties of NSIV sequences, and in particular we find that under appropriate conditions, the limit distribution of the maximum of the sequence $\left(X_1,\ldots,X_n \right)$ has the same type as if the sequence were i.i.d.\
We believe that such results on NSIV sequences did not exist in the EVT literature.
\item The results on the extremal properties of NSIV sequences are applied \emph{to obtain a lower bound on the expected completion time}, when the source disseminates one message, that is, until all the nodes in the tree receive the message.
The lower bound is obtained by aggregating the channels of the upper part of the tree into one mutual channel for all users, which causes a decrease of diversity.
We also derive a matching upper bound.
The upper bound is obtained by increasing the diversity: assuming independent channels for all users.
The bounds are tight in the sense that both upper and lower bounds predict the same scaling law, with a difference only in the constant multiplicative factor of the leading term.
\item In practice, it is important to know \emph{the number of time slots required} in order receive the message at all nodes with a certain probability $1-\epsilon$. Hence, we also bound the CDF of the completing time, and as a result give an explicit expression for this number.
\item We generalize our result to $M$ messages and $K$ child nodes, where $K$ might be a function of $n$.
Furthermore, in some cases, we find tight bounds with the same constant multiplicative leading factor.
\item Finally, we derive a closed-form expression for the distribution of the completion time when
broadcasting a file consisting of $\kappa$ packets to $n$ users, in an unreliable \emph{single hop broadcast} network, for $k = M \log_2 n$ and also for $k = o(\log n)$, $\kappa$ satisfying $\frac{\kappa^2 \log \log n}{\log n}\to 0$.
This extends the known results in the area, valid only for larger values of $k$.
\end{itemize}
The rest of the work is organized as follows.
In \Cref{section2} we give a thorough
description of related works.
\Cref{SystemModelSection} describes the system model and the problem formation.
In \Cref{PreliminariesSection}, we review known EVT results which serve as a basis for this work.
In \Cref{sec:EVTNonStationaty}, we study the extreme value of NSIV sequences. 
In \Cref{MainResultSection}, we obtain the main results, for one message and full binary tree, where in \Cref{sec:ExtentionMmessages}
we extend it to $M$ messages and in  \Cref{sec:ExtentionTreeWithKChild}  we study the case of K-ary trees.
In \Cref{SimulationSection}, we present some numerical result and discuss them.
Finally, \Cref{sec:conclusion} concludes the work.
\section{Related Work}\label{section2}
The analysis of the completion time for broadcast and multicast network models is well studied in the literature, under a wide range of assumptions and constraints.
For example, Farley \cite{N65farley1979minimal} considered the problem of constructing a minimal broadcast network, that is, a network where a message can be broadcasted with in a minimum completion time, regardless the originator, under the constraint that a member can not disseminate a message to more than one node at the same time slot.
The author presented an algorithm to construct such networks with $n$ members.
In \cite{N66farley1980broadcast}, the author extended it to $m$ messages, and determined upper and lower bounds on the completion time in broadcasting $m$ messages throughout a network with $n$ nodes, assuming that a node can either transmit or receive, and an informed node can send to only one adjacent node at a single time slot.
Slater \textit{et al}.\ \cite{N57slater1981information} presented an algorithm that determines the minimum broadcast time to deliver information from an arbitrary node to every other vertex in the tree.
Koh \cite{N58koh1991information} extended it to non-uniform edge transmission times.
Proskurowski \cite{N59proskurowski1981minimum} characterized class of trees, called Minimum Broadcast Trees (MBT), and form an algorithm to determine whether a given tree is MBT.

In all articles mentioned above, the authors assumed lossless channels in each point to point link.
However, this assumption may not be realistic, due to the nature of wired and wireless channels.
When losses are present, in order to guarantee reliability in the network, mechanisms such as \acrfull{arq} or FEC are required.
Starobinski \textit{et al}.\ \cite{N64starobinski2010asymptotically} analyzed the asymptotic performance of data dissemination, focusing on single-hop clusters and multihop cluster chains, assuming that the source is instantly notified when all nodes have received a given packet.
Eryilmaz \textit{et al}. \cite {N96eryilmaz2008delay} studied the delay performance under Possion arrivals, and disseminating (broadcasting) in a single hop network to multiple receivers, under two transmission modes: round robin scheduling and network coding.
In particular, the authors found close forms and asymptotic expressions for the time required to transmit all packets at the head of line file to all the receivers, in a single hop, under these two transmission strategies.
The authors also provided methods for extending the single-hop setting to multiple-hop networks with general topologies.
In \cite{N97ahmed2007scaling}, the authors extended this model by considering delay constraints on the incoming traffic.
Starobinski \textit{et al}.\ \cite{N99stojanovic2009data} also considered broadcasting a file consisting $\kappa$ packets to $n$ users in a single-hop wireless network (see also \cite{N98xie2013network}), and compared the delay performance (completion time), between network coding and cooperative transmission schemes.
In particular, the authors obtained the scaling law of the expected delay under randomized linear coding strategy, for large $n$ and for the cases of fix $\kappa$, $\kappa=\log n$, and where $\kappa$ grows at least as fast as $(\log n)^r$, $r >1$.
Since ARQ with a single source becomes impractical for large amount of users, the authors \cite{N19} exploited the advances of  practical rateless codes (e.g., LT codes \cite{N32LTcode} and Raptor codes \cite{N33RaptorCode}), in order to implement a broadcast scheme which does not require an ACK after each reception, and a node can collect packets from a practically infinite stream.
The paper analysed the completion time when disseminating a file to all receivers located at one hop from the source, by using EVT, including cases with heterogeneous packet loss probability.
Similarly, Eryilmaz \textit{et al}.\ \cite{N100swapna2013throughput} also considered the model of single hop networks, where the source is broadcasting a file consisting of $\kappa$ packets to $n$ users, by performing Random Linear Network Coding (RLNC), for a time-correlated erasure channel model.
The authors analysed the asymptotic throughput and delay, when $\kappa$ increases with $n$, and, in particular, characterized the distribution of the completion time in the cases where $\kappa$ grows faster than $\log n$, i.e.\ $\kappa=\omega (n)$.
Note that besides the cases of constant $\kappa$ and $\kappa=\omega(n)$, the asymptomatic distribution of the completion time is still missing.
In our work, we extend this to the cases where $\kappa$ is a slowly increasing function of $n$, and for $\kappa$ growing logarithmic with $n$.
Moreover, in our main result, we extend the scheme of the single hop to a scheme where the users are \emph{connected via K-ary tree}.
Nonnenmacher \textit{et al}.\ \cite{N63nonnenmacher1997performance} and Bhagwatt \textit{et al}.\ \cite{N62bhagwat1994effect} analysed the performance of reliable multicast trees, assumed end-to-end error control techniques, and in particular, found a recursive formula for the distribution of the number of transmissions until all nodes receive the packet.
Note that in our work we are focusing on the completion time, rather than number of transmissions.
Moreover, Herein, we derived bounds for the \emph{close form} of the expected completion time, which enable us to examine the impart of parameters such as packet loss probability and multicast tree topology.
Mosko \textit{et al}.\ \cite{N88mosko2000analysis} considered FEC for a group of $k$ messages, rather than analyzing the number of transmissions of a particular packet,￼ using $C\left(n,k \right)$￼ erasure codes.
Ghaderi \textit{et al}.\ \cite{N61ghaderi2007network} considered four types of of error control techniques, (end-to-end ARQ, end-to-end FEC, link-by-link ARQ and network coding), for a reliable multicast tree, and evaluated the expected number of transmissions under these four different error control schemes.
In \cite{N60ghaderi2008reliability}, the authors studied the asymptotic analysis of these four types, and also analyzed the performance of broadcasting to a set of $K$ receivers over lossy wireless channels with different reliability mechanisms (ARQ, FEC, and Network Coding).
\subsection*{Extreme Value Theory}
%
As EVT stands at the basis of asymptotic analysis of completion times, in this section, we give an overview of EVT results, starting for the classic results on i.i.d.\ processes, through various enhancements, to more complex stochastic processes.

As noted below, EVT is a branch of statistic dealing with asymptotic distributions of the maximum of random variables, namely $M_n=\max \left(X_1,\ldots,X_n \right)$.
For the i.i.d. case, Gnedenko's theorem \cite{N92gnedenko1943distribution} (see also \cite{haan1970regular}) states that, if after a proper linear normalization, the distribution of $M_n$ converges to some non-degenerate limit distribution $G(x)$, then $G(x)$ belongs to either the \emph{Gumbel}, the \emph{Fréchet} or the \emph{Weibull} family.
Loynes \cite{N8} studied the case where sequences are not i.i.d.\ but have a strong mixing property.
Leadbetter \cite{N9} formulated a long range dependence condition $D \left( u_n \right)$ of the mixing type, yet much weaker than strong mixing, and a local dependence condition $D^{'} \left(u_n \right)$.
In particular, Leadbetter showed that under these conditions, the limit distribution of the maximum is the same as if the stationary sequence were i.i.d.
The author extended the results in \cite{N10} to the case where the local dependence condition $D^{'} \left(u_n \right)$ is omitted, and showed that $D \left( u_n \right)$ is sufficient to guarantee the asymptotic distribution, where $M_n$ has the same \emph{type} as if the sequence were i.i.d., with a modification reflected in the normalizing constants.
Leadbetter described a parameter called the \emph{extremal index}, $ 0 \leq \theta \leq 1$, which connects the extreme value distribution to the amount of dependence in the process.
For a class of stationary sequences which do not satisfy condition $ D^{'}(u_n)$, the author introduced in  \cite{N39leadbetter1989exceedance} a mild local mixing condition called $D^{''}\left( u_n \right)$, and shown that the joint distribution of $\left( X_1,X_2 \right)$, on a stationary sequences, determines whether the extremal index exists, and computed its value.
Chernick \textit{et al}.\ \cite{N38chernick1991calculating} generalized the previous result by defining another local mixing condition called $D^{k}(u_n)$, and showed that under this condition, the extremal index can be calculated (and hence the distribution of $M_n$) with the knowledge of the joint distribution of $k$ consecutive variables. 
Smith \cite{N26smith1992extremal} and Weissman \textit{et al}.\ \cite{N27weissman1995extremal} studied the extremal index, for some stationary sequences.
For the cases where it is hard to find a closed form for $\theta$, an estimation can be found in \cite{N24Smith1994estimating} and \cite{N25hsing1993extremal}.

For stationary normal sequences, Berman \cite{N20} stated that if the covariance sequence $\rho _n$ satisfies $\rho _n\log n \to 0$, then $M_n$  converges in distribution to the type I of extreme value distributions (Gumbel), which is the one which applies to i.i.d.\ normal sequences.
Moreover, it was shown that Berman's condition implies that conditions $D \left(u_n \right)$ and $D^{'} \left(u_n \right)$ are satisfied, for a well know normalizing sequence $u_n$.
Mittal and Ylvisaker \cite{N22} extended it (see also \cite[Chapter 6]{N28leadbetter1983extremes}), and discussed the maxima when $\rho_n \to 0$, but $\rho_n \log n $ is not tending to zero.

EVT has also been extended to non-stationary sequences by Husler \cite{N41husler1986extreme},\cite{N42husler1983mptotic}, who formulated equivalent conditions to those of Leadbetter's, including for non-identical distribution, and also for different boundary level for each $i$, i.e., $u_{n}=u_{n,i}$.

Regarding integer-valued sequences (discrete random variables), the results of the classic EVT do not hold  because most of the discrete distributions (including geometric, \acrfull{nb}, binomial and Poisson), do not belong to the domain of attraction of any of the three types of extreme value distributions.\\
Anderson \cite{N31anderson1970extreme} was the first to introduce the study of extremal behaviour of some i.i.d.\ integer-valued sequences, by defining a class of distributions such that the limiting behaviour is bounded, under suitable conditions, by a non degenerate limit distributions.
In his work, the author also analysed the behaviour of the maximum queue
length for M/M/1 queues.
Based on Andeson's work, several stationary integer-valued models have been studied.
McCormick \textit{et al}.\ \cite{N30mccormick1992asymptotic} analysed the limit distribution of a stationary sequence satisfying Leadbetter's conditions $D(u_n)$ and $D^{'}(u_n)$, and applied it to first order Auto Regressive (AR(1)) NB and geometric processes, using an operation called \emph{binomial thinning} \cite{N93weiss2008thinning}.
Hall \cite{N53hall1996maximum} extended his result to a family of stationary processes, satisfying conditions $D(u_n)$ and $D^{k}(u_n)$, and applied it to Max AR(1) processes.
Serfozo \cite{N49serfozo1988extreme} considered the case of queue lengths in M/G/1 and GI/M/1 systems, and in \cite{N48serfozo1987extreme} considered birth and death processes.
Hall \cite{N50hall2001extremes} extended the result of Rootzen \cite{N47rootzen1986extreme} to Integer-valued Moving Average (INMA), where the marginal distribution tail of the sequence is regularly varying.
This class of moving average models is defined by analogy with the continuous moving average processes, replacing multiplication by binomial thinning. 
Hall \textit{et al}. \cite{N78hall2003extremes} studied the extreme of sub-sampled IVMA models with heavy-tailed innovations, and in \cite{N79hall2004extremal}, the authors extended it to a periodic sub-sampled, i.e., $Y_k=X_{g(k)}$, where  $g(k)$ is an increasing sequence with a periodic pattern.
In \cite{N77hall2008extremes}, the authors considered random generated sub-sampled processes, and in  \cite{N51hall2003extremes}, Hall \textit{et al}.\ studied the limiting distribution of the maximum term of INMA when the innovations have exponential type tails.
In \cite{N52hall2007maximum}, Hall \textit{et al}.\ considered several integer-valued stationary models of MA and max-AR type, where the marginal distribution belongs to the domain of attraction of a \emph{max-semistable law}, which means that instead of $n$ observations, we look at a subsequence of $k_n$ observations, such that $F^{k_n} \left[a_nx+b_n \right]$ converges to a non-degenerate limiting distribution.
Therefore, the authors obtained a well-defined limiting distribution for the maximum term, instead of limiting bounds.
For the same class of max-semistable law, \cite{N80hall2009max} considered the effect of missing values on the extreme value distribution.
Regarding non-stationary sequences, Hall \cite{N54hall2006extremes} analysed the extremal properties of $T$ periodic integer-valued sequences, where the term periodic means that for each choice of integers $i_1 <i_2 <\ldots<i_n$, $\Pr \left(X_{i_1},\ldots,X_{i_n}\right)=\Pr \left(X_{i_1+T},\ldots,X_{i_n+T}\right)$, and extended it 
\cite{N81hall2012maximum} to the class of max-semistable law.
Another family of processes that was studies is called \emph{triangular array}.
In this family, one usually assumes that one or more of the parameters are a function of $n$.
Anderson \textit{et al}.\ \cite{N56anderson1997maxima} studied the limit behaviour for i.i.d.\ Poisson distribution under sufficient conditions, where the mean $\lambda=\lambda_n$.
Nadarajah \cite{N36nadarajah2002asymptotics} extended it to other families of distributions, including NB, where $p=p_n$.

To the best of our knowledge, the study of NSIV sequences is still missing.
In this work, to better analyse the completion time, we derive non-trivial extensions to some of the results above, and give an EVT theorem for such processes.
It is important to note that all the articles mentioned above assume some dependence restrictions, (such as Leadbetter's condition $D\left(u_n \right)$),
  and also local dependence conditions (like $D{'}\left(u_n \right)$ and $D^{''}\left(u_n \right)$), in order to establish their results.
We adopt similar techniques to obtain asymptotic results of the maximum of NSIV sequences. 
\section{Problem Formulation}\label{SystemModelSection}
In this work, we study the limiting behaviour of the completion time when disseminating a file consisting of $M$ packets to all receivers, where all the receivers are connected via a multicast tree.

In particular, assume that the source wishes to disseminate a file consisting of $M$ packets over a multicast network via a full K-ary tree $\mathbf{T}_{n}=\left(V,E \right)$ with $n$ leaves ($|V|=\frac{Kn-1}{K-1}$,$|E|=\frac{K\left(n-1 \right)}{K-1}$), where each $e \in E$  models a point to point channel between a parent and its child, experiencing a packet loss probability $p$, independent between the channels.

We also assume an ACK/NACK mechanism, such that it is sent instantly after each time slot, on a lossless channel (i.e., the packet loss probability for it is zero), with negligible transmission time.
The time axis is slotted, and each packet transmission takes one time slot.
Each node (which is not the source or the end user) waits to receive $M$ different packets, and only afterwards switches to a transmitter mode (assuming half duplex), to send them to all its children.
The node continuous to transmit until all child nodes get the entire file (i.e., $M$ packets), and the process is terminated when all end users have one copy of the file. 
In order to streamline the process, we assume that the nodes use a \emph{rateless code}, similar to \cite{N19}, i.e., each node needs to correctly receive $M$ distinct packets from the infinite stream sent by the parent node in order to recover a file.
Once a certain node receives $M$ different packages, it first decodes it to the original file, and then re-encodes it into a potentially infinite stream, and sends it continuously to all its children, until each child node accumulates $M$ different packages.
The process terminates when all nodes and specifically all leaves can  successful decode the file.
Note that when the file includes only one packet, there is no need to use rateless codes.

Let $T_i$ be the number of time slots needed for leaf $i$ in order to receive $M$ packets.
Since each packet has to traverse $\log_K n$ links before reaching each leaf node, then for $M$ packets, one need $M \log_K n$ successes, one after the other.
Accordingly, the distribution of $T_i$ follows NB with $M \log_K n$ successes with a probability of failure $p$, i.e. $T_i \sim NB(\log_2 n,p)$.

We denote by $M_n$ the completion time to successfully disseminate  the packet to all nodes, i.e., $M_n=\max \left(T_1,\ldots,T_n \right)$.

Our goal is to study $\mathbb{E}\left[M_n \right]$, namely the expected number of time slots that are needed until the packet is sent to all nodes on the tree.
We focus on the case where $n \to \infty$, which practically means that the results hold for sufficiently large $n$.
As mentioned above, we start with the case of $M=1$, and $K=2$, i.e., full binary tree, and in \Cref{sec:ExtentionMmessages,sec:ExtentionTreeWithKChild} we will extend it to general $K$ and $M$, including large $K$ and/or $M$.
\section{Preliminaries}\label{PreliminariesSection}
In this section, we review the relevant studies in EVT, including relevant methods and conditions.
We also characterise properties of the NB distribution, which, as noted earlier, is the marginal distribution of the sequence $\left( T_1,\ldots,T_n \right)$.

Throughout the work, $M_n$ will refer to the maximum of $n$ random variables, where the actual random variables will be clear from the context.
\subsection{Extreme Value Theory}
\label{sec:EVT}
As mentioned above, the main tool to analyse the problem at hand is EVT, which concerns the asymptotic distribution of the maxima of random variables.
EVT is well studied in the literature for a wide variety of stochastic processes.
Here, we briefly review the results of the classic EVT, namely, specify the possible forms of the limit distribution of the maxima in a sequence of continuous valued i.i.d.\ random variables.
Note that as the problem in this work involves \emph{NSIV} sequences, the results below will not be applicable.
Hence, a large part of this work (\Cref{sec:EVTNonStationaty}) is devoted to deriving the needed results.
Yet, we include the results below since they serve as a basis and reference for the rest if the work.
\begin{theorem}\cite[Theorem 1.5.1]{N28leadbetter1983extremes}
\label{theorem:ClassEVT}
Let $\left( X_1,\ldots,X_n \right)$ be an i.i.d.\ sequence, and $M_n=\max  \left( X_1,\ldots,X_n \right)$.
Let $0 \leq \tau \leq \infty$ and suppose that $\lbrace u_n \rbrace $ is a sequence of real numbers such that
\begin{align}\label{eq:limittau}
n \left(1-F \left( u_n \right) \right) \to \tau, \quad as \; n \to \infty.
\end{align}
Then,
\begin{align}\label{eq:limitDistibution}
\Pr \left( M_n \leq u_n \right) \to e^{-\tau}.
\end{align}
\end{theorem}
Note that if $u_n$ is of the linear form of $u_n=x+b_n$, namely, $u_n$ is a shift of $x$ by $b_n$, then $\tau=\tau(x)$.
Furthermore, if $\tau$ has the form $\tau=a^x$, $a \in \mathbb{R}$, then the distribution of $M_n$ converges to a \emph{Gumbel distribution}, with mean $\mu=-\frac{\gamma}{\log a}$, where $\gamma \approx 0.57721$ is the Euler-Mascheroni constant, and $\log$ refers to the natural logarithm (base $e$).
Under mild technical assumptions \cite{BookResnick2007extreme}, the convergence in distribution implies also moment convergence.
Hence
\begin{align}\label{eq:GumbelExpectedValue}
\lim_{n \to \infty} \mathbb{E} \left[  M_n \right]-b_n=\mu.
\end{align}
The reasoning behind \Cref{theorem:ClassEVT} is simple: The continuity of $F$ enables us to assume that the limit in \cref{eq:limittau} exists.
Hence, the probability of one random variable being less than $u_n$ is approximately $1-\frac{\tau}{n}$.
Using the assumption of i.i.d.\ random variables we get that the probability of $n$ random variables being less than $u_n$ is approximately $\left(1-\frac{\tau}{n} \right)^n$.
Taking the limit $n \to \infty$ results in \cref{eq:limitDistibution}.
However, as pointed out above, the sequence $\left(T_1,\ldots,T_n \right)$ is neither independent nor continuous valued.
The assumption of continuous values is essential because the limit in \cref{eq:limittau} may not converge if $F$ is the distribution of a discrete random variable.
Therefore, we say that $F$ does not belong to the \emph{domain of attraction} of $G$, where $G(x)$ is the non degenerate limit CDF.
Anderson \cite{N31anderson1970extreme} was the first to introduce the study of extremal behaviour of some \emph{i.i.d.\ integer-valued distributions}, by defining a class of distributions such that the limit behaviour is bounded, under suitable conditions, by non degenerate limit distributions.
Specifically, the main idea is to define a continues function $F_c(x)$, such that
$F_c(x-1) \leq F[x] \leq F_c(x)$, for all $x$, where both $F_c(x-1)$ and $F_c(x)$ satisfy \cref{eq:limittau} for some $u_n$.
This allows us to derive bounds on the limit distribution of the maximum of some integer-valued processes.
Further details, including how to find $F_c \left(x \right)$, will be presented in the next section.

The next challenge to be dealt with is that the sequence $ \left( T_1,\ldots,T_n \right)$ is correlated.
Several works have considered this topic, for stationary processes (discrete and continuous), for continuous non-stationary processes, and also for random fields.
All were inspired by the works of Leadbetter  in \cite{N9,N10}.
In his papers, Leadbetter gave two main conditions.
The first is a weak ``mixing condition", restricting long range dependence, called \emph{condition} $D \left(u_n \right)$.
There are several versions of this condition, suitable for a variety of stochastic processes.
In our case, since we are dealing with a non-stationary sequence, we present the version of  \cite{N41husler1986extreme} (see also\cite{N42husler1983mptotic}), who studied the extreme value of a non-stationary continuous valued sequence, with the assumption of non-identical marginal distribution $F=F_i$, and boundary level $u_n=u_{n,i}$.  
\begin{condition}{D \left(u_n \right)}
\label{eq:ConditionDun}
The condition $D \left(u_n \right)$ will be said to hold for the sequence $\left(X_1,\ldots,X_n \right)$ and $\lbrace u_n \rbrace$, if for any integers
\begin{align}
1 \leq i_1 < \cdots <i_p <j_1 < \ldots <j_{q} \leq n,
\end{align}
for which $j_1-i_p \geq l$,  $I=\lbrace i_l,l=1,\cdots,p \rbrace$, $J=\lbrace j_l,l=1,\cdots,q \rbrace$, $M \left(I \right)= \lbrace X_i \leq u_n, i \in I \rbrace$ and similarly $M \left(J \right)$,
We have
\begin{align}
\sup_{I,J} \left| \Pr \left(M \left(I \cap J \right) \right) -\Pr \left(M \left(I \right) \right) \Pr \left( M \left( J \right) \right)\right| \leq \alpha_{n,l},
\end{align}
where $\alpha_{n,l_n} \to 0$ as $n \to \infty$ for some sequence $l_n=o \left( n \right)$.
\end{condition}
Note that condition $D \left(u_n \right)$ is significantly weaker than strong mixing, as it considers only \emph{a certain sequence  $u_n$}.
Generally speaking, it ensures that the random variables whose indices are sufficiently separated, are asymptotically independent.

Denote 
\begin{align}
F_{n}^{*}= \sum_{i=1}^{n} \bar{F}_i \left(u_{n,i} \right),
\end{align}
and split the set $\lbrace 1,\ldots,n \rbrace$ into sets of consecutive integers $I_1,I_2,\ldots,I_r$, in the following way:
Let $I_1=\lbrace 1,\ldots,i_1 \rbrace$ be such that
\begin{align}
\sum_{i=1}^{i_1} \bar{F}_i \left(u_{n,i} \right) \leq F_n/r,
\end{align}
and
\begin{align}
\sum_{i=1}^{i_1+1} \bar{F}_i \left(u_{n,i} \right) > F_n/r.
\end{align}
Namely, $i_1$ is chosen as large as possible such that the sum is still smaller that $F_n/r$.
In the same way, $I_2= \lbrace i_{1}+1,\ldots,i_2 \rbrace$, such that
\begin{align}
\sum_{i=i_1+1}^{i_2} \bar{F_i} \left(u_{n,i} \right) \leq F_n/r,
\end{align}
and so on. 
As a result, we get $r$ sets of consecutive integers, such that $i_r \leq n$.
Note that if we assume identical marginal distribution $F$, and identical boundary level $u_n$, for all $i$, we have $|I_1|=|I_2|=\ldots=|I_r|$.
Then it was proved in \cite[Lemma 2.6]{N41husler1986extreme} that if $D \left(u_n \right)$ holds and
\begin{align}
\limsup_{n} F_{n}^{*} \leq \infty \text{ and } \liminf_{n} F_{n}^{*} \geq 0,
\end{align}
then
\begin{align}\label{eq:asymtoticIndependent}
\limsup_{n \to \infty} \left| \Pr \left( M_n \leq u_n \right) -\prod_{l=1}^r  \Pr \left(M \left(I_{l} \right) \right) \right| =0,
\end{align}
for $r$ fixed, and the sets $\lbrace I_l \rbrace$ defined as above.

The second condition is a local condition, restricting the possible of clustering of high level exceedances, and it has various forms.
For example, Leadbetter \cite{N9}  refereed to it as condition $D^{'} \left(u_n \right)$.
\begin{condition}{D^{'} \left(u_n \right)}
The condition $D^{'} \left(u_n \right)$ will be said to hold for the stationary sequence $\left(X_1,\ldots,X_n \right)$ and the sequence $\lbrace u_n \rbrace$ if
\begin{align}\label{eq:ConditionDtag}
\limsup_{n \to \infty} n \sum_{j=2}^{\lfloor n/k \rfloor} \Pr \left(X_1>u_n,X_j>u_n \right) \to 0, \quad as \; k\to \infty,
\end{align}
where $\lfloor * \rfloor$ denotes the integer part.
\end{condition}
Condition $D^{'} \left(u_n \right)$ limits the probability of more than one exceedance of the level $u_n$ among subset of $\lbrace X_1,\ldots,X_{[n/k]} \rbrace$, and it has a critical role in finding the extreme value distribution of stationary sequences whose local dependence is not ``too strong".

It was shown in \cite{N9} that if both conditions $D \left(u_n \right)$ and $D^{'} \left(u_n \right)$ hold for the stationary sequence $\left( X_1,\ldots,X_n \right)$, then the limit  distribution of the normalized $M_n$ behaves the same as if the sequence above were i.i.d., under the same normalizing constants.	
Yet, our sequence is non-stationary (and also integer-valued).
Hence, in \Cref{sec:EVTNonStationaty}, we formulate a suitable condition, such that together with $D \left(u_n \right)$, it enables us to  asymptotically bound the distribution of the maximum of NSIV sequence.
\subsection{The Negative Binomial Distribution}
Let $Y$ be a random variable representing the number of Bernoulli trials needed to get one success, for a probability of failure $p$.
Therefore,
\begin{align}
\Pr \left(Y=k\right)=p^{k-1}(1-p), \quad k=1,2,\ldots
\end{align}
$Y$ has a geometric distribution with parameter $p$, i.e. $Y  \sim \operatorname{G}_0 \left({p}\right)$.
Now, denote the random variable $X=\sum_{i=1}^m Y_i$, i.e., $X$ is a random variable representing the number of Bernoulli trials until the $m$'th
 success occurs, where $m$ is a fixed integer.
$X$ has a \emph{NB Distribution} with parameters $m$ and $p$, i.e., $X \sim NB(m,p)$.\\
For a random variable $X \sim NB(m,p)$, denote the Probability Density Function (PDF) $f[x|m,p]=\Pr \left(X=x \right)$, where  (\cite{N44fogiel1984handbook,N45feller2008introduction})
\begin{align}\label{eq:NBPDF}
f[x|m,p]=\binom{x-1}{m-1}p^{x-m}(1-p)^m \quad x=m,m+1,\ldots
\end{align}
and the expected value $\mathbb{E}\left[X \right]$ is given by
\begin{align}\label{ExpectedvalueNegativeBinomial}
\mathbb{E}\left[X\right]=\frac{m}{1-p}.
\end{align}
Let  $F[x|m,p]$ be the Cumulative Distribution Function (CDF), i.e., $F[x|m,p]=\Pr \left(X \leq x \right):$
\begin{align}\label{CDFNegativeBinomial}
F[x|m,p]=I_{1-p}(m,x-m+1),
\end{align}
where $I_{x}(a,b)$ is the regularized beta function, that is,
\begin{align}
I_{x}(a,b)
&=\frac{B(x;a,b)}{B(a,b)}
\\ \nonumber
&=1-\frac{B(1-x;b,a)}{B(a,b)},
\end{align}
and $B(x;a,b)$ is the incomplete beta function, while $B(a,b)$ is the complete beta function, i.e.,
\begin{align}
&B(x;a,b)=\int_{0}^x t^{a-1}(1-t)^{b-1} dt,
\\ \nonumber
&B(a,b)=\int_{0}^1 t^{a-1}(1-t)^{b-1} dt.
\end{align}
$B(a,b)$ has several other forms, including
\begin{align}\label{betaComplete}
B(a,b) & =\frac{\Gamma(a)\Gamma(b)}{\Gamma(a+b)},
\end{align}
where $\Gamma(x)$ is the Gamma function, which follows the recurrence relation
\begin{align}\label{eq:GammaRecurrenceRelation}
\Gamma \left(x+1 \right)=x  \Gamma \left(x \right).
\end{align}
Furthermore, if $x$ is a positive integer, then $\Gamma(x)=\left(x-1 \right)!$.\\
Also, $B(x;a,b)$ has other forms, including
\begin{align}\label{betaInComplete1}
B(x;a,b)=x^a \sum_{i=0}^{b-1}(-1)^i \binom{b-1}{i}\frac{x^i}{a+i}
\end{align}
and (see \cite[pp.994]{N46abramowitz1972handbook})
\begin{align}\label{betaInComplete2}
 B(x;a,b)=\frac{x^a(1-x)^b}{a} \mathop{F\/}\nolimits\!\left(a+b,1;a+1;x\right).
\end{align}
The hypergeometric function $\mathop{F\/}\nolimits\!\left(a,b;c;z\right)$ is defined by the Gauss series
\begin{align}\label{eq:gaussianSeries}
\mathop{F\/}\nolimits\!\left(a,b;c;z\right)=1+\frac{ab}{c}z+ \frac{a \left(a+1\right)b\left( b+1 \right)}{c \left(c+1 \right)2!}z^2+\ldots
\end{align}
\section{New Extreme Value Results for Non-Stationary Integer-Valued Sequences}\label{sec:EVTNonStationaty}
In this section, we study the limiting distribution of the maximum term of NSIV, in order to apply it in the next section to obtain bounds on $M_n$.
As mentioned earlier, the results of the classic EVT mainly concern continuous random variables.
Anderson \cite{N31anderson1970extreme} was the first to introduce the study of extremal behaviour of some i.i.d.\ integer-valued distributions, by defining a class of distributions such that the limiting behaviour is bounded, under suitable conditions, by a non-degenerate limit distribution. 
Based on Andeson's work, several stationary integer-valued models have been proposed 
\cite{N49serfozo1988extreme,N48serfozo1987extreme,N50hall2001extremes,N51hall2003extremes,N53hall1996maximum,N30mccormick1992asymptotic}, however, recall that in our the case, the sequence is not stationary.
Hence, in this section, we derive the limit distribution of a maxima of a NSIV sequences, under some restricting conditions.

Following Anderson's work \cite{N31anderson1970extreme}, define a continuous distribution function $F_c$ as follows:
For $F \in \mathbb{D}$ (namely, $F$ belongs to the domain of attraction of one of the nondegenarate limit distributions), and for each positive integer $n$, let
\begin{equation}
h[n]=-\log(1-F[n]).
\end{equation}
Now, define $h_c$ to be the extension of $h$ by linear interpolation, i.e., for $x \geq 1$
\begin{equation}
h_c(x)=h[\lfloor x\rfloor ] +(x-\lfloor x\rfloor)(h[\lfloor x+1\rfloor ]-h[\lfloor x\rfloor ],
\end{equation}
where $\lfloor x \rfloor$ is the integer part of $x$.
Then, set a continuous CDF, $F_c(x)=1-e^{-h_c(x)}$.
Note that $F_c(x-1) < F[x] \leq F_c(x)$, and for any integer $n$, $F_c(n)=F[n]$.
Now, the choice of $b_n$ should be such that
\begin{align}\label{eq:choiceOfBn}
\bar{F}_c(b_n)=\frac{1}{n},
\end{align}
where $\bar{F}$ denotes $1-F$.
As mentioned above, in order to establish the extreme value result, one needs to show that both the long range dependence condition (mixing) and  the local dependence condition hold.\\
Here We will use the technique developed in \Cref{PreliminariesSection}, and split $\lbrace 1,\ldots,n \rbrace $ into set of consecutive integers $I_1,\ldots,I_{r_n}$ where
\begin{align}\label{eq:PartionOfn}
&I_{1}=\lbrace 1,2,\ldots,k_n \rbrace
   \\ \nonumber &
 I_{2}=\lbrace k_n+1,\ldots,2k_n \rbrace 
    \\ \nonumber &
    \vdots
    \\ \nonumber &    
 I_{r_n}=\lbrace (r_n-1)k_n+1,\ldots,r_n k_n \rbrace.
\end{align}
Namely, each set includes exactly $k_n$ indices, thus $r_n=\lfloor \frac{n}{k_n} \rfloor$.
Regarding the long range dependence, we will use condition $D \left(u_n \right)$ as presented in \Cref{eq:ConditionDun}, which implies that we can assume that \cref{eq:asymtoticIndependent} holds.
However, here, the relation between $n$ and $r$ has to satisfy certain conditions, hence, we will obtain a somewhat more general version of \cref{eq:asymtoticIndependent} result.
The next lemma proves that if condition $D \left(u_n \right)$ holds, for NSIV sequences, then \cref{eq:asymtoticIndependent} holds, yet $r$ may also be increasing sequence $r_n$.
\begin{lemma}
\label{lemma:AsymtoticIndependet}
Assume $D \left( u_n \right)$ is satisfied by the NSIV sequence $ \left(X_1,\ldots,X_n \right)$, with identical marginal distribution, and for some identical boundary level $u_n$ such that
\begin{align}\label{eq:limittausup}
\limsup_{n \to \infty} n \Pr \left( X>u_n \right) = \tau.
\end{align} 
Let $r_n$ be a sequence satisfying
\begin{align}\label{eq:ConditionOnRforaasymtoticIndependent}
r_n\alpha_{n,l_n} \to 0 \text{ and }r_n\frac{l_n}{n} \to 0,
\end{align}
as $n \to \infty$, where $l_n, \alpha_{n,l_n}$ are given in condition $D \left( u_n \right)$.
Then, \cref{eq:asymtoticIndependent} holds for $r=r_n$.
\begin{proof}
See Appendix \ref{appendix:ProofAsymtoticIndependet}.
\end{proof}
\end{lemma} 
For the local dependence restriction, we now present condition $D_{k_n}^{'}\left(u_n \right)$ as follows.
\begin{condition}{D_{k_n}^{'}\left(u_n \right)}\label{ConditionD_k_n}
Let $r_n= \lfloor n/k_n \rfloor$ satisfy \cref{eq:ConditionOnRforaasymtoticIndependent}, and let $I_1,\ldots,I_{r_n}$ be disjoint sets of $\lbrace 1,\ldots,n \rbrace$, as defined in \cref{eq:PartionOfn}.
Condition $D_{k_n}^{'}\left(u_n \right)$ will be said to hold for a non-stationary sequence $\left(X_1,\ldots,X_n \right)$ and sequence $\lbrace u_n \rbrace$ if
\begin{align}
\max_{\substack{I_l:  \\ 1 \leq l \leq r_n}} \sum_{i<j \in I_l}\Pr \left( X_i>u_n,X_j>u_n \right) \leq \alpha^{'}_{n,r_n},
\end{align}
where
\begin{align}
 \limsup_{n \to \infty} r_n \alpha^{'}_{n,r_n}=0.
\end{align}
\end{condition}
Note that $D_{k_n}^{'}\left(u_n \right)$ is stronger, yet simpler, compared to the condition $D^{'}$ presented in \cite{N41husler1986extreme} and \cite{N42husler1983mptotic}, in the sense that $D_{k_n}^{'}\left(u_n \right)$ herein is harder to achieve, yet easier to check.
In \cite{N41husler1986extreme,N42husler1983mptotic}, the condition $D^{'}$ allows to omit a negligible number of elements from the set $I_l$.
More detailed explanations will be given after the proof of the next theorem.

In the next theorem, which is the main result on this section, we prove the existence of a limit distribution of the NSIV sequence  $\left(X_1,\ldots,X_n \right)$, under conditions $D \left(u_n \right)$ and $D_{k_n}^{'}\left(u_n \right)$.
\begin{theorem}\label{EVT_for_nonstationary_integer_valued}
For the NSIV sequence $\left(X_1,\ldots,X_n \right)$ with identical marginal distribution function $F[x]$, let $M_n=\max \left(X_1,\ldots,X_n \right)$.
Then, if conditions $D \left(u_n \right)$ and $D_{k_n}^{'}\left(u_n \right)$ hold for $u_n=x+b_n$ satisfying
\begin{align}\label{FirstCondition}
& \lim_{n \to \infty} n\bar{F}_c(u_n) \to \tau
 \\ \nonumber
 & \lim_{n \to \infty} n\bar{F}_c(u_n-1) \to \tau^{'},
\end{align}
for $0 < \tau <\tau^{'} < \infty$, then
\begin{align}\label{GemeralExtremeValueDiscrete}
&\limsup_{n \to \infty} \Pr \left( M_n \leq u_n \right) \leq e^{-\tau},
 \\ \nonumber &
\liminf_{n \to \infty} \Pr \left( M_n \leq u_n \right) \geq e^{-\tau^{'}}.
\end{align}
\begin{proof}
The proof is based on the proof in \cite[Theorem 3.4.1]{N28leadbetter1983extremes}, yet note that the latter was proved for \emph{stationary} sequences, with a \emph{continuous marginal} $F$.
Here, we extended it to a discrete marginal $F$, by bounding it with a continuous marginal $F_c$, and, also, generalize it to non-stationary sequences.\\
First, for the sets $I_1,\ldots,I_{r_n}$ defined above, recall that
\begin{align}
M \left(I_l \right)= \lbrace X_i \leq u_n, \forall i \in I_l \rbrace, \quad l=1,\ldots,r_n.
\end{align}
Namely, the event $M \left(I_l \right)$ implies that the maximum from the set $I_l$ is less than $u_n$.
Now, let $M^C \left(I_l \right)$ be the complementary event, hence
\begin{align}
M^C \left(I_l \right)=\bigcup_{j \in I_l} \lbrace X_j>u_n \rbrace.
\end{align}
Thus, by the Bonferroni inequality, we have
\begin{align}
 \sum_{j \in I_l} \Pr \left( X_j>u_n \right)-\sum_{ i < j \in I_l} \Pr \left( X_i>u_n,X_j>u_n \right) 
\leq \Pr \left(M^C \left(I_l \right) \right) \leq \sum_{j \in I_l} \Pr \left( X_j>u_n \right).
\end{align}
Writing $S_{l}=\sum_{ i < j \in I_l} \Pr \left( X_i>u_n,X_j>u_n \right) $, and noting that $\Pr \left( X_j>u_n \right)$ is identical for each $j$,
we have
\begin{align}
1-k_n  \Pr \left( X_j>u_n \right)  \leq \Pr  \left( M \left(I_l \right) \right) 
\leq 1-k_n  \Pr \left( X_j>u_n \right)+S_{l}.
\end{align}
Since
\begin{align}
\bar{F}_c \left(u_n \right) \leq \Pr \left( X_j>u_n \right)\leq \bar{F}_c \left(u_n-1 \right),
\end{align}
then
\begin{align}\label{237}
 1-k_n  \bar{F}_c \left(u_n-1  \right)  \leq Pr \left( M \left(I_l \right) \right) 
 \leq 1-k_n  \bar{F}_c \left(u_n  \right)+S_{l}.
\end{align}
Also note that by condition $D^{'}_{k_n}\left(u_n \right)$, $S_{l}=o\left(\frac{1}{r_n} \right)$ for each $l$.
Thus, we can write \cref{237} as follows
\begin{align}\label{eq:forEachl}
 1-\frac{n \bar{F}_c \left(u_n-1\right)}{n/k_n} \leq Pr \left( M \left(I_l \right) \right)
\leq 1-\frac{n \bar{F}_c \left(u_n \right)}{n/k_n}+o \left(\frac{1}{r_n} \right).
\end{align}
Multiply \cref{eq:forEachl} over each $I_l$.
Since we assume that $D \left(u_n \right)$ holds, we can use \cref{eq:asymtoticIndependent}, i.e.,
\begin{align}
 \prod_{l=1}^{r_n}Pr \left( M \left(I_l \right) \right)=\Pr \left(M_n \leq u_n \right) +o(1),
\end{align}
hence, we have
\begin{align}\label{Eq31}
\left( 1-\frac{n  \bar{F}_c \left(u_n-1 \right)}{n/k_n} \right)^{r_n}
\leq \Pr \left(M_{n} \leq u_n \right) +o(1)
\leq \left(1-\frac{n \bar{F}_c \left(u_n \right)}{n/k_n}+o \left(\frac{1}{r_n} \right) \right)^{r_n},
\end{align}
and recall that $r_n=\lfloor \frac{n}{k_n} \rfloor$.
Now, in order to complete the proof, we use the following claim.
\begin{claim}\label{claim1}
Let $a_m$ be a sequence such that $a_m \to a$, as $m \to \infty$.
Then
\begin{align}
\lim_{m \to \infty} \left( 1+\frac{a_m}{m} \right)^m =e^a.
\end{align}
We give the justification with more details in Appendix \ref{Appendix:LimitTwiceExplanation}.
\end{claim}
Using the claim with $m=\frac{n}{k_n}$, once with $a_m=n  \bar{F}_c \left(u_n-1 \right)$, and once with $a_m=n\bar{F}_c \left(u_n \right)+o(1)$, taking $n \to \infty$, we have
\begin{align}\label{Eq32}
 e^{-\tau^{'}} \leq \liminf \Pr \left( M_n \leq u_n \right) \leq \limsup \Pr \left( M_n \leq u_n \right) \leq e^{-\tau}.
\end{align}
Which completes the proof.
\end{proof}
\end{theorem}
We may deduce at once from \Cref{EVT_for_nonstationary_integer_valued} that under conditions $D \left(u_n \right)$ and $D_{k_n}^{'}\left(u_n \right)$, the limiting distribution of $M_n$ is the same as that which would apply if $\left(X_1,\ldots,X_n \right)$  were i.i.d.\ 

Observe also that \Cref{EVT_for_nonstationary_integer_valued} does not apply to all non-stationary sequences, since we assume identical marginal distributions.
Moreover, as noted before, condition $D_{k_n}^{'}\left(u_n \right)$ may be too strict, such that there might be a sequence that fails to satisfy this condition, yet the limit distribution of the maximum is precisely the same (including the same normalizing constants), as if the sequence were i.i.d.
For example: consider the sequence $\left( X_1,\ldots,X_n \right)$, where in one of the sets $I_l$, there are $\sqrt{k_n}$ elements (out of $k_n$ elements) who are identical (i.e., $X_1=X_2=\cdots=X_{\sqrt{k_n}}$), while the rest are i.i.d.
Then, one might check that $D_{k_n}^{'}\left(u_n \right)$ fails, yet we have $n-\sqrt{k_n} \approx n$ independent elements, which implies that it is possible to obtain the extreme value result as in \Cref{EVT_for_nonstationary_integer_valued}.
Hall \cite{N41husler1986extreme,N42husler1983mptotic} addressed this issue, and formulated a mild condition, such that it removes "unimportant" elements from the set $I_l$.\\
Another possible extension, similar to the stationary case \cite{N39leadbetter1989exceedance,N38chernick1991calculating}, is to study extreme value of processes who have a non negligible local dependence.

In the next section, we apply our result to obtain bounds on the distribution of $M_n$ in our problem.
\section{Asymptotic Analysis of the Completion Time}
\label{MainResultSection}
As we can see from \Cref{EVT_for_nonstationary_integer_valued}, $D\left(u_n \right)$ is essential in order to characterise the asymptotic distribution of $M_n$.
In our case, the correlation between two sets of random variables on the sequence $\left( T_1,\ldots,T_n \right)$ is expressed by their mutual channels.
For instance, the pair $ \left( T_1,T_{m} \right)$ share in common $\log_2 n- \lceil \log_2 m \rceil$ ($\log_2 \frac{n}{m}$ if $m$ is a power of 2) channels.
Therefore, in most cases, the number of mutual channels increases with $n$.
As a result, the dependence between two sets is not negligible, even for large $n$, and consequently, $D\left(u_n \right)$ \emph{does not hold}.
Nevertheless, we derive upper and lower bounds for the expected completion time, i.e., $\mathbb{E} \left[ M_n \right]$, where the lower bound is achieved by modifying the topology of the tree, such that we can analyse it using the tools we derived ,and the upper bound is achieved by considering independent channels.
\subsection{Lower bound}
\label{sec:lowerBound}
\begin{figure}[t] 
\centering   
  \includegraphics[width=0.7\textwidth]{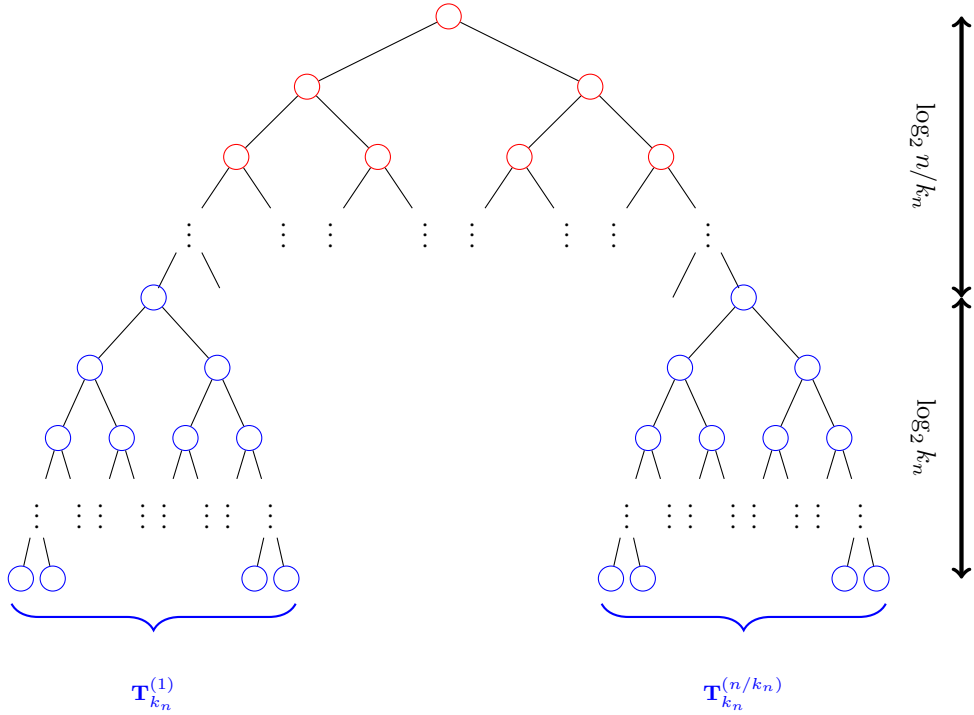}
\caption{Illustration of the partition of $\mathbf{T}_n$. The red part is one tree $\mathbf{T}_{\frac{n}{k_n}}$, and the blue part includes $\frac{n}{k_n}$ independent replications of $\mathbf{T}_{k_n}$}
  \label{fig:BinaryTree}
\end{figure}

In order to derive a lower bound, we split $\mathbf{T}_n$ vertically into two parts at level $\log_2 k_n$, where $\log_2 k_n \in 1,2,\ldots,\log_2 n-1$, such that the first part is a binary dissemination tree, $\mathbf{T}_{\frac{n}{k_n}}$, and the second part includes $\frac{n}{k_n}$ replications of sub-trees $\mathbf{T}_{k_n}$, denoted by $\mathbf{T}_{k_n}^{(1)},\mathbf{T}_{k_n}^{(2)},\ldots,\mathbf{T}_{k_n}^{(n/k_n)}$ (see \cref{fig:BinaryTree}). 
\begin{figure}[t]
    \centering
    \includegraphics[width=0.8\textwidth]{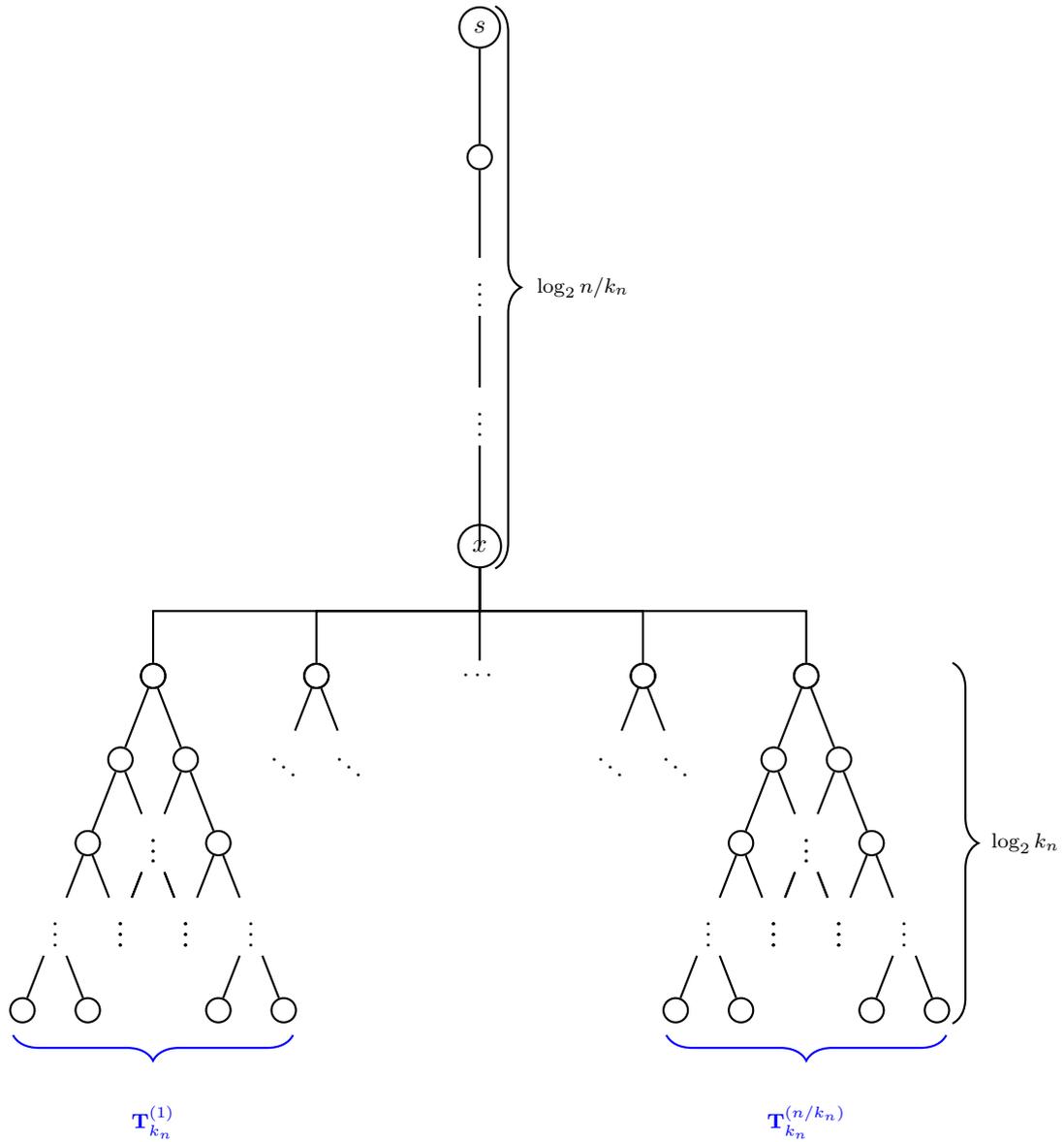}
\caption{Illustration of the construction of the lower bound.
Here we can see that a message should pass through $\log_2 n/k_n$ nodes, represented by a random variable $W$ until it gets to node $X$, and afterwards, the message is disseminated to each one of the subtrees $\mathbf{T}_{k_n}^{(1)},\mathbf{T}_{k_n}^{(2)},\ldots,\mathbf{T}_{k_n}^{(n/k_n)}$.}
\label{fig:LowerBinaryTree}
\end{figure}

Let $Y_j^{(i)}$ be a random variable representing the number of time slots it takes to forward one packet from the root of the subtree $\mathbf{T}_{k_n}^{(i)}$ to the leaf $j$, $1\leq j \leq k_n$.
In addition let $Y^{(i)}= \max \left( Y_1^{(i)},Y_2^{(i)},\ldots,Y_{k_n}^{(i)} \right)$, thus, $Y^{(i)}$ represents the completion time of the subtree $\mathbf{T}_{k_n}^{(i)}$.
Clearly, the number of time slots it take to forward one packet from the root of the tree $\mathbf{T}_n$ to one of the end-user is the time it takes to forward the packet to the root of corresponding subtree $\mathbf{T}_{k_n}^{(i)}$, plus the time it takes to forward it from the root of $\mathbf{T}_{k_n}^{(i)}$ to the leaf.
Formally, if we denote by $W_i$, $1 \leq i \leq n/k_n$, the number of time slots needed to forward the packet to some intermediate node in $\mathbf{T}_n$ at level $\log_2 k_n$, then the completion time needed to disseminate a message to all nodes\ of $\mathbf{T}_n$ can be represent as
\begin{align}\label{eq:MnOriginal}
M_n=\max_i \left(W_i+Y^{(i)}; 1\leq i \leq n/k_n \right).
\end{align}
We construct the lower bound as follows:
First, we will calculate the distribution of $Y=\max \left(  Y^{(1)}, Y^{(2)},\ldots, Y^{(n/k_n)} \right)$, by using EVT, which enables us to find the expected value of it.
Then, in the next stage, we will add the expected value of $W$, where $W$ is a random variable, which represents the time slots it takes to forward the message from the root to a node $X$  (see \cref{fig:LowerBinaryTree}), which is located at distance of  $\log_2 n/k_n$ hops from it, thus $W \sim NB \left(\log_2 n/k_n,p \right)$.
In fact, the only change that has been made is that we unite all the paths with length of $n/k_n$ edges, starting from the root, into a single path, from $S$ to $x$.
This leads to\emph{ decrease in the channels diversity}, or in other words, increase the correlation of $\left(T_1,\ldots,T_n \right)$, which may give an insight about the acceptability of the lower bound. 

Our first result, which is a basic lower bound,  follows from the observation that for each realization, we have $T_i \leq \max \left(T_1,\ldots,T_n \right)$, for each $i$.
Hence, we obtain a trivial lower bound on $\mathbb{E} \left[ M_n \right]$, by just considering the expected value of a single NB random variable.
Therefore, by \cref{ExpectedvalueNegativeBinomial}, we have the following claim.
\begin{claim}
\begin{align}\label{eq:TrivialLowerBound}
\mathbb{E} \left[ M_n \right] \geq \frac{\log_2 n}{1-p}.
\end{align}
However, as we will see in the simulation results, this bound can be quite loose, and tighter bounds are necessary.
\end{claim}
Formally, define $M_n^l=W+Y$.
Next, we show that $\mathbb{E} \left[ M_n \right] \geq \mathbb{E} \left[ M_n^l \right]$, and afterwards, we obtain an asymptotic expression of $\mathbb{E} \left[ M_n^l \right]$.
\begin{lemma}\label{ProveLowerBound}
\begin{equation}
\mathbb{E} \left[ M_n \right] \geq \mathbb{E} \left[ M_n^l \right].
\end{equation}
\begin{proof}
Denote $i^{*}=\argmax_{i} \left(  Y^{(i)}; 1 \leq i \leq n/k_n \right)$, then
\begin{align}
\mathbb{E} \left[ M_n \right]&=
\mathbb{E} \left[ \max_i \left(W_i+Y^{(i)} \right) \right]
 \\ \nonumber &
 \geq 
 \mathbb{E} \left[ W_{i^{*}}+Y^{(i^{*})} \right]
  \\ \nonumber &
 =
  \mathbb{E} \left[ W + Y \right]
    \\ \nonumber &
 = \mathbb{E} \left[ M_n^l \right],
\end{align}
where the inequality results from the fact that $ \max_i \left( W_i+Y^{(i)} \right) \geq W_j+Y^{(j)}$ for each $j$, including $j=i^{*}$.
\end{proof}
\end{lemma}
\begin{figure}[t]
\centering
  \includegraphics[width=0.6\textwidth]{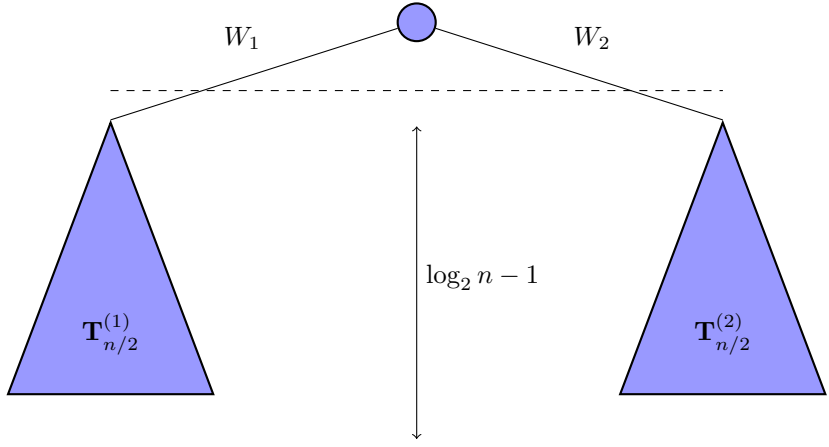}
\caption{Example of the partition of $\mathbf{T}_{n}$ into two identical subtrees, $\mathbf{T}_{n/2}^{(1)}$ and $\mathbf{T}_{n/2}^{(2)}$, which are connected via a single channel, $W_1$ and $W_2$, respectively.}
\label{fig:Nolabel2}
\end{figure}
An intimidate conclusion from the proof of \Cref{ProveLowerBound} is that the bound would be tight if $i^{*}=\argmax_i \left(W_i+Y^{(i)} \right)$. In other words, the end user that was the maximum from all subtrees $\mathbf{T}_{k_n}^{(1)},\mathbf{T}_{k_n}^{(2)},\ldots,\mathbf{T}_{k_n}^{(n/k_n)}$ is also the maximum from $\mathbf{T}_{n}$.
Consequently, in order to tighten the bound as much as possible, we would like to choose $k_n$ as large as possible.
For example, if $k_n=\frac{n}{2}$, we get two identical subtrees $\mathbf{T}_{n/2}^{(1)}$ and $\mathbf{T}_{n/2}^{(2)}$ (see \cref{fig:Nolabel2}), that are connected to the root of $\mathbf{T}_{n}$ by a single channel, $W_1$ and $W_2$ respectively.
Here, we can figure out that we have the tightest possible lower bound.
However, as mentioned earlier, the distribution of $Y$ will be evaluated via EVT, and specifically, by \Cref{EVT_for_nonstationary_integer_valued} we derived in the previous section.
Hence, we need to further find restrictions on $k_n$, such that it would be possible to apply \Cref{EVT_for_nonstationary_integer_valued}, in order to derive a limit distribution for $Y$.

We would like to mention that if $k_n$ is a constant, the sequence $ \left( Y_1,\ldots,Y_n \right)$ meets the definition in Hall \cite{N54hall2006extremes}.
Namely, for each subset $\lbrace Y_{l_1},\ldots,Y_{l_n} \rbrace$, the distribution of $ \left( Y_{l_1},\ldots,Y_{l_n} \right)$ and $ \left( Y_{l_1+T},\ldots,Y_{l_n+T} \right)$ are identically distributed, for a final period $T=k_n$.
Accordingly, the extreme value distribution for the sequence is easily found.
However, as discussed earlier, $k_n$ should be as large as possible, and, in particular, such that $k_n \to \infty$ as $n \to \infty$.

Conditions $D \left(u_n \right)$ and $D_{k_n}^{'}\left(u_n \right)$ are essential in order to derive the asymptotic distribution of the normalized $Y$.
Regarding condition $D \left(u_n \right)$, note that by the construction of the lower bound, $Y_{j_1}^{(i_1)}$ and $Y_{j_2}^{(i_2)}$ are independent if $i_1 \neq i_2$ (since they do not belong to the same subtree).
Therefore, we say that the sequence $\lbrace Y_{j}^{(i)}  \rbrace$ (we write from now, for simplicity, $\left( Y_1,\ldots,Y_n \right)$) is $k_n$ dependent, in the sense that we can divide the sequence $\left( Y_1,\ldots,Y_n \right)$ into sub-sequences with size  $k_n$, which are independent of each other.
As a result, similar to \Cref{sec:EVTNonStationaty}, let divide the indices into $r_n$ sets, $I_1,\ldots,I_{r_n}$, where $r_n=\frac{n}{k_n}$.
Then, we have
\begin{align}
\Pr \left(Y \leq u_n \right) = \prod_{l=1}^{r_n}Pr \left( M \left(I_l \right) \right).
\end{align}
In particular, one can see that \cref{eq:asymtoticIndependent} holds.
Hence, if $r_n$ is an increasing sequence, condition $D^{'}_{k_n} \left(u_n \right)$ is \emph{sufficient to guarantee} (for a proper choice of $u_n$) asymptotic distribution of the normalized $Y$.
Furthermore,  note that by the construction of $\mathbf{T}_{k_n}$, it can be verified that a sufficient condition for $D_{k_n}^{'}\left(u_n \right)$ is $n \sum_{i=2}^{k_n} \Pr \left( X_1>u_n,X_i>u_n \right) \to 0$, as $n \to \infty$, which is exactly the condition $D^{'} \left(u_n \right)$, presented in \cref{eq:ConditionDtag}.
Moreover, $\left( Y_1,Y_i \right)$ are independent, for $i> \frac{k_n}{2}$, (since they do not share mutual channels).
Hence, if the assumption in \cref{FirstCondition}  holds, we have:
\begin{align}
n \sum_{i=k_n/2}^{k_n} \Pr \left( X_1>u_n,X_i>u_n \right) 
&
=n \sum_{i=k_n/2}^{k_n}   O\left( \frac{1}{n^2} \right)
\\ \nonumber &
= n \frac{k_n}{2}  O\left( \frac{1}{n^2} \right) \to 0,
\end{align} 
for $k_n=o\left(n \right)$.
Hence, for the sequence $\left( Y_1,\ldots,Y_n \right)$, condition $D_{k_n}^{'}\left(u_n \right)$ may be written as
\begin{align}\label{ConditionD_k_nV2}
D_{k_n}^{'}\left(u_n \right): \limsup n \sum_{i=2}^{k_n/2} \Pr \left( X_1>u_n,X_i>u_n \right) =0.
\end{align}

Note that since we are dealing with a practical information dissemination problem, namely, $n$ is finite in real life, we should consider also the rate of convergence of the result in \Cref{EVT_for_nonstationary_integer_valued}, and in particular, the choice of $u_n$ which affects on the convergence in \cref{FirstCondition}.
Moreover, the size of $k_n$ (and thus also $r_n$) also has a great impact on it, and this can be seen in the transition from \cref{Eq31} to \cref{Eq32}, and also by the convergence of \cref{ConditionD_k_nV2}. Namely, there is a direct relationship between the choice of $k_n$ and the rate of convergence, that is, \emph{as $k_n$ grows, then the convergence rate slows down}. 
As a result, one can see that there is a \emph{trade off} between the tightness of the lower bound and the rate of convergence.
This will be illustrated via simulation results in \cref{SimulationSection}.

Next, we find $u_n$ to satisfy \cref{FirstCondition} and restrictions on $k_n$ such that $D_{k_n}^{'}\left(u_n \right)$ would hold.
First, define a continuous CDF
\begin{align}\label{chi_cdf}
 \bar{F}_c\left(x|m_n,p \right)
 = \frac{B(x-m_n+1,m_n)}{ B(p;x-m_n+1,m_n)},
\end{align}
for $x>m_n-1$, and by the definition of $F_c$ we have
\begin{align}
 \bar{F}_c\left(x|m_n,p \right) \leq  \Pr \left(X>x \right) \leq \bar{F}_c\left(x-1|m_n,p \right),
\end{align}
for some random variable $X \sim NB \left(m_n,p \right)$.
The first lemma finds an asymptotic approximation for $\bar{F}_c\left(x|m_n,p \right)$, that will be needed in order to find an appropriate $b_n$ satisfying \cref{eq:choiceOfBn}, and also to find an asymptotic approximation for the joint distribution of $\left( Y_1,Y_i \right)$.
\begin{lemma} \label{General}
Let
\begin{align}\label{eq:phiValue}
 \phi \left(n \right)=\alpha \log_{\frac{1}{p}} n+\left( m_n -1 \right) \log_{\frac{1}{p}} \left(  \frac{1-p}{p}  \Psi_{\alpha}\left(m_n \right) \right)
 - \log_{\frac{1}{p}} (m_n -1)! +x,
\end{align}
where
\begin{align}\label{eq:PsiFunction}
 \Psi_{\alpha}\left(m_n \right) = \alpha \log_{\frac{1}{p}} n +\left( m_n -1 \right) \log_{\frac{1}{p}} \left(  \frac{1-p}{p} \right),
\end{align}
for $0 <\alpha<1$, (we write for simplicity $\Psi_{1}=\Psi$).
Assume the integer sequence $m_n=o\left(\log n \right)$ is a slowly increasing function of $n$, (and note that $  \Psi_{\alpha} \left(m_n \right)$ a function of $m_n$, and not for $n$).
In addition, let
 \begin{align}\label{Assumptions}
\epsilon \left(n \right)=\frac{\prod_{j=0}^{m_n-1}\left(\phi(n)-j \right)}{\left(  \Psi_{\alpha} \left(m_n \right) \right)^{m_n}}.
\end{align}
Then, we have
\begin{align}
\bar{F}_c\left(\phi \left(n \right)|m_n,p \right) = \frac{p^x}{n^{\alpha}}\epsilon(n) \left( 1+O \left( \frac{m_n}{\log n} \right) \right).
\end{align}
Note that $\phi (n)$ depends on $x$.
Moreover, if $\epsilon(n)  \to 1$, as $n \to \infty$,  we have
\begin{align}
\bar{F}_c\left(\phi \left(n \right)|m_n,p \right) = \frac{p^x}{n^{\alpha}} \left(1+o(1) \right).
\end{align}
\begin{proof}
See Appendix \ref{AppendixProofOfLemmaGeneral}.
\end{proof}
\begin{remark}
To understand how to choose $m_n$ such that \cref{Assumptions} is satisfied, note that
\begin{align}
\frac{\prod_{j=0}^{m_n-1}\left(\phi(n)-j \right)}{\left(  \Psi_{\alpha}\left(m_n \right)  \right)^{m_n}} & \leq  \left( \frac{\phi(n)}{  \Psi_{\alpha}\left(m_n \right)} \right)^{m_n} 
\\ \nonumber &
=\left(1+\frac{\Delta(n)}{ \Psi_{\alpha}\left(m_n \right) } \right)^{m_n},
\end{align}
where
\begin{align}
\Delta(n)=\left( m_n -1 \right) \log_{\frac{1}{p}} \left(  \Psi_{\alpha}\left(m_n \right) \right) - \log_{\frac{1}{p}} (m_n -1)! +x.
\end{align}
Using the fact that $1+x \leq e^x$, we can write
\begin{align}
 \frac{\prod_{j=0}^{m_n-1}\left(\phi(n)-j \right)}{\left(  \Psi_{\alpha}\left(m_n \right)  \right)^{m_n}} 
&
\leq \exp \left(\frac{\Delta(n)}{ \Psi_{\alpha}\left(m_n \right) } \right)^{m_n} 
\\ \nonumber &
 =\exp \left( \frac{ O \left(  m_n  \log \left(   \Psi_{\alpha}\left(m_n \right) \right) \right)m_n}{  \Psi_{\alpha}\left(m_n \right) } \right).
\end{align}
Hence, a sufficient condition for choosing $m_n$ is
\begin{align}\label{eq:sufficentCondtiionMn}
\lim_{n \to \infty} \frac{m_n^2 \log \log n}{\log n} = 0.
\end{align}
\end{remark}
\begin{remark}
By the proof of \Cref{General}, it holds for any $x$ satisfying \cref{Assumptions}, hence it can be concluded that a sufficient condition for $x$ is that $x= O \left(m_n \log \log n \right)$.
Furthermore, if $x<0$, but still $\phi \left( n \right)=O \left(\log n \right)$, then
\begin{align}
 \phi \left(n \right) = \hat{\phi} \left(n \right) +x \leq \hat{\phi} \left(n \right).
\end{align}
Hence, if $\hat{\phi} \left(n \right)$ satisfies \cref{Assumptions}, we get that for large enough $n$,
\begin{align}
\bar{F}_c(\phi(n)|m_n,p) \leq \frac{p^x}{n^{\alpha}}.
\end{align}
This will be used in order to show the asymptotic approximation of the joint distribution.
\end{remark}
\end{lemma}
\begin{proposition}\label{ConditionA}
 Let $u_n(x)=x+b_n$  (for simplicity $u_n$), where
\begin{align}\label{ab_n}
 b_n=\log_{\frac{1}{p}} n+ \left( \log_2 k_n -1 \right) \log_{\frac{1}{p}} \left(  \frac{1-p}{p}  \Psi \left( \log_{2} k_n \right) \right)
 - \log_{\frac{1}{p}} (\log_2 k_n -1)!.
\end{align}
We have the following result, 
\begin{align}\label{limitBnStatement}
\lim_{n \to \infty} n \bar{F}_c \left( u_n | \log_2 k_n,p \right) =p^x.
\end{align}

\begin{proof}
The proposition follows immediately from \Cref{General}, with $\alpha=1$, and $m_n=\log_2 k_n$.
\end{proof}
\end{proposition}
An immediate conclusion from \Cref{ConditionA} is that if $\left( Y_1,\ldots,Y_n \right)$ would be i.i.d., then \cref{GemeralExtremeValueDiscrete} would hold for $\tau=p^x$ and $\tau^{'}=p^{x-1}$.
We should pay attention that for constant $\log_2 k_n$, this lemma was already found in  \cite{N19}, and also for a slightly different $b_n$ in \cite{N30mccormick1992asymptotic}.
However, here we extend it to the case where $\log_2 k_n$ may also be an increasing sequence of $n$, where $\log_2 k_n$ satisfies \cref{Assumptions}.
As we saw earlier, a sufficient condition for $k_n$ is
\begin{align}\label{eq:suffientConditionLogKn}
\lim_{n \to \infty} \frac{\left( \log_2 k_n \right)^2 \log \log n}{\log n}=0.
\end{align}
Consequently, the results of \cite{N19} remains true also for the case where the file consisting of $\kappa$ packets, where $\kappa$ is an \emph{increasing sequence} of $n$ satisfying $\frac{\kappa^2 \log \log n}{\log n} \to 0$, as $n \to \infty$.
In the next subsection, we will extend it for the of $\kappa=\log_2 n$, and in \Cref{sec:ExtentionMmessages}, we will generalize it for $\kappa=M \log_2 n$, for some constant $M$.

Observe that for $\log_2 k_n=1$, we get that the sequence $\left(Y_1,\ldots,Y_n \right)$ is i.i.d.\ following geometric random variable.
Thus, by \Cref{EVT_for_nonstationary_integer_valued}, it is possible to obtain asymptotic approximation of $Y$, namely for large enough $n$,
\begin{align}
e^{-p^{x-1}} \leq \Pr \left( Y \leq b_n+x \right) \leq e^{-p^x},
\end{align}
with $b_n= \log_{\frac{1}{p}} n$.
Therefore, the distribution of $Y-b_n$ is bounded from both sides by the CDF of a Gumbel distribution.\footnote{Note that this is result is a well known results in the literature.}
Hence, according to \cref{eq:GumbelExpectedValue},
\begin{align}\label{eq:expectedValueOFgeometric}
 \mathbb{E} \left[ Y \right] \geq b_n-\frac{\gamma}{\log p},
\end{align}
which implies that we can derive an immediate lower bound on $\mathbb{E} \left[ M_n^l \right]$ as follows.
\begin{claim}\label{claima}
\begin{align}
\mathbb{E} \left[ M_n \right] \geq \left( \frac{1}{1-p}-\frac{1}{\log_2 p } \right)\log_2 n-\frac{\gamma}{\log p}-\frac{1}{1-p}.
\end{align}
\begin{proof}
\begin{align}\label{eq:TrivialLowerBoundV2}
\mathbb{E} \left[ M_n^l \right] 
&=\mathbb{E} \left[ Y+W \right]
 \\ \nonumber &
=\mathbb{E} \left[ Y \right]+\mathbb{E} \left[ W \right]
 \\ \nonumber &
\geq \log_{\frac{1}{p}} n -\frac{\gamma}{\log p}+\frac{\log_2 (n)-1}{1-p}
 \\ \nonumber &
= \left( \frac{1}{1-p}-\frac{1}{\log_2 p } \right)\log_2 n-\frac{\gamma}{\log p}-\frac{1}{1-p}.
\end{align}
\end{proof}
The lower bound in the claim above has a scaling law of $\left( \frac{1}{1-p}-\frac{1}{\log_2 p } \right)\log_2 n$.
However, as discussed earlier, this bound can be improved by setting $k_n$ as large as possible, which will cause dependence between the variables in the sequence $\left(Y_1,\ldots,Y_n \right)$.
Hence, next, we show that condition $D_{k_n} \left(u_n \right)$ holds with $k_n$ satisfying \cref{eq:suffientConditionLogKn}, hence we can further tighten our bound.
\end{claim}
First, in the next lemma, we find an asymptotic bound for the joint distribution.
This bound will be used in order to prove that condition $D_{k_n} \left(u_n \right)$ holds.
\begin{lemma}\label{JointDistributionLemma}
For $i=1,2,\ldots,\log_2 k_n-1$, we have
\begin{multline}
\Pr \left(Y_1>u_n,Y_{2^i} >u_n \right) 
 \\ 
\leq  2 \frac{p^x}{n} \left( C \Psi \left( \log_2 k_n \right) \right)^{\log_2 k_n-1} \left( C \frac{4}{1-p}  \log_2 k_n \right)^{i} 
+\frac{1}{n^{2\alpha}}\left(1+o \left( 1 \right) \right)
\end{multline}
for some $\frac{1}{2} < \alpha <1$, where 
\begin{align}
C= \max \left( \frac{1}{•\log_{\frac{1}{p}} n },\frac{1}{\Psi \left( \log_2 k_n \right) }\right).
\end{align}
\begin{proof}
See Appendix \ref{AppendixProofJointDistibution}.
\end{proof}
\end{lemma}
We may now give the key result on $ \left(Y_1, \ldots,Y_n \right)$, which will enable us to reason on its properties for large $n$.
\begin{lemma}\label{ConditionDtag}
The sequence  $\left( Y_1,\ldots,Y_n \right)$ satisfies condition $D^{'}_{k_n} \left( u_n \right)$.
\begin{proof}
Let
\begin{align}
\alpha_n= n \sum_{i=2}^{k_n/2} \Pr \left( Y_1>u_n,Y_i>u_n \right) 
\end{align}
Then, by \cref{ConditionD_k_nV2}, condition $D^{'}_{k_n} \left( u_n \right)$ is satisfied if $\alpha_n \to 0$, as $n \to \infty$.\\
As mentioned before, we may write the joint distribution $ \Pr \left(Y_1,Y_i \right)$ as $ \Pr \left(W+Z_1,W+Z_2 \right)$, where $W$ is a random variable, representing the shared branches of the tree (dependent channels) of $Y_1,Y_i$ and $Z_1,Z_2$ are the independent branches (channels) of $Y_1,Y_i$, respectively.\\
Therefore, we partitioning all the pairs $ \left( Y_1,Y_i \right),i \in \left\lbrace 2, \ldots, k_n/2 \right\rbrace $, according to the number of shared branches.
The number of pair with $\log_2 k_n -i$ shared branches and $i$ independent branches is $2^{i-1}$.\\
As a result, we can write  $\alpha_{n}$ as
\begin{align}
 \alpha_{n}= & n  \sum_{i=2}^{k_n/2}  \Pr \left(Y_1>u_n,Y_i >u_n \right)
 \\ \nonumber =&
				n \sum_{i=1}^{\log_2 k_n-1}  2^{i-1} \Pr \left(Y_1>u_n,Y_{2^i}>u_n \right)
\end{align}
Using \Cref{JointDistributionLemma}, we have
\begin{align}
\alpha_n \leq & \frac{n}{2} \sum_{i=1}^{\log_2 k_n-1}  2^i \cdot
\left( \frac{2}{n} \left( C \Psi \left( \log_2 k_n \right) \right)^{\log_2 k_n-1} \left( C \frac{4}{1-p}  \log_2 k_n \right)^{i} +\frac{1}{n^{2\alpha}} \right) 
  \\ \nonumber = &
   \left( C \Psi \left( \log_2 k_n \right) \right)^{\log_2 k_n-1} 
   \sum_{i=1}^{\log_2 k_n-1}\left( C \frac{8}{1-p}  \log_2 k_n \right)^{i}+\frac{1}{2}\frac{1}{n^{2 \alpha-1}} \sum_{i=1}^{\log_2 k_n-1} 2^i
\end{align}
Let $g(n)=  C \frac{8}{1-p}  \log_2 k_n$, and note that $g(n)=O \left( \frac{\log k_n}{ \log n} \right)$.
Hence, by \cref{eq:suffientConditionLogKn}, $g(n) \to 0$ as $n \to \infty$.
Then
\begin{align}\nonumber
 \alpha_{n} \leq &     \left( C \Psi \left( \log_2 k_n \right) \right)^{\log_2 k_n-1}
   \sum_{i=1}^{\log_2 k_n-1}\left( g(n) \right)^{i}+\frac{1}{2}\frac{k_n}{n^{2 \alpha-1}} 
   \\ \nonumber = &
 \left( C \Psi \left( \log_2 k_n \right) \right)^{\log_2 k_n-1}
   \frac{g(n)}{1-g(n)} \left(1-g(n) \right)^{\log_2 k_n}
      \\ \nonumber  & 
   +\frac{1}{2}\frac{k_n}{n^{2 \alpha-1}}  
  \left( C \Psi \left( \log_2 k_n \right) \right)^{\log_2 k_n-1}    \frac{g(n)}{1-g(n)}+\frac{1}{2}\frac{k_n}{n^{2 \alpha-1}}  
 \end{align}
Now, as $n \to \infty$, clearly $    \frac{g(n)}{1-g(n)} \to 0$.
In addition, note that $  \left( C \Psi \left( \log_2 k_n \right) \right)^{\log_2 k_n-1} \to 1$.
Moreover, note that, $\frac{k_n}{n^{2 \alpha-1}} \to 0$ if 
\begin{align}
\log k_n -(2 \alpha-1) \log n \to -\infty,
\end{align}
and as noted before, by the choice of $k_n$, $\frac{\log k_n}{\log n} \to 0$, then together with $\alpha>\frac{1}{2}$, we obtain $\frac{k_n}{n^{2 \alpha-1}} \to 0$.
Hence, $\alpha_n \to 0$ and the lemma follows.
\end{proof}
\end{lemma}
We can now give the main result in this section, which bounds the expected completion time in the dissemination tree $\mathbf{T}_n$.
\begin{theorem}\label{theorem:ComplentionTimeLower}
The expected completion time $\mathbb{E} \left[ M_n \right]$ needed to broadcast one packet to all receivers in the dissemination tree is bounded by
\begin{align}
\mathbb{E} \left[ M_n \right] \geq \frac{\log_2 n/k_n}{1-p} -\frac{\gamma}{\log p}+b_n,
\end{align}
where $b_n$ was defined in \cref{ab_n}, and $k_n$ satisfies \cref{eq:suffientConditionLogKn}.
\begin{proof}
Since \Cref{ConditionDtag} and \Cref{ConditionA} hold for the sequence $Y=\max \left(Y_1,\ldots,Y_n \right)$, then, by \Cref{EVT_for_nonstationary_integer_valued}, for sufficiently large $n$,
\begin{align}
e^{-p^{x-1}} \leq \Pr \left( Y \leq u_n \right) \leq e^{-p^x}.
\end{align}
Therefore, as indicated previously, $Y-b_n $ is bounded by a non-degenerate distribution following the Gumbel law, hence by \cref{eq:GumbelExpectedValue}
\begin{align}\label{eq:boundsOfEY}
b_n+\frac{\gamma}{\log p} \leq \mathbb{E}\left[ Y \right] \leq  b_n-\frac{\gamma}{\log p}+1,
\end{align}
and in particular
\begin{align}
\mathbb{E}\left[ Y \right] \geq  -\frac{\gamma}{\log p}+b_n.
\end{align}
Hence,
\begin{align}\label{eq:LowerBoundResult}
\mathbb{E}\left[ M_n \right] \stackrel{(a)}{\geq }  & \mathbb{E}\left[ M_n^l \right]
\\ \nonumber \stackrel{(b)}{=} & 
\mathbb{E} \left[ W + Y \right]
 \\ \nonumber  = & 
 \mathbb{E} \left[ W \right]+ \mathbb{E} \left[ Y \right]
  \\ \nonumber \stackrel{(c)}{\geq } & 
  \frac{\log_2 n/k_n}{1-p} +b_n-\frac{\gamma}{\log p}
   \\ \nonumber  = & 
  \left( \frac{1}{1-p}-\frac{1}{\log_2 p} \right) \log_2 n + o \left(\log n \right),
\end{align}
where $(a)$ is by \Cref{ProveLowerBound} and $(b)$ is according to the construction of the lower bound.
In $(c)$, we used the expected value of NB (\cref{ExpectedvalueNegativeBinomial}) for $W$ and for $Y$.
\end{proof}
\end{theorem}
\Cref{theorem:ComplentionTimeLower} presents a lower bound on the expected completion time in the dissemination tree for large enough $n$.
The choice of $k_n$ has a crucial effect of the convergence of the CDF of $Y$, and as a result, also on $\mathbb{E} \left[ Y \right]$.
In addition, note that asymptotically, $k_n$ has no effect on the the multiplicative factor of the leading term, $\log n$.
However, as noted above, the bound is tighter as $k_n$ increases, and in particular, tighter than the bound defined in \Cref{claima}, which refers to $k_n=1$.
\subsection{Upper bound}
\label{sec:UpperBound}
In this section, we provide an upper bound on $\mathbb{E} \left[ M_n \right]$, by considering an independent sequence of NB random variables with $\log_2 n$ successes.
We study the distribution of the maximum of that sequence, and as a result, we also find an upper bound of the tail distribution of $M_n$.
Namely, for each $\epsilon>0$, define the probability $1-\epsilon$ to be that after $T$ time slots, the message reaches to all nodes in $\mathbf{T}_n$.
Then we find an upper bound on the completion time $T$ as a function of $\epsilon$.

\begin{figure}
        \centering
        \begin{subfigure}[b]{0.6\columnwidth}
        \begin{center}
  \includegraphics[width=0.8\textwidth]{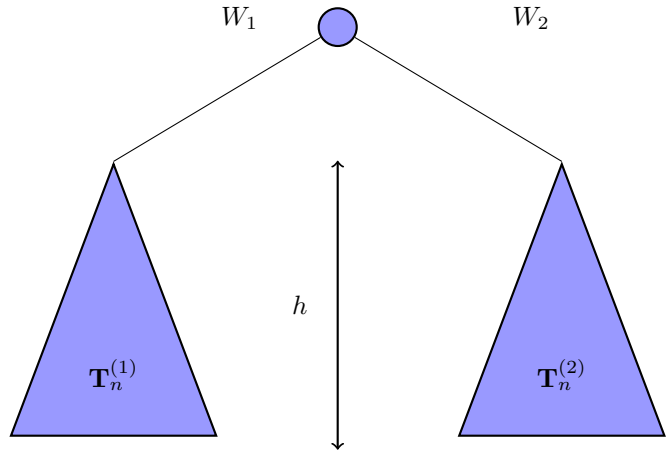}
                \caption{Illustration of the dissemination tree $\mathbf{T}_{2n}$, which consists of two identical subtrees $\mathbf{T}_n^{(1)}$ and $\mathbf{T}_n^{(2)}$, with completion times $M_{n}^{(1)}$ and $M_{n}^{(2)}$, respectively.
Each one of the subtrees is connected to $\mathbf{T}_{2n}$ via a single channel, where the time it takes to forward a message through is denoted by $W_1$ and $W_2$, respectively.
                }
\label{subfig:Upper_Tree}
  \end{center} 
        \end{subfigure}%
        \hfill
        ~ 
        \begin{subfigure}[b]{0.6\columnwidth}
              \begin{center}  
  \includegraphics[width=0.8\textwidth]{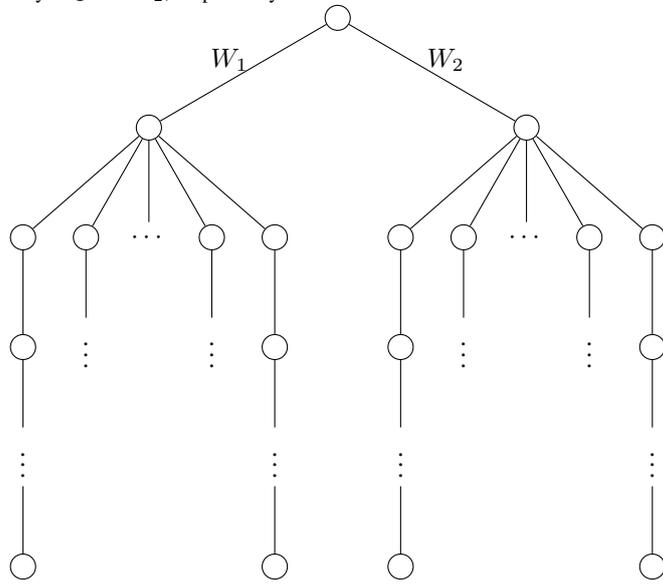}
                \caption{Illustration of $2n$ independent paths, each has $h$ hops that merge into $W_1$ and $W_2$.}
                   \label{subfig:Upper_middle}
                  \end{center} 
        \end{subfigure}
        ~ 
          \qquad
        \begin{subfigure}[b]{0.6\columnwidth}
        \begin{center}
\centering
    \includegraphics[width=1.2\textwidth]{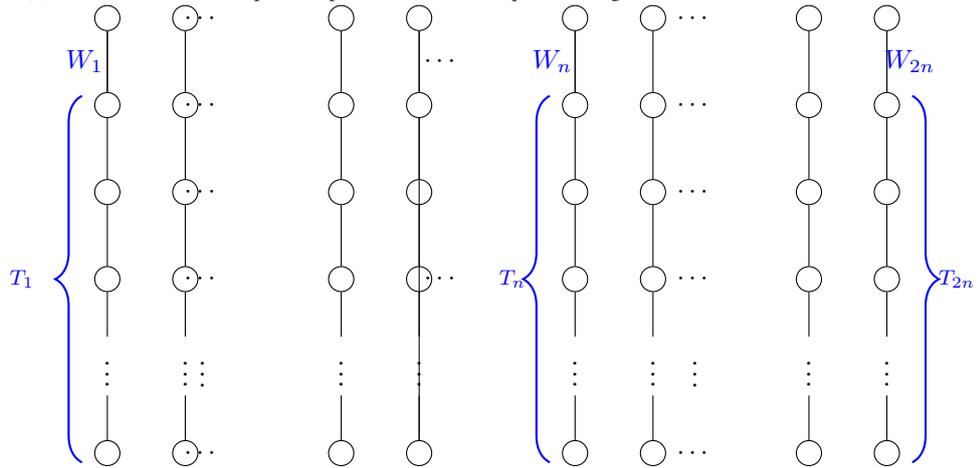}
 \caption{Illustration of $2n$ independent paths, each has $h+1$ hops, where the time to forward a message through the first hop is denote by $\lbrace W_i \rbrace_{i=1}^{2n}$, and the remaining $h$ hops are denoted by $\lbrace T_i \rbrace_{i=1}^{2n}$.}
                            \label{subfig:Upper_IID}  
             \end{center}          
        \end{subfigure}
        \caption{Phases of the induction in the proof of \Cref{lemmaUpperBound}.}       
        \label{fig:IllustationUpperBoundProof}
\end{figure}

\begin{figure}[t]  
\centering
    \includegraphics[width=0.7\textwidth]{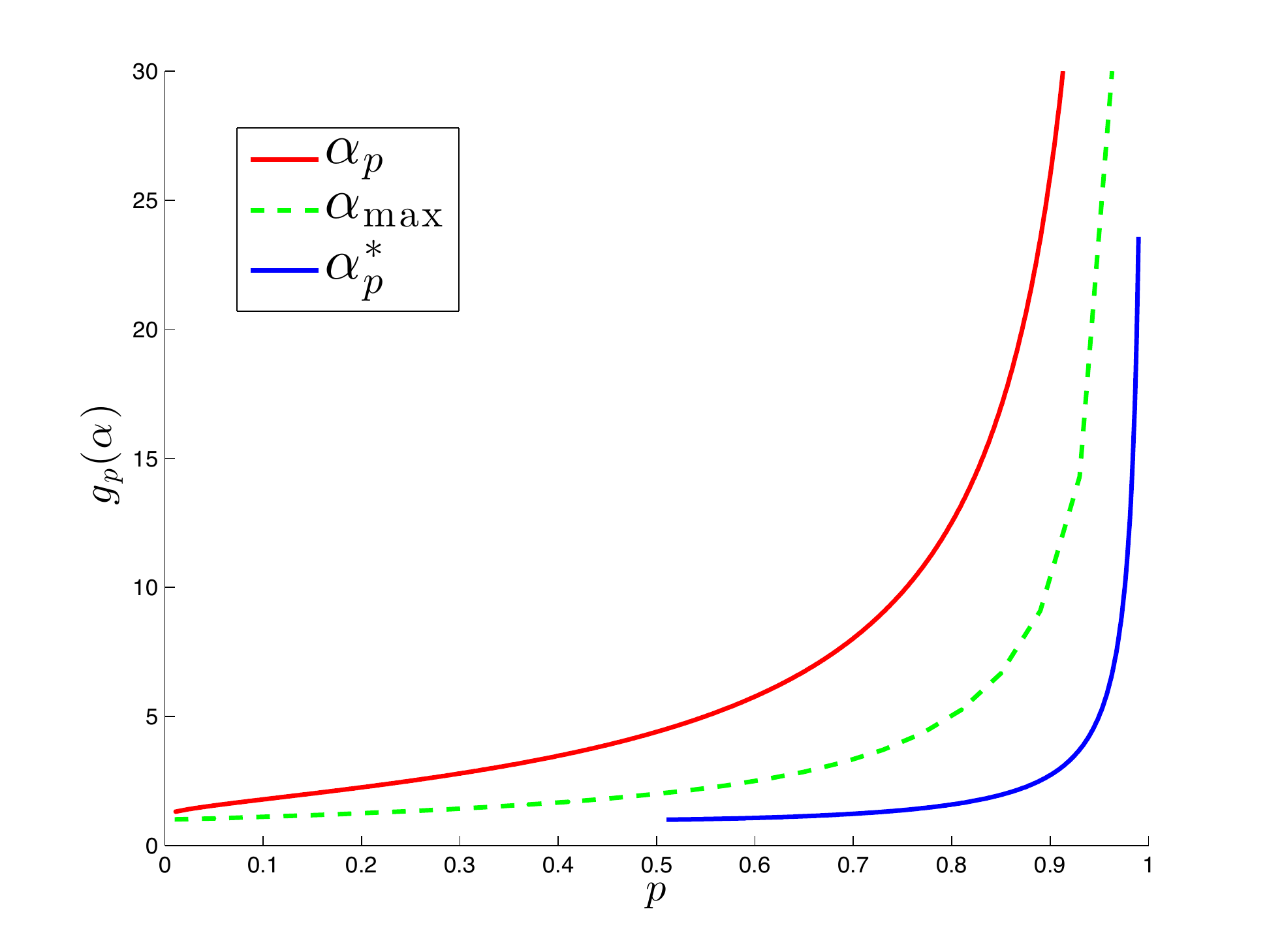}
    \caption{Plot of $g_p\left(\alpha\right)$, with the roots $\alpha_{p}$, $\alpha_{p}^{*}$  and $\alpha_{\max}$, where $\alpha_{\max}=\frac{1}{1-p}$, as a function of $p$.}
        \label{fig:RootGraph}
\end{figure}

Let $ \left( \hat{T_1},\ldots,\hat{T_n} \right)$ be i.i.d.\ random variables with a marginal distribution function $F \left[x|\log_2 n ,p \right]$, i.e.  $\hat{T_i} \sim NB\left(\log_2 n ,p \right)$, and let $\hat{M}_n=\max \left( \hat{T_1},\ldots,\hat{T_n} \right)$.
In the following lemma, we make precise the intuitive result that the maximum over the leaves in the tree $\mathbf{T}_n$ is upper bounded by the maximum of i.i.d.\ variables with the same marginal distribution.
\begin{lemma}\label{lemmaUpperBound}
\begin{align}
 \mathbb{E} \left[ M_n \right] \leq \mathbb{E}  \left[ \hat{M}_n \right].
\end{align}
\begin{proof}
We prove the lemma by induction on the height of $\mathbf{T}_n$.
Let $h=\log_2 n$ be the height of $\mathbf{T}_n$.
For the base case, i.e.\ h=1, clearly $\Pr \left( M_n  \leq x \right) = \Pr \left( \hat{M}_n \leq x \right)$, and thus  $\mathbb{E} \left[ M_n \right] \leq \mathbb{E}  \left[ \hat{M}_n \right]$.
For the induction step, assume that the claim holds for some $h$, namely  $\mathbb{E} \left[ M_n \right] \leq \mathbb{E}  \left[ \hat{M}_n \right]$.
We prove it for $h+1$, i.e. $\mathbb{E} \left[ M_{2n} \right] \leq \mathbb{E}  \left[ \hat{M}_{2n} \right]$.
As mentioned above, $M_{2n}$ can be represented recursively, namely, the dissemination time of $\mathbf{T}_{2n}$ is the maximum between the dissemination time of two independent subtrees $\mathbf{T}_n^{(1)}$ and $\mathbf{T}_n^{(2)}$, plus the time it takes to forward the message to the root of each one of these subtrees.
Formally, let  $W_1$ and $W_2$ be i.i.d.\ geometric distributions, and $M_{n}^{(1)},M_{n}^{(2)}$ be the dissemination time of each one of $\mathbf{T}_n^{(1)}$ and $\mathbf{T}_n^{(2)}$, respectively (see \cref{subfig:Upper_Tree}).
Then
\begin{align}
\mathbb{E} \left[ M_{2n} \right] &= \mathbb{E} \left[ \max \left( W_1+M_{n}^{(1)},W_2+M_{n}^{(2)} \right) \right]
\\ \nonumber & \leq 
\mathbb{E} \left[ \max \left( W_1+\hat{M}_{n}^{(1)},W_2+\hat{M}_{n}^{(2)} \right) \right],
\end{align}
where in the last step we used the induction step (see \cref{subfig:Upper_middle}).
Thus, in order to complete the proof, it is sufficient to show that $\mathbb{E} \left[ \max \left( W_1+\hat{M}_{n}^{(1)},W_2+\hat{M}_{n}^{(2)} \right) \right] \leq \mathbb{E} \left[ \hat{M}_{2n} \right]$.
Now, for this part of the proof, we use the same technique we used in \Cref{ProveLowerBound}, that is, $\hat{M}_{2n} =\max \left( W_i+\hat{T_i}; \; 1 \leq i \leq 2n \right)$, where $\lbrace W_i\rbrace_{i=1}^{2n}$ is an i.i.d.\ sequence following a geometric distribution, and $\lbrace T_i \rbrace_{i=1}^{2n}$ is also i.i.d., but $T_i \sim NB \left(h,p \right)$, (see \cref{subfig:Upper_IID}).
Now, denote $l_1=\argmax \left( \hat{T_i}; 1\leq i \leq n \right)$ and $l_2=\argmax \left( \hat{T_i}; n+1\leq i \leq 2n \right)$.
Therefore
\begin{align}
 \hat{M}_{2n}&=\max \left( W_i+\hat{T_i}; \; 1 \leq i \leq 2n \right)
 \\ \nonumber & =
\max  \left(  \max_{1 \leq i \leq n} \left( W_i+\hat{T_i} \right),\max_{n < j \leq 2n} \left( W_j+\hat{T_j} \right) \right)
 \\ \nonumber & \geq 
 \max \left(  W_{l_1}+\hat{T_{l_1}}, W_{l_2}+\hat{T_{l_2}}  \right)
  \\ \nonumber & =
   \max \left( W_1+\hat{M}_{n}^{(1)},W_2+\hat{M}_{n}^{(2)} \right),
\end{align}
which completes the proof.
\end{proof}
\end{lemma}
By \Cref{lemmaUpperBound},  $\mathbb{E} \left[ \hat{M}_n \right]$ is an upper bound on $\mathbb{E} \left[ M_n \right]$. Therefore,
we now derive a closed-form expression for  $\Pr \left( \hat{M}_n \leq u_n \right) $, and use this distribution to find $\mathbb{E}  \left[ \hat{M}_n \right]$.
Note that, as mentioned earlier, the extremes of $n$ random variables with NB marginal distribution have been already studied by \cite{N19,N30mccormick1992asymptotic}, for a fix number of successes, while here we consider $\log_2 n$ successes.
At first, similar to the previous section, we define a continuous random variable with CDF $F_c(x)$, such that for any real $x$, $F_c(x-1) \leq F[x] \leq F_c(x)$, and for an integer $x$, $F_c(x)=F[x]$.
The next step is to find a sequence $b_n$, which  satisfies $\bar{F}_c (b_n) = \frac{1}{n}$.\\
Since
\begin{align}
\bar{F}[x|m,p]= \Pr \left(X>x \right)=\sum_{k=x+1}^{\infty} \Pr \left(X=k \right)
\end{align}
But, with the expression above, it is not clear how to define $F_c \left(x \right)$.
Although we can use \cref{chi_cdf}, here we find $F_c$ in a slightly different way.
In the next lemma, we present a closed form for $F[x|m,p]$ when $x$ has the form $x=a m +d_m$.
\begin{lemma}\label{CLoseFormExpressionCDFNegativeBinomial}
Let $X \sim NB\left(m,p \right)$, and define the sequence $\widetilde{b}_m=am+d_m$, where $a$ is a constant satisfying $a>\frac{1}{1-p}$ and $d_m$ is a negative sequence such that $d_m=o \left(m \right)$. Then, for sufficiently large $m$,
\begin{align}
\Pr \left(X > \widetilde{b}_m \right)  = C_m \Pr \left( X=\widetilde{b}_m+1 \right)
\end{align}
 where $C_m \to C$, as $m \to \infty$, and $C=\frac{1}{1-\frac{a}{a-1}p}$.
\begin{proof}
See Appendix \ref{ProofCLoseFormExpressionCDFNegativeBinomial}.
\end{proof}
\end{lemma}
\Cref{CLoseFormExpressionCDFNegativeBinomial} shows that the tail distribution $\Pr \left(X > \widetilde{b}_m \right)$, under an appropriate condition on $\widetilde{b}_m$, is the PDF at $\widetilde{b}_m+1$, multiplied by a constant.
The PDF is defied by \cref{eq:NBPDF}.
Note that for any integers $n$ and $k$, $\binom{n}{k}=\frac{\Gamma(n+1)}{\Gamma(k+1)\Gamma(n-k+1)}$.
Hence, an immediate consequence of \Cref{CLoseFormExpressionCDFNegativeBinomial} is that it is possible to define a continuous function $F_c$, where
\begin{align}\label{Fc}
\bar{F}_c \left(\widetilde{b}_m \right)= C_m \frac{\Gamma(\widetilde{b}_m+1)}{\Gamma(m) \Gamma(\widetilde{b}_m-m+2)}(1-p)^m p^{\widetilde{b}_m-m+1}.
\end{align}
In the next lemma, we find an asymptotic approximation for 
$\bar{F}_c \left(\widetilde{b}_m \right)$.
\begin{lemma}\label{lemma259}
Let $\widetilde{b}_m=am+d_m$, where $d_m$ satisfies $\frac{d_m^2}{m} \to 0$ as $m \to \infty$.
Then, for sufficiently large $m$,
\begin{align}
\bar{F}_c \left(\widetilde{b}_m \right)
=C_m \left(\frac{a^a}{\left(a-1 \right)^{a-1}}p^{a-1}(1-p) \right)^m  \delta^{d_m} \frac{1}{\sqrt{m}} \frac{1}{a-1} \sqrt{\frac{\delta p}{2 \pi}}
\left(1+O \left(\frac{d_m^2}{m} \right) \right),
\end{align}
where $\delta=\frac{a}{a-1}p$. 
\begin{proof}
See Appendix \ref{ProofLemma259}.
\end{proof}
\end{lemma}
As mentioned before, we are eventually interested in studying the extreme value of random variables with marginal $F[x|\log_2 n,p]$, hence, later we will assign $m=\log_2 n$, and clearly $b_n=\widetilde{b}_{\log_2 n}$, namely $b_n$ has the form $b_n=a \log_2 n+o\left(\log n\right)$.
In order to find an appropriate constant $a$ and sequence $d_m$ such that $\bar{F}_c (b_n) = \frac{1}{n}$, we will use two steps:
At first we find $a$ satisfying
\begin{align}\label{MustSatisfy}
\frac{a^a}{\left(a-1 \right)^{a-1}}p^{a-1}(1-p)= \frac{1}{2}
\end{align}
subject to $a>\frac{1}{1-p}$, and then find $d_m$ such that
\begin{align}
\delta^{d_m} \frac{1}{\sqrt{m}}\frac{C_m}{(a-1)} \sqrt{\frac{\delta p}{2 \pi}}=1.
\end{align}
However, it is not clear whether the solution for \cref{MustSatisfy} exists, and in particular if it satisfies the condition present in \Cref{CLoseFormExpressionCDFNegativeBinomial}, i.e. the solution must satisfy that it is larger than $\frac{1}{1-p}$.
In the next proposition, we argue that there exists one and only one solution for $a$ in  \cref{MustSatisfy}, satisfying $a>\frac{1}{1-p}$.
\begin{proposition}\label{lemma:OneSolRoot}
For any $\alpha>1$, define 
\begin{align}\label{eq:gPAlpha}
g_p(\alpha)=\
& 2 \frac{\alpha^{\alpha}}{\left( \alpha-1\right)^{ \alpha-1}} p^{\alpha-1}\left(1-p \right)-1.
\end{align}
Then, for each $0<p<1$, there exists exactly one root for $g_p(\alpha)$, $\alpha_p$, which satisfies $\alpha_p>\frac{1}{1-p}$. 
\begin{proof}
See Appendix \ref{Appendix:ExaclyOneRoot}.
\end{proof}
\label{Exsist_root}
\end{proposition}
In \Cref{fig:RootGraph}, we plot the roots of $g_p \left(\alpha \right)$: $\alpha_p$, and $\alpha_p^{*}$, together with the maximum value, i.e $\alpha_{\max}=\frac{1}{1-p}$.
It can be seen from the graph that $\alpha_{p} > \frac{1}{1-p} > \alpha_{p}^{*}$, for all $0 < p < 1$, and in particular, there is exactly one root satisfying $\alpha_{p} > \frac{1}{1-p}$. Hence, we can conclude that there exists a unique solution for  \cref{MustSatisfy} subject to $a>\frac{1}{1-p}$, which implies  that $b_n$ has the form $b_n=\alpha_p \log_2 n+o\left(\log n \right)$.
The next lemma presents an explicit expression of $b_n$.
\begin{lemma}\label{limitConditionIID}
Let $F_c \left(x \right)$ be the CDF of a continuous random variable as defined in \cref{Fc},
for real $x>\log_2 n-1$, and define
\begin{align}\label{bnIID}
b_n=\alpha_p\log_2 n+ \log_{\delta} \sqrt{ \log_2 n} +\beta,
\end{align}
where $\alpha_p$ is the root of $g_p \left(\alpha\right)$,
and\nomenclature{$\beta$}{Some constant, defined in \cref{gammaSolution}.}
\begin{align}\label{gammaSolution}
  \delta=\frac{\alpha_p}{\alpha_p-1}p,
  \quad
\beta=\log_{\delta} \left( \sqrt{\frac{2 \pi}{\delta p} } \frac{\alpha_p-1}{C}  \right).
\end{align}
Then
\begin{align}
\lim_{n \to \infty} n\bar{F}_c\left(b_n+x \right)=\delta^x.
\end{align}
\begin{proof}\
By \Cref{lemma259},
\begin{align}
 &n\bar{F}_c\left(b_n+x\right) 
   \\ \nonumber
=&
 n C_n \left(\frac{\alpha_p^{\alpha_p}}{\left(\alpha_p-1 \right)^{\alpha_p-1}}p^{\alpha_p-1}(1-p) \right)^{\log_2 n} 
  \delta^{\log_{\delta} \sqrt{ \log_2 n} +\beta+x} \frac{1}{(\alpha_p-1)\sqrt{\log_2 n}} \sqrt{\frac{\delta p}{2 \pi}}\left(1+ o \left( 1 \right) \right)
  \\ \nonumber
=&
n \left(\frac{g_p \left(\alpha_p\right)+1}{2} \right)^{\log_2 n} \delta^{\beta+x} \frac{C_n}{\alpha_p-1} \sqrt{\frac{\delta p}{2 \pi}}\left(1+ o \left( 1 \right) \right)
  \\ \nonumber
=&
 \delta^{\beta+x} \frac{C_n}{\alpha_p-1} \sqrt{\frac{\delta p}{2 \pi}}\left(1+ o \left( 1 \right) \right)
  \\ \nonumber  
=&
\delta^x \frac{C_n}{C}
\left(1+ o \left( 1 \right) \right) \to \delta^x,
\end{align}
as $n \to \infty$.
\end{proof}
\end{lemma}
The next theorem gives bounds on the limit distribution of $\hat{M}_n$.
\begin{theorem}\label{TheoremExsistenseIID}
Let $u_n=b_n+x$, for $b_n$ defined in \cref{bnIID}. Then,
\begin{align}\label{exsistenseAndesros}
&\limsup_{n \to \infty} \Pr \left( \hat{M}_n \leq u_n \right) \leq e^{-\delta^x}
 \\ \nonumber &
\liminf_{n \to \infty} \Pr \left(  \hat{M}_n \leq u_n \right) \geq e^{-\delta^{x-1}}
\end{align}
\begin{proof}
recall that $F_c (x-1 ) \leq F \left[ x|\log_2 n,p \right] \leq F_c (x) $, therefore
\begin{align}
\Pr \left( \hat{M}_n \leq u_n \right) &= F \left[ u_n|\log_2 n,p \right]^n
 \\ \nonumber &
\leq F_c \left(u_n \right)^n
 \\ \nonumber &
= \left( 1-\frac{\delta^x}{n}+o\left(\frac{1}{n} \right) \right)^n \to e^{-\delta^x}
\end{align}
where the last step follows from \Cref{limitConditionIID}.
In the same way,
\begin{align}
\Pr \left( \hat{M}_n \leq u_n \right) &= F \left[ u_n|\log_2 n,p \right]^n
 \\ \nonumber &
\geq F_c \left(u_n-1 \right)^n
 \\ \nonumber &
= \left( 1-\frac{\delta^{x-1}}{n}+o\left(\frac{1}{n} \right) \right)^n \to e^{-\delta^{x-1}}
\end{align}
\end{proof}
\end{theorem}
\begin{remark}
\Cref{TheoremExsistenseIID} may be also used to extend the model an unreliable single-hop broadcast networks, as mention earlier, where $\kappa=\log_2 n$.
In particular, the theorem reveals that the number of redundant packets required to guarantee file completion by all receivers with probability bounded between $e^{-\delta^x}$ and $e^{-\delta^{x-1}}$ is $b_n+x$.
In \Cref{sec:ExtentionMmessages}, we will generalize it for $\kappa=M \log_2 n$, for some constant $M$.
\end{remark}
By \Cref{TheoremExsistenseIID}, the CDF of $\hat{M}_n-b_n$ is bounded by two CDFs of continuous random variables, each of which follows the Gumbel distribution.
Denote them as $T_u$ and $T_l$.
It can be verified that
\begin{align}
&\mathbb{E} \left[T_l \right]= -\frac{\gamma}{\log \delta}
 \\ \nonumber 
 &\mathbb{E} \left[T_u \right]=1-\frac{\gamma}{\log \delta},
\end{align} 
hence
\begin{align}\label{upperLowerBoundIIDCase}
 b_n-\frac{\gamma}{\log \delta}\leq \mathbb{E} \left[\hat{M}_n \right] \leq b_n-\frac{\gamma}{\log \delta}+1.
\end{align}
Now using \Cref{lemmaUpperBound}, we get the upper bound, i.e.
\begin{align}\label{eq:upperBoundResult}
\mathbb{E} \left[M_n \right] \leq b_n-\frac{\gamma}{\log \delta}+1.
\end{align}
\subsection{Summary}
\begin{figure}[t]
  \centering
    \includegraphics[width=0.7\textwidth]{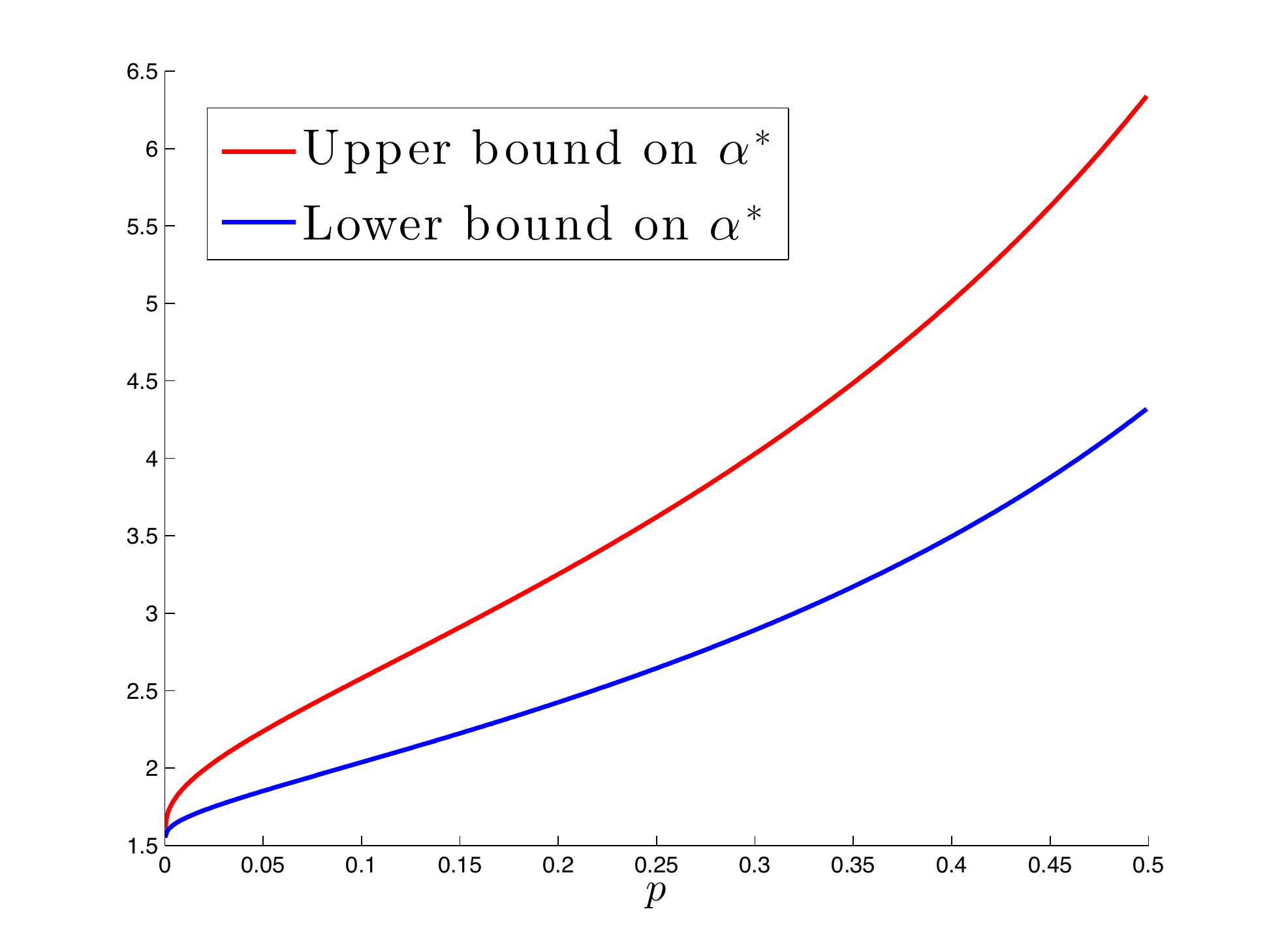}
    \caption{The upper and lower bounds on $\alpha^{*}$ as given in \cref{eq:boundsofAlpha}, where $\alpha^{*}$ was defined in \cref{eq:DominantComponent}.}
      \label{fig:UpperLoweAnaliticGraph}
\end{figure}

In this subsection, we summarize the results of the two previous subsections.
Joining the result of the lower bound in \Cref{sec:lowerBound}, and in particular  \cref{eq:LowerBoundResult}, and the result from the  upper bound from \Cref{sec:UpperBound} \cref{eq:upperBoundResult}, we have
\begin{align}\label{eq:conclutionOfUpperLowerBoundEq}
&\mathbb{E} \left[ M_n \right] \geq \left(\frac{1}{1-p}-\frac{1}{\log_2 p}\right) \log_2 n +o\left(\log n \right) 
 \\ \nonumber 
& \mathbb{E} \left[ M_n \right] \leq  \alpha_p \log_2 n +o\left(\log n \right),
\end{align}
where $\alpha_p$ is the roof of $g_p \left(\alpha \right)$, given in \cref{eq:gPAlpha}.
Since it is clear that the bounds differ only in the multiplicative constant factor of the dominant component ($\log n$), define
\begin{align}\label{eq:DominantComponent}
\alpha^{*}= \lim_{n \to \infty} \frac{\mathbb{E} \left[ M_n \right]}{\log n},
\end{align}
then, clearly by \cref{eq:conclutionOfUpperLowerBoundEq}
\begin{align}\label{eq:boundsofAlpha}
\left(\frac{1}{\left( 1-p\right) \log 2}-\frac{1}{\log p}\right)  \leq \alpha^{*} \leq \frac{\alpha_p}{\log 2} .
\end{align}
In \cref{fig:UpperLoweAnaliticGraph} we give the constants $\frac{1}{(1-p) \log 2}-\frac{1}{\log p}$ of the lower bound and $\frac{\alpha_p}{\log 2}$ of the upper bound on $\alpha^{*}$ for $0 < p \leq 0.5$.
We can see that the bounds are tight for small enough $p$.

In addition, by \Cref{TheoremExsistenseIID} and \Cref{lemmaUpperBound}, we conclude that for large enough $n$, 
\begin{align}
\Pr \left(M_n \leq x \right) \leq e^{-\delta^{x-b_n}}.
\end{align}
Therefore, for $\epsilon>0$, define the probability $1-\epsilon$ that after $T$ time slots, the message reached all nodes in $\mathbf{T}_n$.
Then, $1-\epsilon \leq  e^{-\delta^{T-b_n}}$, which implies that 
\begin{align}\label{TailUpperBound}
T \leq & b_n +\log_{\delta} \left(-\log \left(1- \epsilon \right) \right) 
 \\ \nonumber 
 =& b_n +\log_{\delta}\epsilon+O \left(\epsilon \right).
\end{align}
Namely, the time it takes to disseminate a single message to all nodes in $\mathbf{T}_n$ (with probability of $1- \epsilon$), \emph{is no more than} $b_n +\log_{\delta} \left(-\log \left(1- \epsilon \right) \right)$.

Simulation results of the CDF of $M_n$, and also $\mathbb{E}\left[M_n \right]$ with the proposed bounds, are available in \Cref{SimulationSection}.
In the next two sections, we generalize our model to the case where source send $M$ messages, and where $\mathbf{T}_n$ is K-ary tree.
\section{Completion time with $M$ messages}
\label{sec:ExtentionMmessages}
Thus far, we have discussed the case of disseminating a single message in the network.
However, in a practical system, it is most likely that one would need to disseminate several messages.
Fortunately, the methods above can easily extend to the general case.
For simplicity, we now assume a full binary tree, but, this result can be readily generalized to $K$ child nodes as well.

Regarding the lower bound, there is no change in the structure as given in \Cref{sec:lowerBound}, i.e., we divide $\mathbf{T}_n$ vertically at level $\log_2 k_n$, where there is only a single path which transmits the message from the source to the roots of $\mathbf{T}_{k_n}^{(1)},\mathbf{T}_{k_n}^{(2)},\ldots,\mathbf{T}_{k_n}^{(n/k_n)}$, (see \cref{fig:LowerBinaryTree}).
The time it takes to forward $M$ messages one after another through $\log_2 \frac{n}{k_n}$ hops is denoted by $W$.
Hence, $W \sim NB \left(M \log_2 \frac{n}{k_n},p \right)$.
In addition, the asymptotic approximation of the normalised $Y$ (where $Y$ denoted the maximum of the completion times in $\mathbf{T}_{k_n}^{(1)},\mathbf{T}_{k_n}^{(2)},\ldots,\mathbf{T}_{k_n}^{(n/k_n)}$), remains unchanged, expect that $\log_2 k_n$ is multiplied by a constant $M$.
Therefore, by \Cref{General}, for $m_n=M \log_2 k_n$, we get that $n F_c \left(b_n+x \right) \to p^x$, for
\begin{multline}
 b_n=\log_{\frac{1}{p}} n+ \left( M \log_2 k_n -1 \right) \log_{\frac{1}{p}} \left(  \frac{1-p}{p}  \Psi \left( M \log_{2} k_n \right) \right)
 - \log_{\frac{1}{p}} \left( M\log_2 k_n -1 \right)!
\end{multline}
In fact, since $M$ is a constant (with respect to $n$), the result holds for any $k_n$ satisfying \cref{eq:suffientConditionLogKn}.
Moreover, from the same reason, it can be verified that \Cref{ConditionDtag,JointDistributionLemma} hold for $M \log_2 k_n$ instead of $\log_2 k_n$, and therefore 
\begin{align}
\mathbb{E} \left[ Y \right] \geq b_n-\frac{\gamma}{\log p},
\end{align}
and
\begin{align}\label{eq:LowerBoundGeneralM}
\mathbb{E} \left[ M_n \right] & \geq  \mathbb{E} \left[ M_n^l \right]
 \\ \nonumber &
=\mathbb{E} \left[ W+Y \right]
 \\ \nonumber &
  \geq \frac{M \log_2 n/k_n}{1-p}+b_n-\frac{\gamma}{\log p}
   \\ \nonumber &
 =\left( \frac{M}{\left( 1-p \right) \log 2}-\frac{1}{\log p} \right) \log n+o\left(\log n \right).
\end{align}

\begin{figure}[t]
\centering
\includegraphics[width=0.7\textwidth]{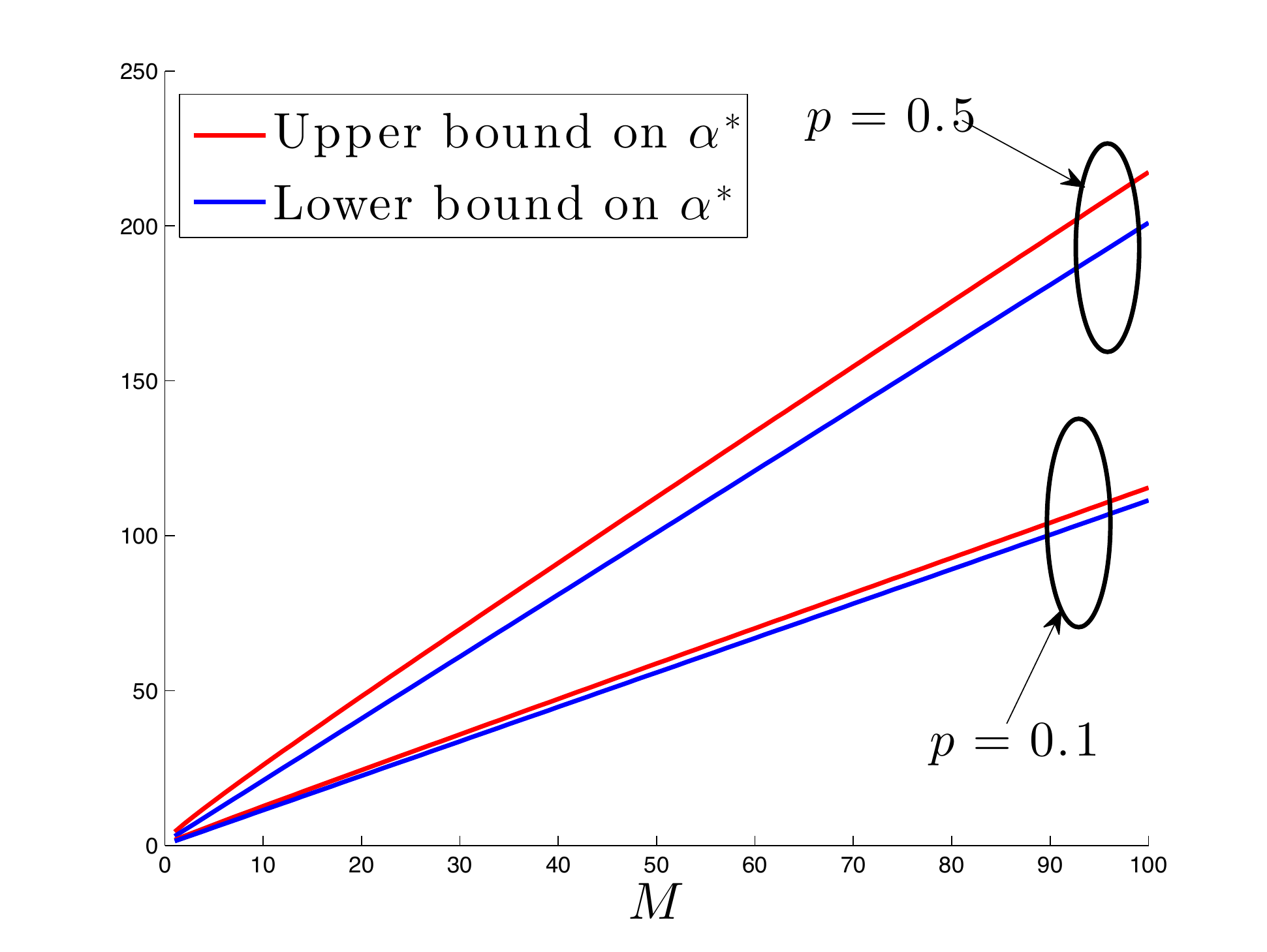}
\caption{Upper an  lower bounds on $\alpha^{*}$, given in \cref{eq:UpperBoundGeneralM,eq:LowerBoundGeneralM}, ($\alpha^{*}$ was defined \cref{eq:DominantComponent})  as a function of $M$, for $p=0.1,0.5$.}
      \label{fig:scaleLawM}
\end{figure}

Regarding the upper bound, similar to \Cref{sec:UpperBound}, define $\left( \hat{T}_1,\ldots, \hat{T}_n \right)$ to be an i.i.d.\ sequence with NB distribution following $M \log_2 n$ successes, and define $\hat{M}_n= \max \left( \hat{T}_1,\ldots, \hat{T}_n \right)$.
Then, using 
\Cref{CLoseFormExpressionCDFNegativeBinomial,lemma259}, we can derive an asymptotic approximation for the CDF, with $m=M \log_2 n$.
As a result, \Cref{limitConditionIID} and \Cref{TheoremExsistenseIID} can be applied with the following $b_n$.
\begin{align}
b_n&=\alpha_{p,M}M \log_2 n+\log_{\delta} \sqrt{M \log_2 n}+\beta,
\end{align}
$\beta$ and $\delta$ were defined in \cref{gammaSolution},
and $\alpha_{p,M}$ is the solution for 
\begin{align}
\frac{a^a}{\left(a-1 \right)^{a-1}}p^{a-1}(1-p)=2^{-\frac{1}{M}}.
\end{align}
Finally, we argue that there is only one solution for $\alpha_{p,M}$, satisfying $\alpha_{p,M} >\frac{1}{1-p}$.
This can be seen by the proof of \Cref{lemma:OneSolRoot}, which is easily generalized by assigning $2^{\frac{1}{M}}$ instead of 2.
As a result, we have
\begin{align}\label{eq:UpperBoundGeneralM}
\mathbb{E} \left[ M_n \right] & \leq  \mathbb{E} \left[ \hat{M}_n \right]
 \\ \nonumber &
  \leq b_n-\frac{\gamma}{\log \delta}+1
   \\ \nonumber &
 = \frac{\alpha_{p,M}M}{\log 2} \log n +o \left(\log n \right).
\end{align}
Hence, by \cref{eq:LowerBoundGeneralM,eq:UpperBoundGeneralM}, we obtain
\begin{align}\label{eq:conclutionMmessages}
 \left( \frac{M}{\left( 1-p \right) \log 2}-\frac{1}{\log p} \right) \leq  \alpha^{*} \leq \frac{\alpha_{p,M}M}{\log 2},
\end{align}
were $\alpha^{*}$ was defined in \cref{eq:DominantComponent}.\\
\cref{fig:scaleLawM} depicts the leading multiplicative constant, of the upper and lower bounds, as outlined in \cref{eq:conclutionMmessages}.
We can conclude that both of these bounds grow with the same scale, as $M$ increase.
Indeed, the next lemma limits the solution for $\alpha_{p,M}$, to the area between $\frac{1}{1-p}$, and $\frac{1}{1-p}+\epsilon_M$, where $\epsilon_M \to 0$, as $M \to \infty$.
As a result, both upper and lower bounds have $\frac{1}{(1-p)\log 2}$ as the leading multiplicative constant.
\begin{lemma}\label{lemma:EVTLargeM}
Let,
\begin{align}
g_p(\alpha|M)=\
& 2^{\frac{1}{M}} \frac{\alpha^{\alpha}}{\left( \alpha-1\right)^{ \alpha-1}} p^{\alpha-1}\left(1-p \right)-1,
\end{align}
And denote the root of $g_p(\alpha|M)$ as $\alpha_{p,M}$.
Then, for sufficiently large $M$, we have
\begin{align}
\frac{1}{1-p} < \alpha_{p,M} \leq \frac{1}{1-p}+\epsilon_M,
\end{align}
where $\epsilon_M>0$ is a slowly decreasing function, such that $\epsilon_M =\Omega \left(\frac{1}{\sqrt{M}} \right) $, as $M \to \infty$.
Namely, $\epsilon_M  \to 0 $ slower than $\frac{1}{\sqrt{M}}$. 
\begin{proof}\label{Lemma:MmessageSolution}
See Appendix \ref{appendix:ProofTheorem6}.
\end{proof}
\end{lemma}
An immediate conclusion from \Cref{lemma:EVTLargeM} is that for large $n$ and large $M$, $\mathbb{E} \left[ \hat{M}_n \right]$ scales as $\frac{M\log_2 n}{1-p}$.
As a result, since both upper and lower bounds have the same scaling law, we obtain the completion time of $M_n$ scales as  $\frac{M\log_2 n}{1-p}$, i.e.,
\begin{align}
\mathbb{E} \left[ M_n \right]=\frac{M\log_2 n}{1-p}+o\left(M\log n \right).
\end{align}
In addition, recall that the arrival time of some certain end user $i$ follows NB with $M\log_2 n$ successes.
Hence, $\mathbb{E} \left[ T_i \right]=\frac{M \log_2 n}{1-p}$,
which implies that the expected completion time to disseminate $M$ messages to $n$ end-users through $\mathbf{T}_n$, \emph{scales as the same as the time to forward it to a single terminal node } with a hop count of $\log_2 n$. 
\section{Completion time with $K$ child nodes}
\label{sec:ExtentionTreeWithKChild}
In this chapter, we generalize our problem, to the case where each node in the multicast tree has exactly $K$ child nodes. 
We distinguish between two cases:
First, when $K$ is a constant, and second, where $K$ is an increasing sequence with $n$.
For both cases, we assume that the multicast tree has $n$ end users (leaves).
The height of the full K-ary tree with $n$ leaves is
\begin{align}
h_n=\log_K n=\frac{\log n}{\log K}.
\end{align}
\subsection{Constant K}\label{sec:KIsConstant}
\begin{figure}[t]
\centering
    \includegraphics[width=0.7\textwidth]{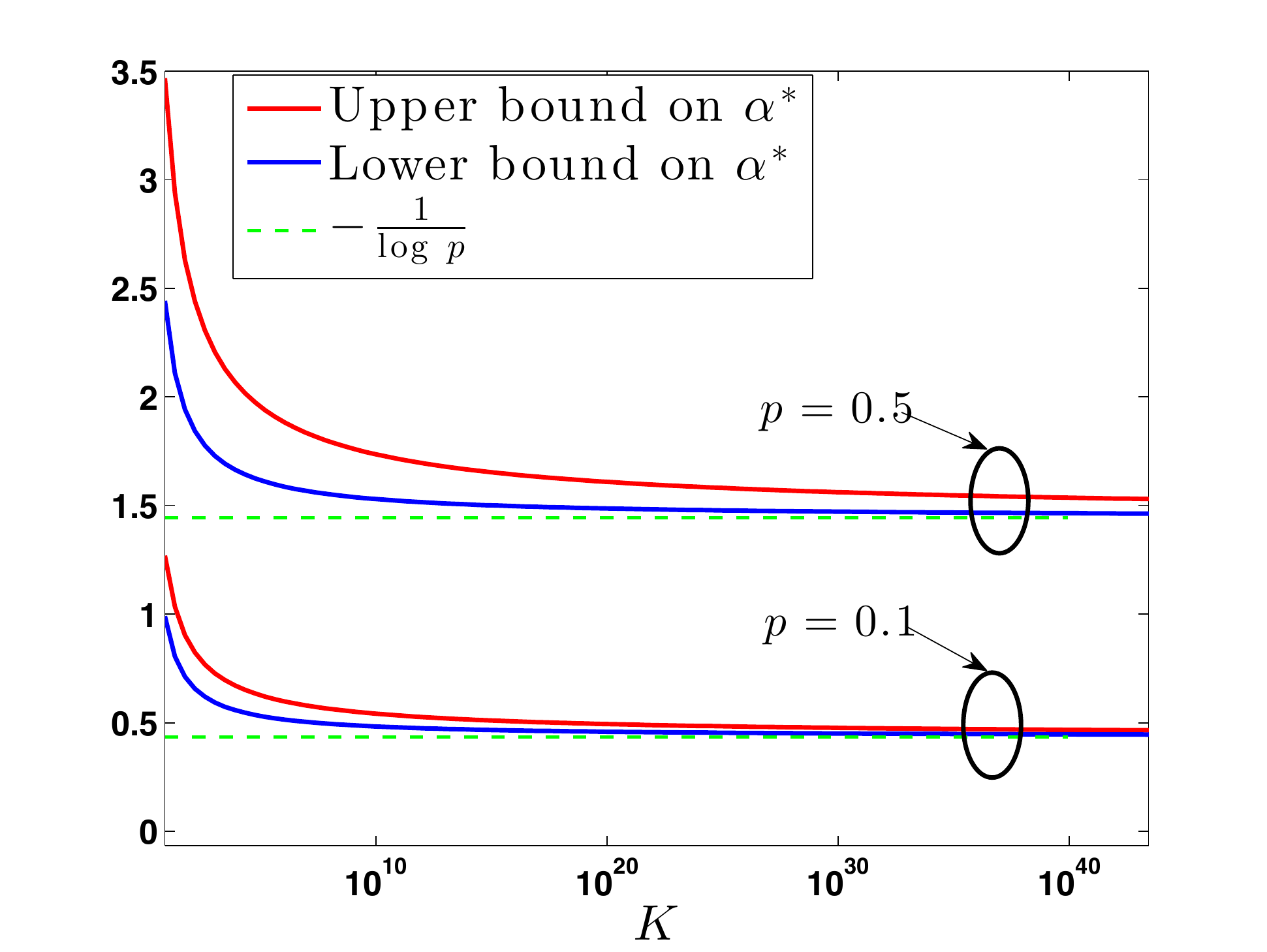}
    \caption{Graph of the the upper and lower bounds on $\alpha^{*}$, as a function of $K$, with exponential scale, for $p=0.1,0.5$. The coefficient of the lower bound is $\frac{1}{log K}\frac{1}{1-p}-\frac{1}{\log p}$ and the upper bound is $ \frac{\alpha_{p,K}}{\log K}$.
   We can see that both these bounds, are converge (relatively slow) to $-\frac{1}{\log p}$.}
      \label{fig:upperLowerBoundAnaliticFunctionOfK}
\end{figure}

We start with the case where $K$ is constant.
We will consider here the limit as $K$ grows, however only in the sense that $n \to \infty$ is taken first.
Regarding the upper bound, we would like to find an asymptotic approximation for the distribution of $\hat{M}_n$, i.e., the maximum of $n$ i.i.d.\ random variables, each following a NB distribution with $h_n$ successes.
Hence, we can use \Cref{CLoseFormExpressionCDFNegativeBinomial,lemma259} in order to derive an asymptotic approximation for the CDF, with $m=h_n$.
Consequently, \Cref{limitConditionIID} and \Cref{TheoremExsistenseIID} can be applied with
\begin{align}\label{eq:GeneralBnIIDCase}
b_n&=\alpha_{p,K} h_n+\log_{\delta} \sqrt{h_n}+\beta 
\\ \nonumber &
=\frac{\alpha_{p,K}}{\log K } \log n+o \left( \log n \right),
\end{align}
where $\beta$ and $\delta$ were defined in \cref{gammaSolution},
but the generalized solution for $\alpha_{p,K}$ is no longer \cref{MustSatisfy}, but
\begin{align}\label{MustSatisfy2}
\frac{a^a}{\left(a-1 \right)^{a-1}}p^{a-1}(1-p)= \frac{1}{K}.
\end{align}
Finally, we argue that there is only one solution for $\alpha_{p,K}$, satisfying $\alpha_{p,K} >\frac{1}{1-p}$.
This can be seen by the proof of \Cref{lemma:OneSolRoot}, which can be generalized easily by substituting $K$ instead of 2.
As a result, 
\begin{align}\label{eq:UpperBoundGeneralK}
\mathbb{E} \left[ M_n \right] & \leq  \mathbb{E} \left[ \hat{M}_n \right]
 \\ \nonumber &
  \leq b_n-\frac{\gamma}{\log \delta}+1
   \\ \nonumber &
 = \frac{\alpha_{p,K}}{\log K}\log n+o \left(\log n \right).
\end{align}
Regarding the lower bound, there is no change in the structure as given in in \Cref{sec:lowerBound}, except that the partition is done at level $\log_K k_n \in 1,2,\ldots, \log_K n-1$.
Hence $W \sim NB \left( \log_K \frac{n}{k_n},p \right)$.
In addition, the asymptotic approximation of the normalised $Y$ remains unchanged, expect that $\log_2 k_n$ is replaced by $\log_K k_n$.
Namely, for sufficiently large $n$,
\begin{align}
e^{-p^{x-1}} \leq \Pr \left(Y \leq u_n \right) \leq e^{-p^x},
\end{align} 
for $u_n=b_n+x$, where
\begin{align}\label{eq:b_nLowerBoundGeneralK}
 b_n=\log_{\frac{1}{p}} n+ \left( \log_K k_n -1 \right) \log_{\frac{1}{p}} \left(  \frac{1-p}{p} \Psi \left( \log_{K} k_n \right) \right)
 - \log_{\frac{1}{p}} (\log_K k_n -1)!
\end{align}
For further details see Appendix \ref{Appendix:ExtentionKChildLowerBound}.\\
Therefore, we write the lower bound as
\begin{align}\label{eq:expectedValueLowerBlouldGeneralK}
\mathbb{E} \left[M_n^l \right] & \geq \frac{\log_{K} n/k_n}{1-p}-\frac{\gamma}{\log p}+b_n
\\ \nonumber &
= \left(\frac{1}{\left(1- p \right)\log K}-\frac{1}{\log p} \right) \log n +o \left(\log n \right)
\end{align}
Note that for a fix $n$, we expect that $\mathbb{E} \left[ M_n \right]$ would decrease as $K$ increase, since the number of successes required for each end-user (i.e. $\frac{\log n}{\log K}$) is decreased as a function of $K$.
As a result, we expect that $\alpha^{*}$ would also decrease as $K$ increase.
Indeed, for the same reason, this claim holds also for the i.i.d.\ case.
Hence, we also expect that the upper bound on $\alpha^{*}$ (i.e., $\frac{\alpha_{p,K}}{\log K}$) would decrease.
For the lower bound, it is easy to see in \cref{eq:expectedValueLowerBlouldGeneralK} that the lower bound on $\alpha^{*}$ indeed decreases as a function of $K$.
In \cref{fig:upperLowerBoundAnaliticFunctionOfK} we plot these bounds on $\alpha^{*}$ as a function of $K$  for $p=0.1$ and $p=0.5$.
For the lower bound, since the multiplicative constant of the leading term is $\frac{1}{log K}\frac{1}{1-p}-\frac{1}{\log p}$, it is easy to see that for $K \to \infty$ the expression goes to $-\frac{1}{\log p}$.
However, we can also see that the upper bound tends to $-\frac{1}{\log p}$ as well, for large $K$, which implies that these bound are getting tighter as $K$ increases.
Note that $-\log p$ is a  limit inferior for $\alpha^{*}$, since each end-user has at least one an exclusive channels.
Thus, as we seen in \Cref{sec:lowerBound}, the completion time of disseminating to $n$ nodes, scales as $\log_{\frac{1}{p}} n$.

\subsection{$K$ growing with n.}
Next, we consider the case where $K$ (in fact $K_n$) is monotonically increasing with $n$.
We distinguish between two cases: the first is where the height of the tree (denoted by $h$) is constant, and in the second it is monotonically increasing with $n$.
\subsubsection{Constant h}
As mentioned above, we can bound $\mathbb{E} \left[ M_n \right]$ from above by the expected value of the maximum of an i.i.d.\ sequence with the same marginal distribution.
Since $h$ is constant, then clearly by  \Cref{ConditionA}, for $\log_2 k_n=h$,
\begin{align}
 \mathbb{E} \left[ M_n \right] & \leq b_n+\frac{\gamma}{\log p}+1
  \\ \nonumber &
 = \log_{\frac{1}{p}}n+o \left(\log n \right),
\end{align}
 where
\begin{align}\label{b_nForConstantH}
b_n=\log_{\frac{1}{p}} n+ \left( h -1 \right) \log_{\frac{1}{p}} \left(  \frac{1-p}{p}  \Psi \left( h \right) \right)
 - \log_{\frac{1}{p}} (h-1)!
\end{align}
Furthermore, \Cref{EVT_for_nonstationary_integer_valued} guarantees that if both local and long range dependence conditions are satisfied, then it is possible to obtain tight bounds (which differ only by 1) for $\mathbb{E} \left[ M_n \right]$.\\
Note that $\mathbf{T}_n$ has $K_n$ child nodes, each having $n/K_n$ leaves, which are independent of each other.
Hence, $\mathbf{T}_n$ can be considered as $K_n$ dependence (namely, the set with the indices $\lbrace 1,\ldots,K_n-1 \rbrace, \lbrace K_n,\ldots,2K_n-1 \rbrace,\ldots$ are independent), which implies that  the mixing condition holds.
However, unfortunately,  $D_{k_n}^{'}\left(u_n \right)$ does not hold.
For more details, see Appendix \ref{Appendix:NonExsistenceConditionDun}.
Therefore, it is impossible to ensure a tight bounds for $\mathbb{E} \left[ M_n \right]$ using \Cref{EVT_for_nonstationary_integer_valued}.\\
Nevertheless, similar to the lower bound as presented in \Cref{sec:lowerBound}, denote $M_n^l$ to be the maximum of $n$ random variables that follow a NB distribution with $h$ successes, where the first $h-1$ channels are in common for all users, (denoted as $W$), whereas the last channel is independent for each one of the users, denoted as $\left(Y_1,\ldots,Y_n \right)$, and $Y=\max \left(Y_1,\ldots,Y_n \right)$, where $Y_i$ are i.i.d.\ geometric random variables.
As a result, we have
\begin{align}
 \mathbb{E} \left[ M_n^l \right]
& = \mathbb{E} \left[ W \right]+\mathbb{E} \left[Y \right]
  \\ \nonumber &
 \geq \frac{h-1}{1-p}+\log_{\frac{1}{p}} n-\frac{\gamma}{\log p}
   \\ \nonumber &
 =  \log_{\frac{1}{p}} n+o \left(\log n \right),
\end{align}
where $\mathbb{E} \left[ W \right]$ is the expected valued for a NB with $h-1$ successes, and $\mathbb{E} \left[Y \right]$ was already found in \cref{eq:expectedValueOFgeometric}.\\
In conclusion, for constant $h$, although it is not possible to obtain the exact expression, we may still find very tight upper and lower bounds, such that both scale as $\log_{\frac{1}{p}} n$, namely both have the same constant multiplicative factor.
Hence, the scaling law is known exactly.
Note that, unlike previous cases, when $h$ is constant, the correlation between any two sets of random variables, is not increasing since the number of the correlated channels (i.e., channels that are mutual for both sets) is not increased, which causes the scaling law to behave as the i.i.d.\ case.
\subsubsection{$h_n$ is an increasing sequence}
\label{sec:ExtentionsK_nH_n}
Similar to the previous subsection, the lower bound is also valid in this case.
For the upper bound, note that by \Cref{General}, and specifically by \cref{eq:sufficentCondtiionMn}, the result of the previous subsection can be applied also for $h_n$, as long as it satisfies
\begin{align}\label{eq:SuffientConditionHn}
\frac{h_n^2 \log \log n}{\log n} \to 0,
\end{align}
as $n \to \infty$.
When \cref{eq:SuffientConditionHn} is not satisfied, we found that it is possible to apply \Cref{TheoremExsistenseIID} on the distribution of $\hat{M}_n$, for
\begin{align}
b_n=\alpha_{p,K_n} \log n+\log_{\delta} \sqrt{\log n \log K_n}+\beta,
\end{align} 
where
\begin{align}
\delta=\frac{\alpha_{p,K_n} -\frac{1}{log K_n}}{\alpha_{p,K_n} }, \quad \beta=\log_{\delta} \left(  \frac{a-\frac{1}{\log K_n }}{C}\sqrt{\frac{2 \pi}{\delta p}} \right)
\end{align}
and $\alpha_{p,K_n}$ is the solution for 
\begin{align}\label{eq:GeneralizedalphaP}
 \log K_n^{\frac{1}{\log K_n}} \frac{a^a}{\left( a-\frac{1}{\log K_n} \right)^{ a-\frac{1}{\log K_n}}} p^{a-\frac{1}{\log K_n}}\left(1-p \right)^{\frac{1}{\log K_n}} =e^{-1}.
\end{align}
Therefore,
\begin{align}
\mathbb{E} \left[ M_n \right] & \leq  \mathbb{E} \left[ \hat{M}_n \right]
 \\ \nonumber &
  \leq b_n-\frac{\gamma}{\log \delta}+1
   \\ \nonumber &
 = \alpha_{p,K_n}\log n+o \left(\log n \right).
\end{align}
For further details see Appendix \ref{Appendix:ExtensionIIDKnChildren}.
\section{Simulation Results}\label{SimulationSection}
In this section, we present some simulation results in order to examine important aspects of the analytical result given thus far, such as convergence rates and the tightness of the bounds.
The simulation clearly depict the tightness of the bounds given, and their ability to closely approximate the correct completion time under various system parameters.
In some cases, however it is clear that the number of nodes should be large before the asymptotic  kicks in. 

\begin{figure*}%
\centering
\begin{subfigure}{0.5\columnwidth}
\centering
  \includegraphics[width=1.2\textwidth,right]{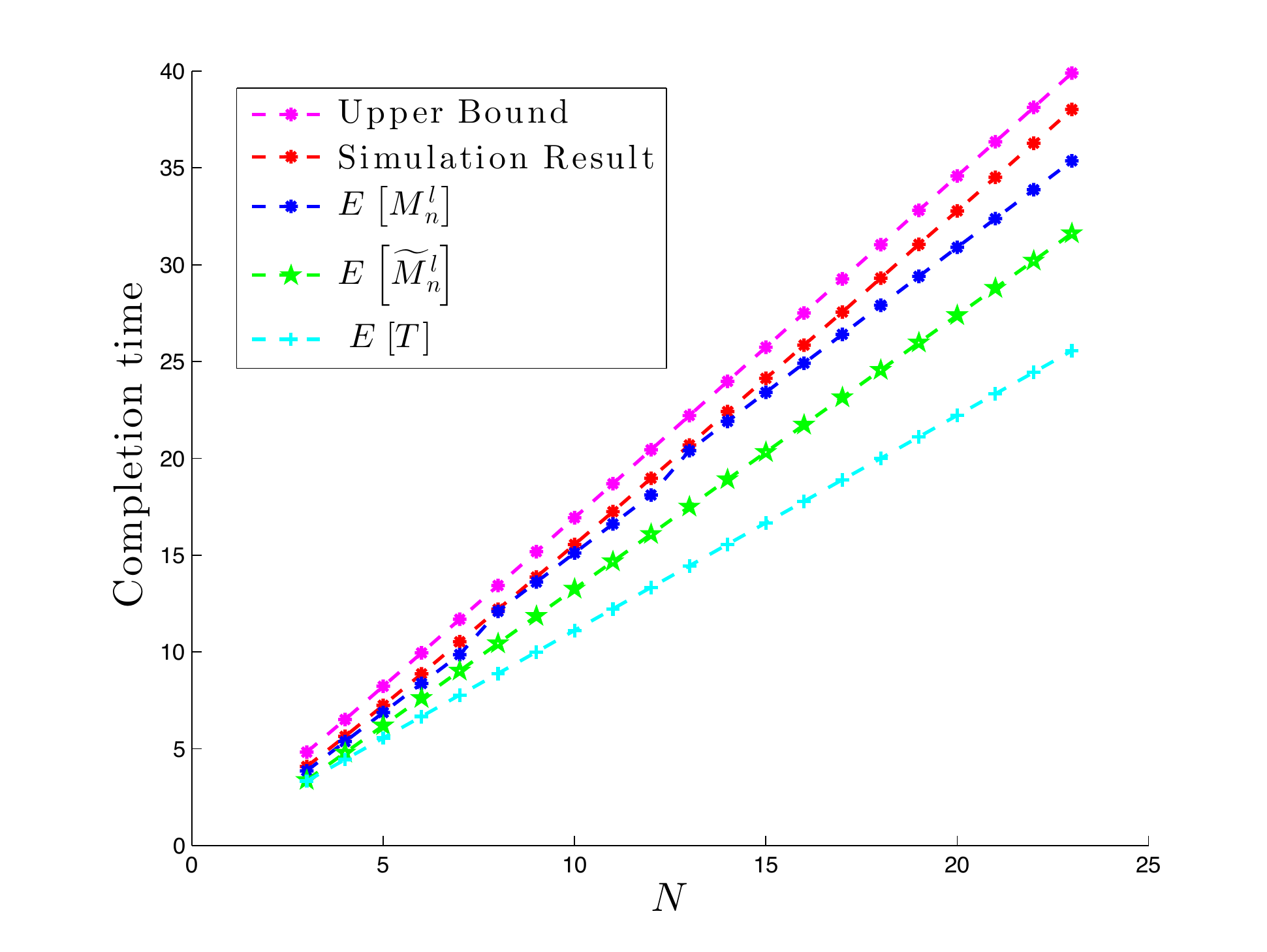}
\caption{$p=0.1$}%
\label{subfig:simulationExpectedValuep01V2}%
\end{subfigure}\hfill%
\begin{subfigure}{0.5\columnwidth}
\centering
  \includegraphics[width=1.2\textwidth,left]{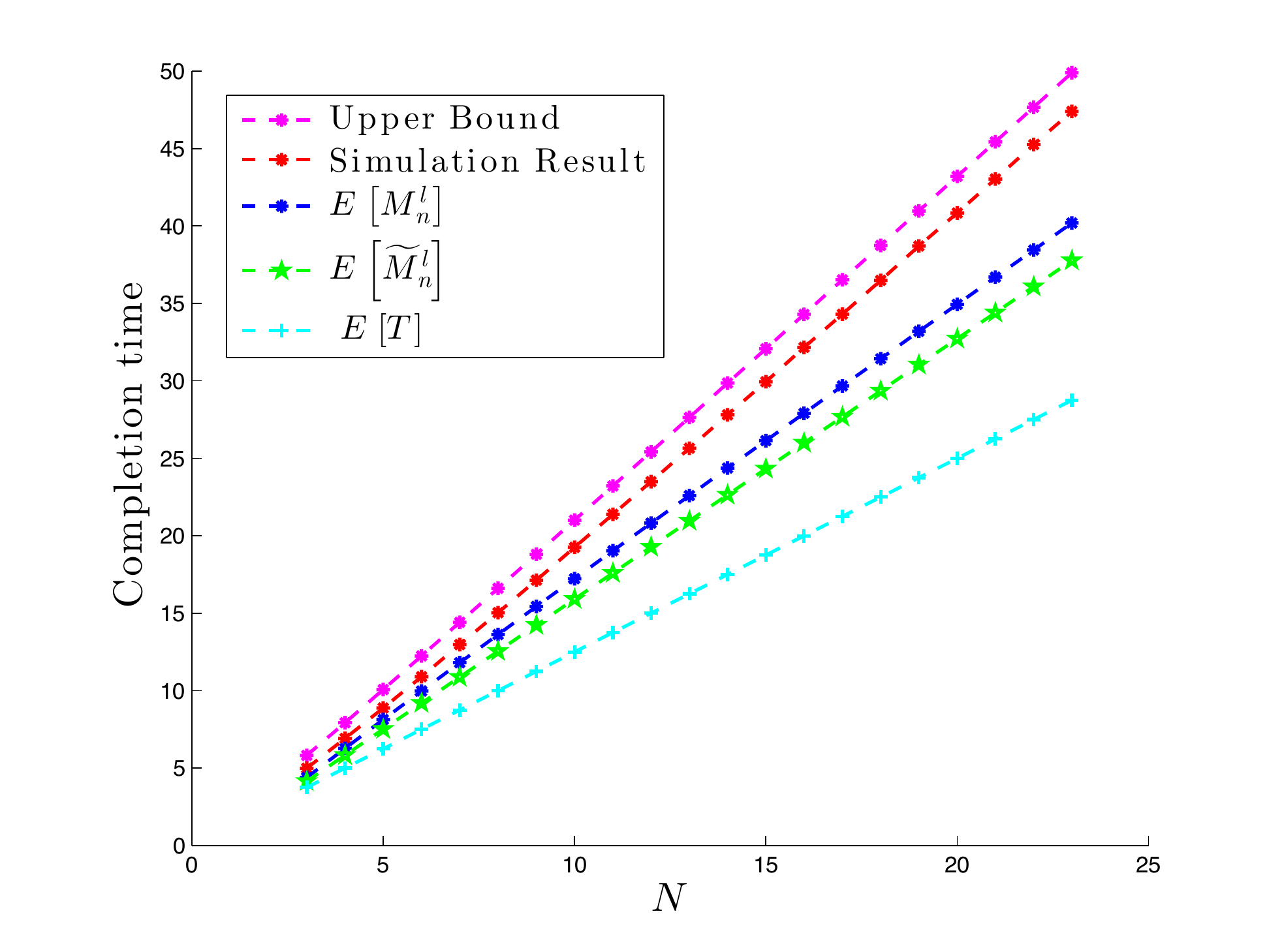}
\caption{$p=0.2$}%
\label{subfig:simulationExpectedValuep02V2}%
\end{subfigure}\hfill%
\begin{subfigure}{0.5\columnwidth}
\centering
  \includegraphics[width=1.2\textwidth]{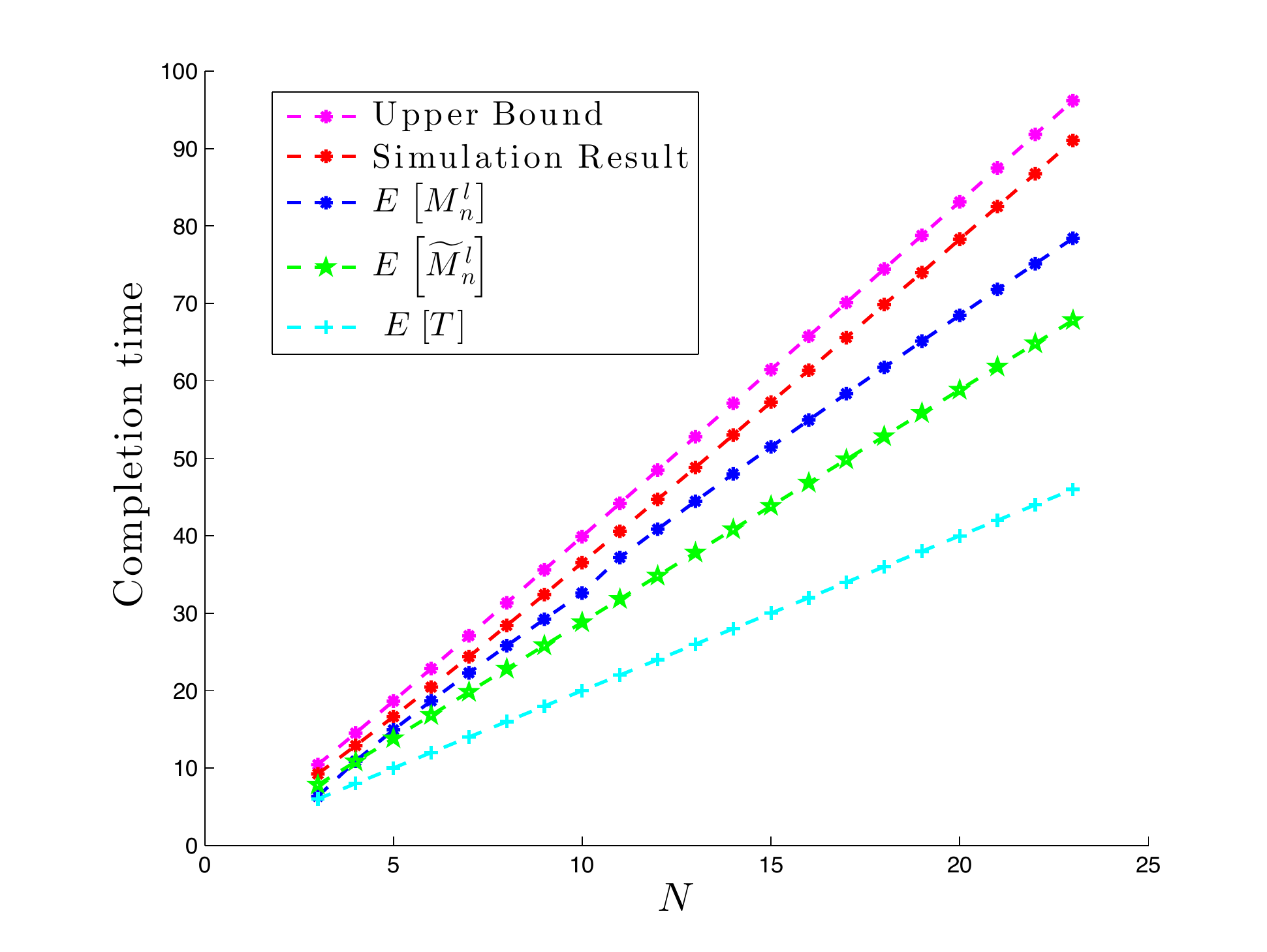}
\caption{$p=0.5$}%
\label{subfig:simulationExpectedValuep05V2}%
\end{subfigure}%
\caption{Simulations for $\mathbb{E} \left[ M_n \right]$, $p=0.1,0.2,0.5$ (averaging $10^4$ realizations), together with the upper bound ($\alpha_p=1.78807,2.25379,4.4035$ respectively), as given in \cref{eq:upperBoundResult}, and three different lower bounds: the simple one written in the graph as $\mathbb{E} \left[ T \right]$ and outlined in \cref{eq:TrivialLowerBound};
The second one, $\mathbb{E} \left[ \widetilde{M}_{n}^{l} \right]$ corresponding to \cref{eq:TrivialLowerBoundV2};
Finally,  $\mathbb{E} \left[ M_{n}^{l} \right]$ as in \cref{eq:LowerBoundResult}, where for $p=1$, (\cref{subfig:simulationExpectedValuep01V2}), we have chosen $\log_2 k_n=4$ for $N=3,\ldots,10$ and $\log_2 k_n=7$ for $N=11,\ldots,23$.
For $p=0.2$ (\cref{subfig:simulationExpectedValuep02V2}), $\log_2 k_n=4$ for all $N$.
Finally, for $p=0.5$ (\cref{subfig:simulationExpectedValuep05V2}), $\log_2 k_n=4$ for $N=3,\ldots,10$ and $\log_2 k_n=7$ for $N=11,\ldots,23$.} 
\label{fig:simulationExpectedValuep010205}
\end{figure*}

The first simulation was done for the main result, that is, the expected value of the completion time, $\mathbb{E} \left[ M_n \right]$.
In particular, we considered a multicast network, with a structure of a full binary tree $\mathbf{T}_n$, with $n=2^N$, $N=4, \ldots,23$, and for  $p=0.1,0.2,0.5$, and we checked the expected completion time to disseminate a single packet over $\mathbf{T}_n$.
We evaluated the suggested completion time through extensive MATLAB simulations, and the results are given in \cref{fig:simulationExpectedValuep010205}.
For each $N$ and $p$, we averaged over $10^4$ realizations, where on each observed value, we found the maximum of $2^N$ random variables, each of which is the sum of $N$ geometric random variables with probability of failure $p$, where each geometric random variables represents the numbers of trials needed to forward the packet through a certain channel.
We compared the result to the upper bound, i.e. $\mathbb{E} \left[ \hat{M}_n \right]$, given in \cref{eq:upperBoundResult}, and the analytic result of the three lower bounds, as proposed earlier in \cref{eq:TrivialLowerBound,eq:TrivialLowerBoundV2,eq:LowerBoundResult}.
Recall that for the lower bound proposed in \cref{eq:LowerBoundResult}, we have a degree of freedom, by the choice of $k_n$, where large $k_n$ results in a tighter bound, but the convergence may be slower.
Hence, in the simulations below, $k_n$ was chosen for the tightest bound  possible.
However, to show the dependence on $k_n$, we conclude this section  with simulations that specifically address this issue, and give bounds of different values of $k_n$.

\begin{figure}[t]
\centering
    \includegraphics[width=0.7\textwidth]{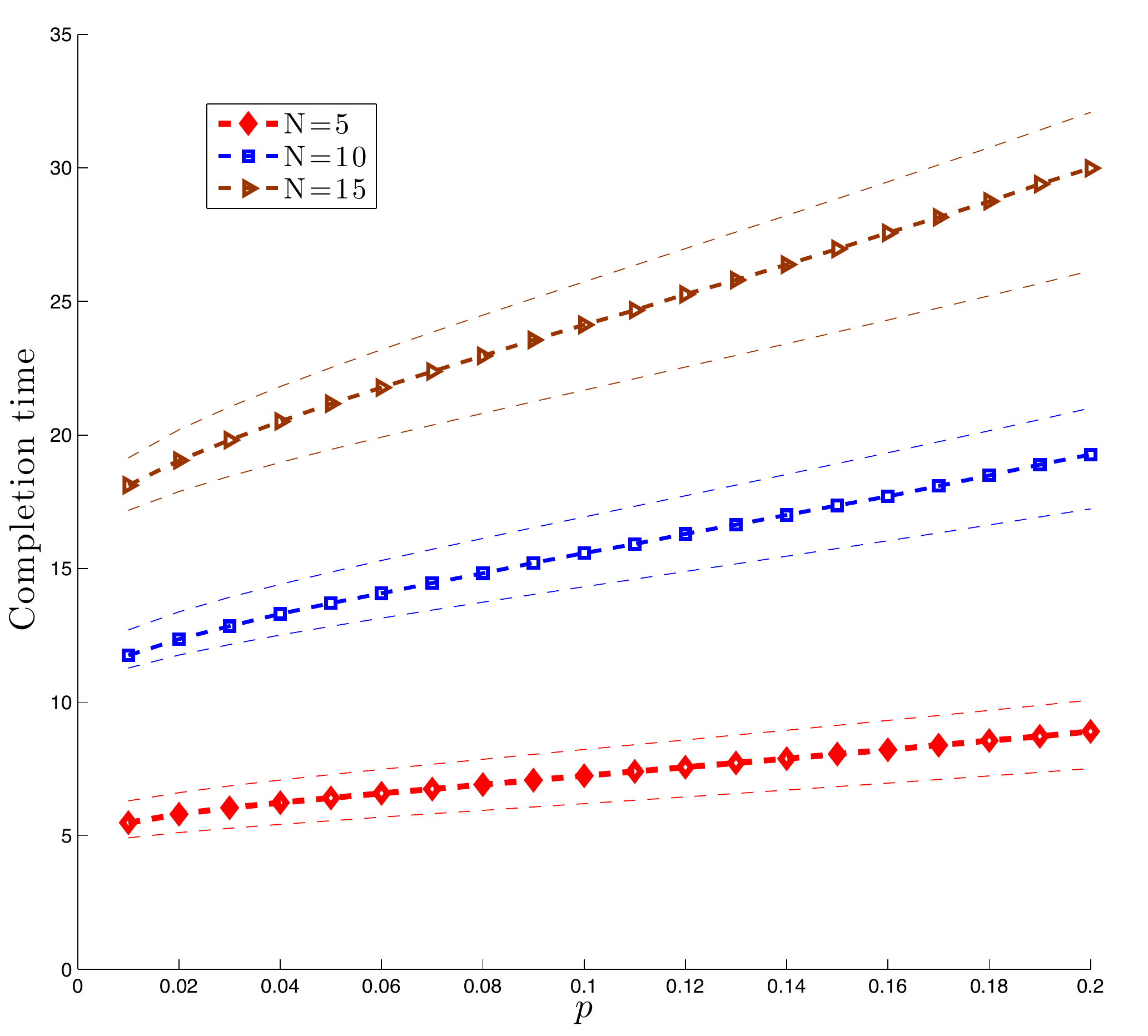}
    \caption{Simulations for $\mathbb{E} \left[ M_n \right]$, for $N=5,10,15$ as a function of $p$ (averaging 5000 realizations), together with the upper lower bounds for $\log_2 k_n=4$, given in \cref{eq:upperBoundResult,eq:LowerBoundResult}, respectively.
The bounds for any $N$ are the corresponding dashed lines.}
        \label{fig:simulationExpectedValue_p_function}
\end{figure}

In addition, we also plot in  \cref{fig:simulationExpectedValue_p_function} a simulation for $N=5,10,15$ as a function of $p$, for $0 \leq p \leq 0.2$.
As shown in these graphs, the bounds hold for each $N$, including relatively small $N$.
Moreover, one can see that the gap between the simulations results of $\mathbb{E} \left[ M_n \right]$ and the upper bound which is based on the i.i.d.\ scaling law, is increases as $N$ increase.
This implies that it is indeed not possible to apply \Cref{EVT_for_nonstationary_integer_valued} for the problem at hand, as \Cref{EVT_for_nonstationary_integer_valued} would imply that the extreme value distribution of the correlated sequence should converge to the same limiting distribution as if the sequence were i.i.d.

\begin{figure*}%
\centering
\begin{subfigure}{0.5\columnwidth}
  \includegraphics[width=1.2\textwidth,right]{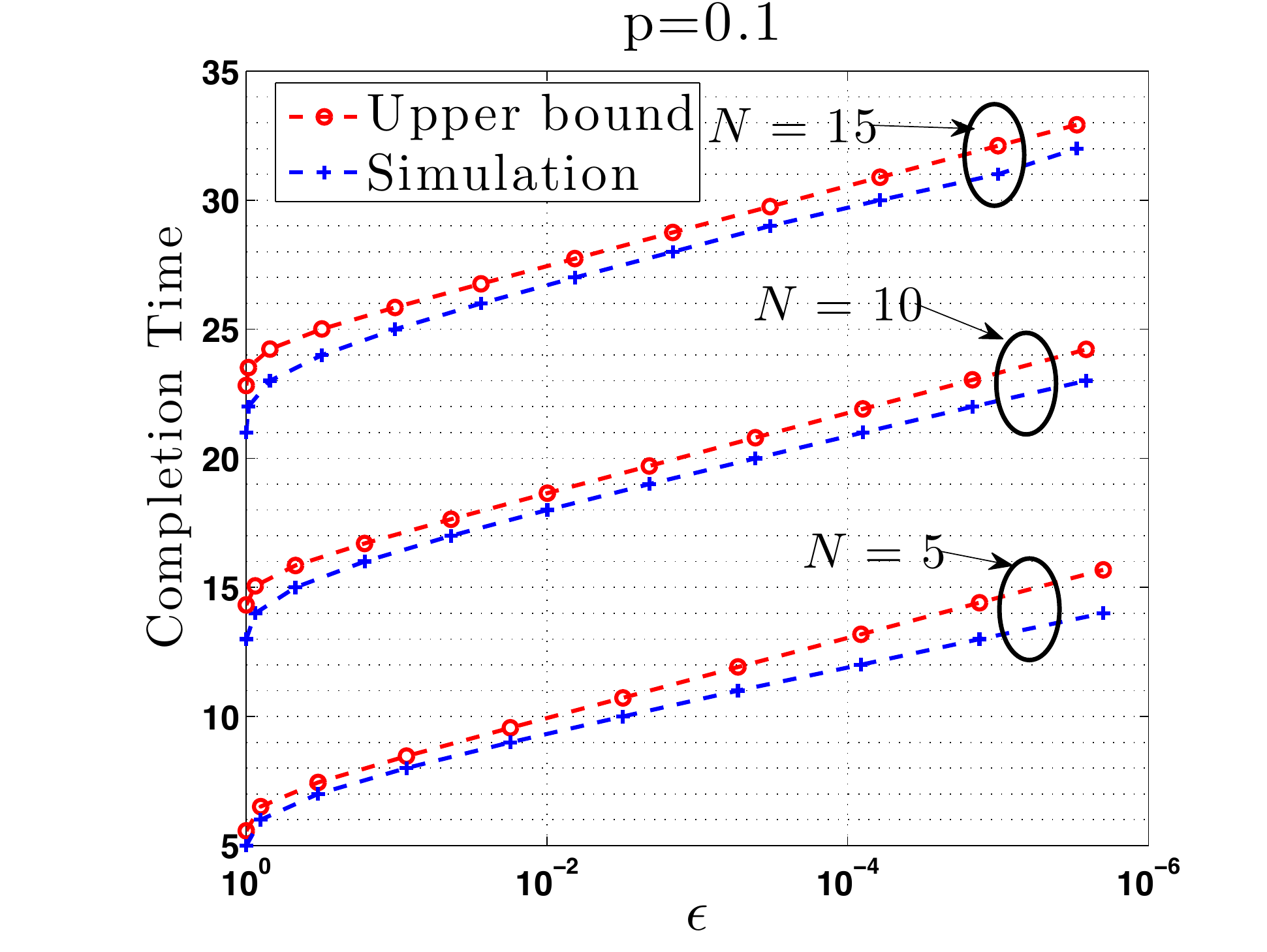}
        \label{fig:TailDistibutionp01}%
\end{subfigure}\hfill%
\begin{subfigure}{0.5\columnwidth}
  \includegraphics[width=1.2\textwidth,left]{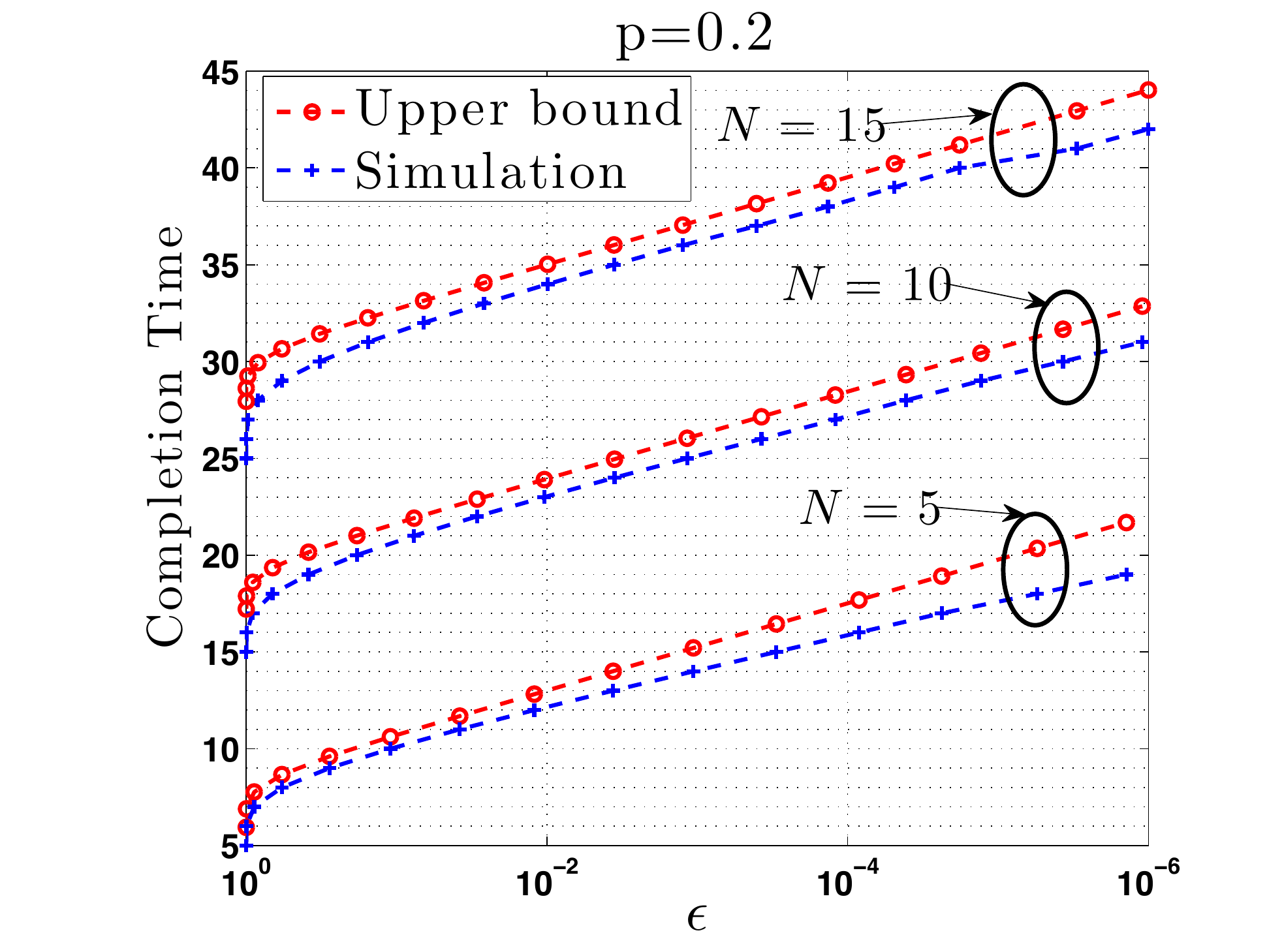}
        \label{fig:TailDistibutionp02}
\end{subfigure}\hfill%
\begin{subfigure}{0.5\columnwidth}
  \includegraphics[width=1.2\textwidth]{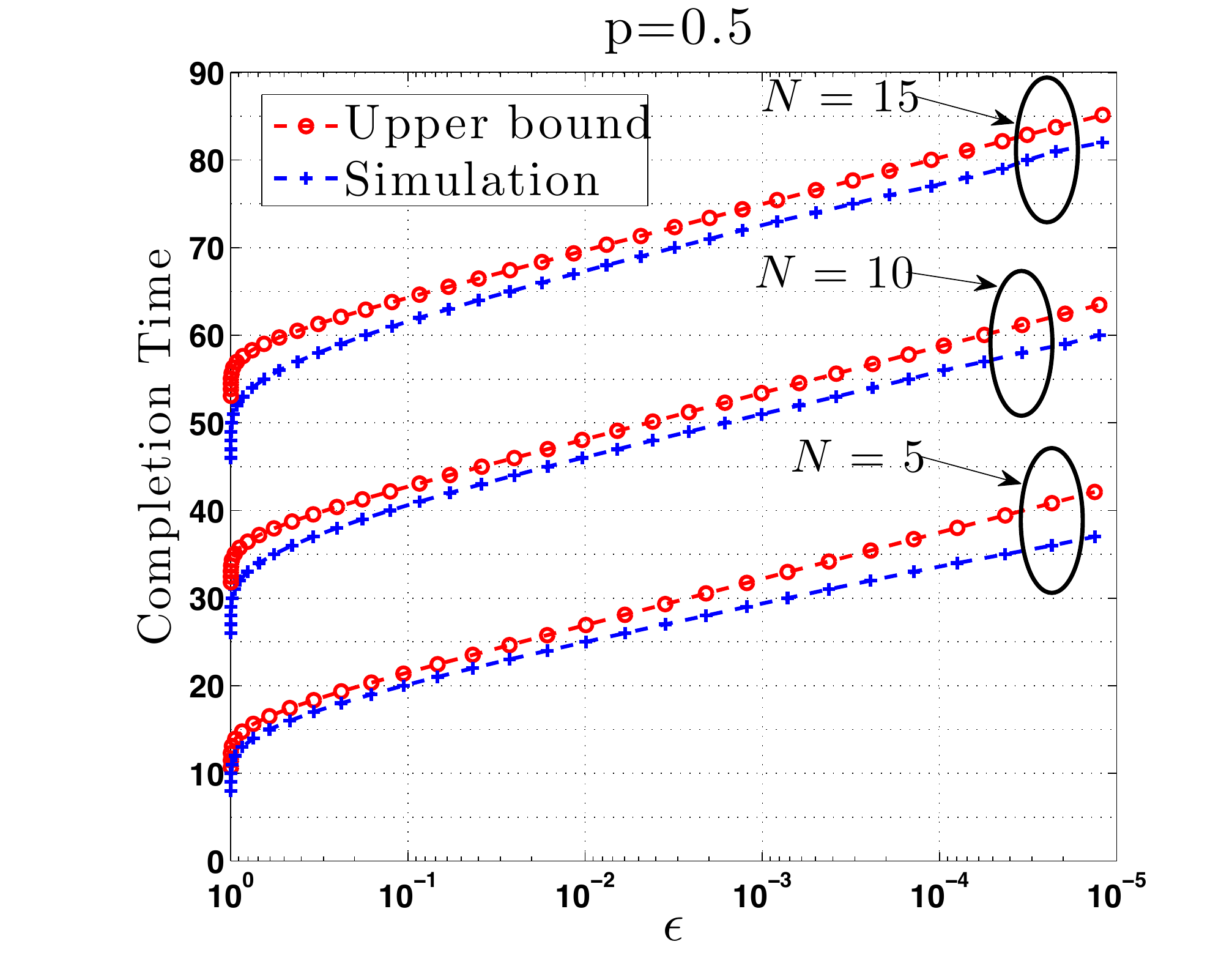}
        \label{fig:TailDistibutionp05}
\end{subfigure}%
\caption{Simulations for the number of time slots $T$ required until all the nodes received one packet in $\mathbf{T}_{2^N}$ with probalility $1-\epsilon$, as a function of $\epsilon$, for $p=0.1,0.2,0.5$, and for $N=5,10,15$, with the upper bound given in  \cref{TailUpperBound}.}
\label{fig:TailDistibutionp010205}
\end{figure*}

Further result that was examined is the gap between the completion time $T$, and its upper bound, as a function of the probability that a message is successively received by all nodes in $\mathbf{T}_n$, as we found in \cref{TailUpperBound}.
The result is given in  \cref{fig:TailDistibutionp010205}, on a log scale.
Note that the linearity of the graph results from the fact that $\log_{\delta} \left(-\log \left(1- \epsilon \right) \right) =\log_{\delta} \epsilon +O \left(\epsilon \right)$.
The construction of the simulation is made by calculating the empirical distribution function of $M_n$, from the $10^6$ generated random samples data, for $n=2^N$, $N=5,10,15$ and for $p=0.1,0.2,0.5$.
In general, we can conclude from the graph that the simulations and the upper bound are roughly close to each other, for each $N$ and $p$.
Furthermore, we can see that for large $N$, i.e., $N=10$ and $N=15$, the gap  remains stable, and in particular does not increase, even for very small $\epsilon$, while for relatively small $N$, i.e., $N=5$ there in a moderate increase in the gap, as $\epsilon$ decreases\\
An additional conclusion, is that the slop of the upper bound (and hence also for the simulation result) is only negligibly effect by $N$, yet is obviously affected by $p$.
This can easily be seen from \cref{TailUpperBound}, i.e., the expression $\log_{\delta} \left(-\log \left(1- \epsilon \right) \right) $ is a function of $\epsilon$ and $p$, but not of $N$.

\begin{figure*}%
\centering
\begin{subfigure}{0.5\columnwidth}
\includegraphics[width=1.2\textwidth,right]{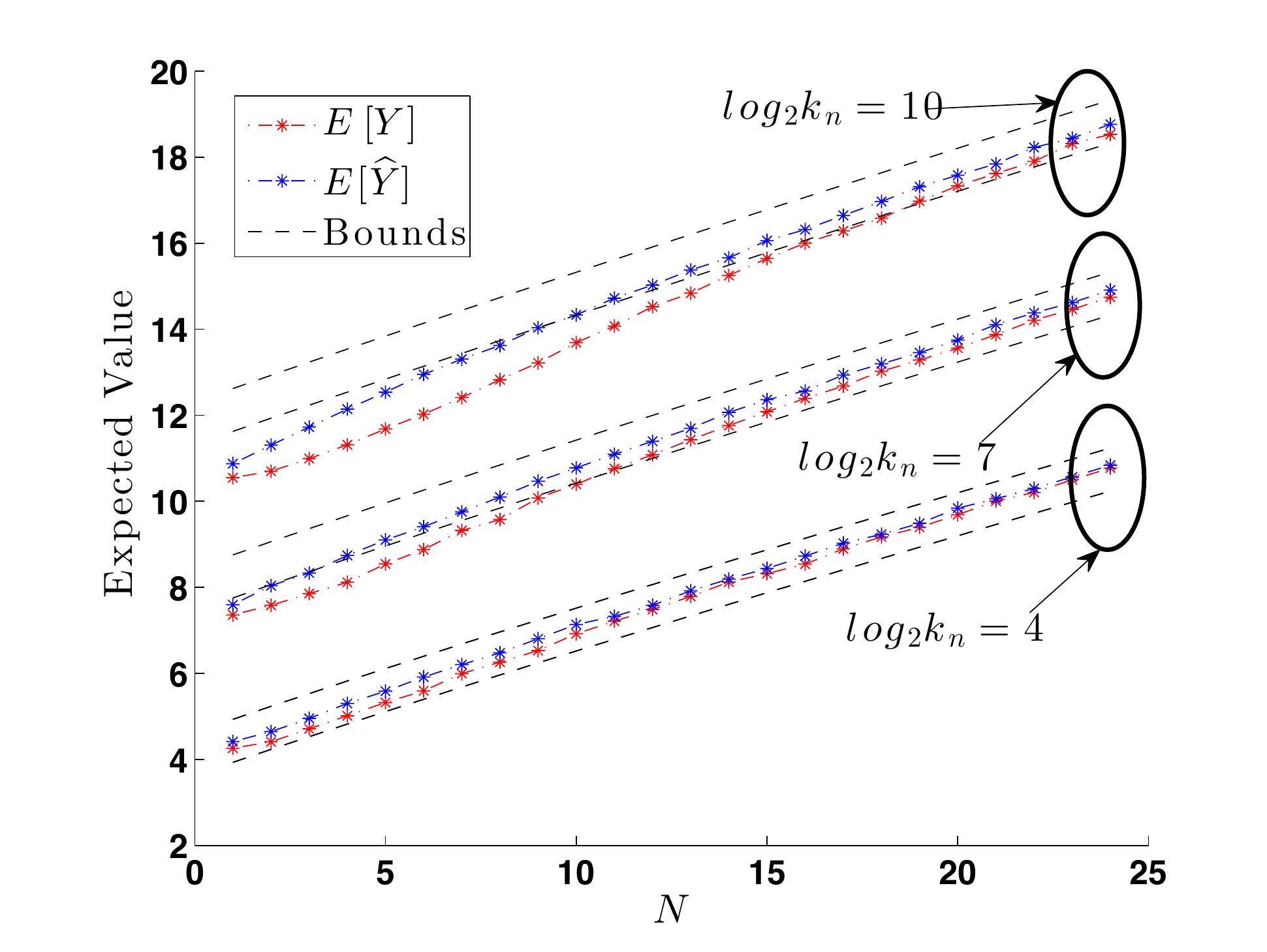}
\caption{$p=0.05$}%
\label{subfig:GraphLowerBoundp01}%
\end{subfigure}\hfill%
\begin{subfigure}{0.5\columnwidth}
    \includegraphics[width=1.2\textwidth,left]{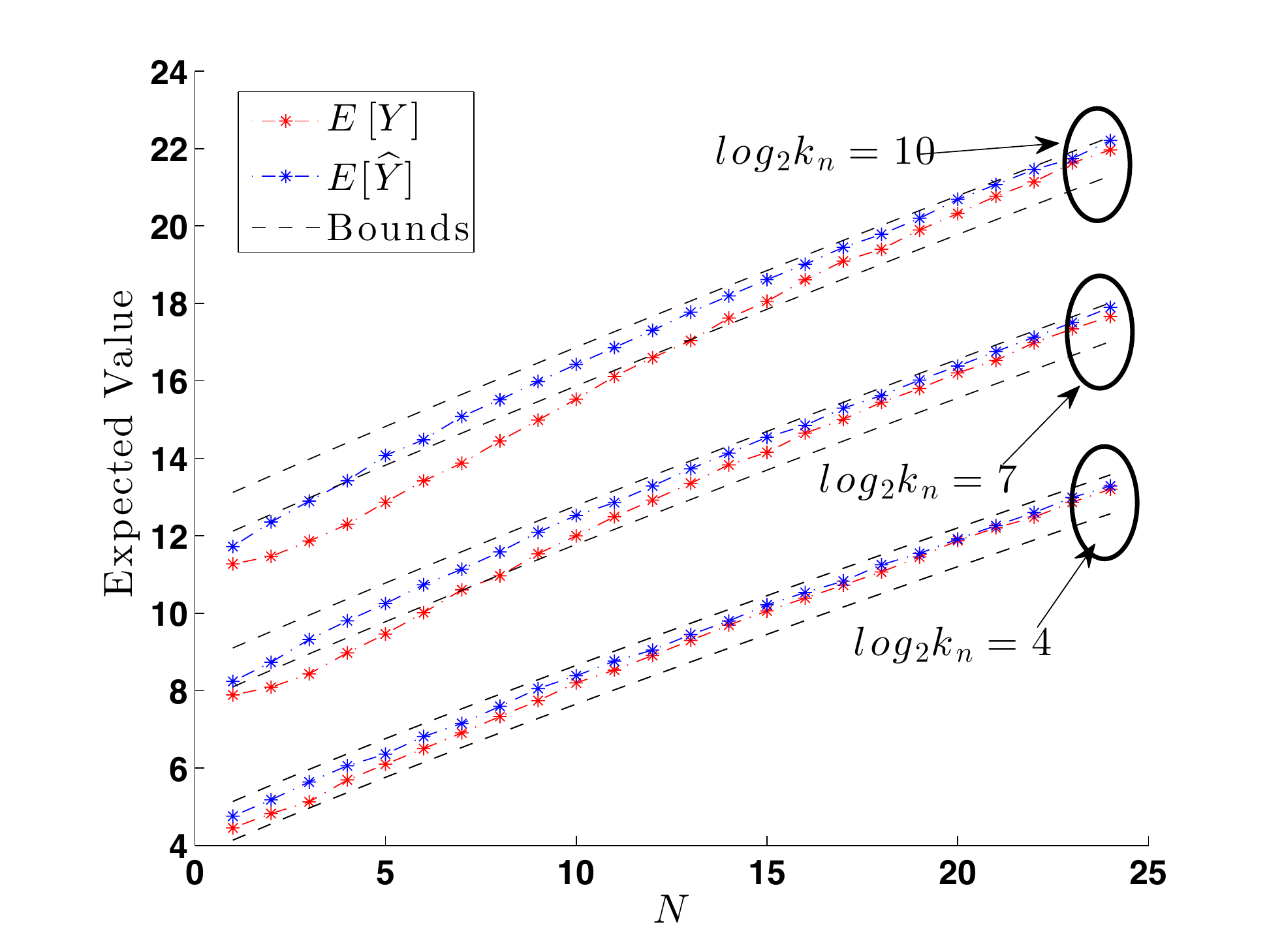}
\caption{$p=0.1$}%
\label{subfig:GraphLowerBoundp02}%
\end{subfigure}\hfill%
\begin{subfigure}{0.5\columnwidth}
      \includegraphics[width=1.2\textwidth,right]{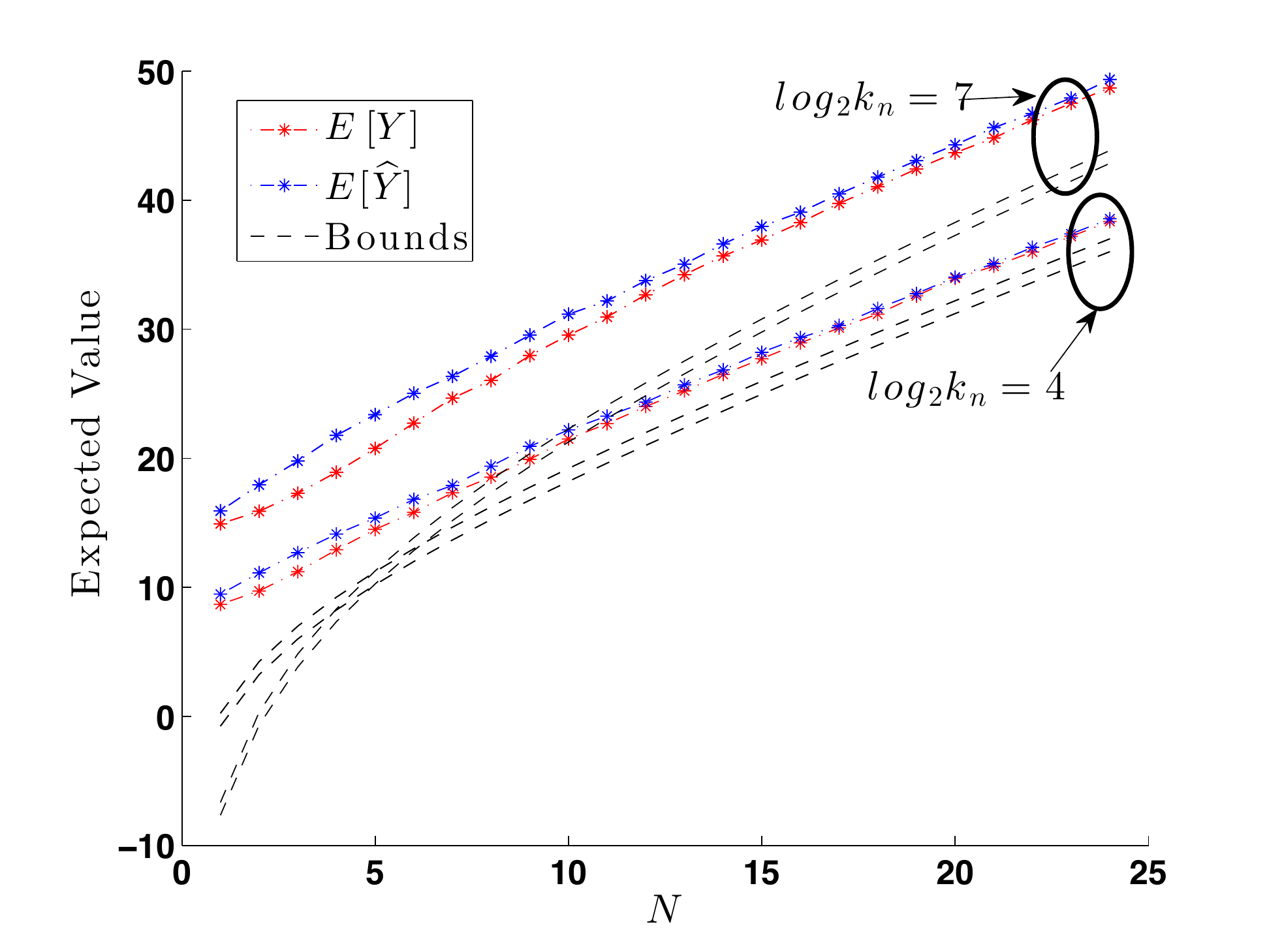}
\caption{$p=0.5$}%
\label{subfig:GraphLowerBoundp05}%
\end{subfigure}
\begin{subfigure}{0.5\columnwidth}
    \includegraphics[width=1.2\textwidth,left]{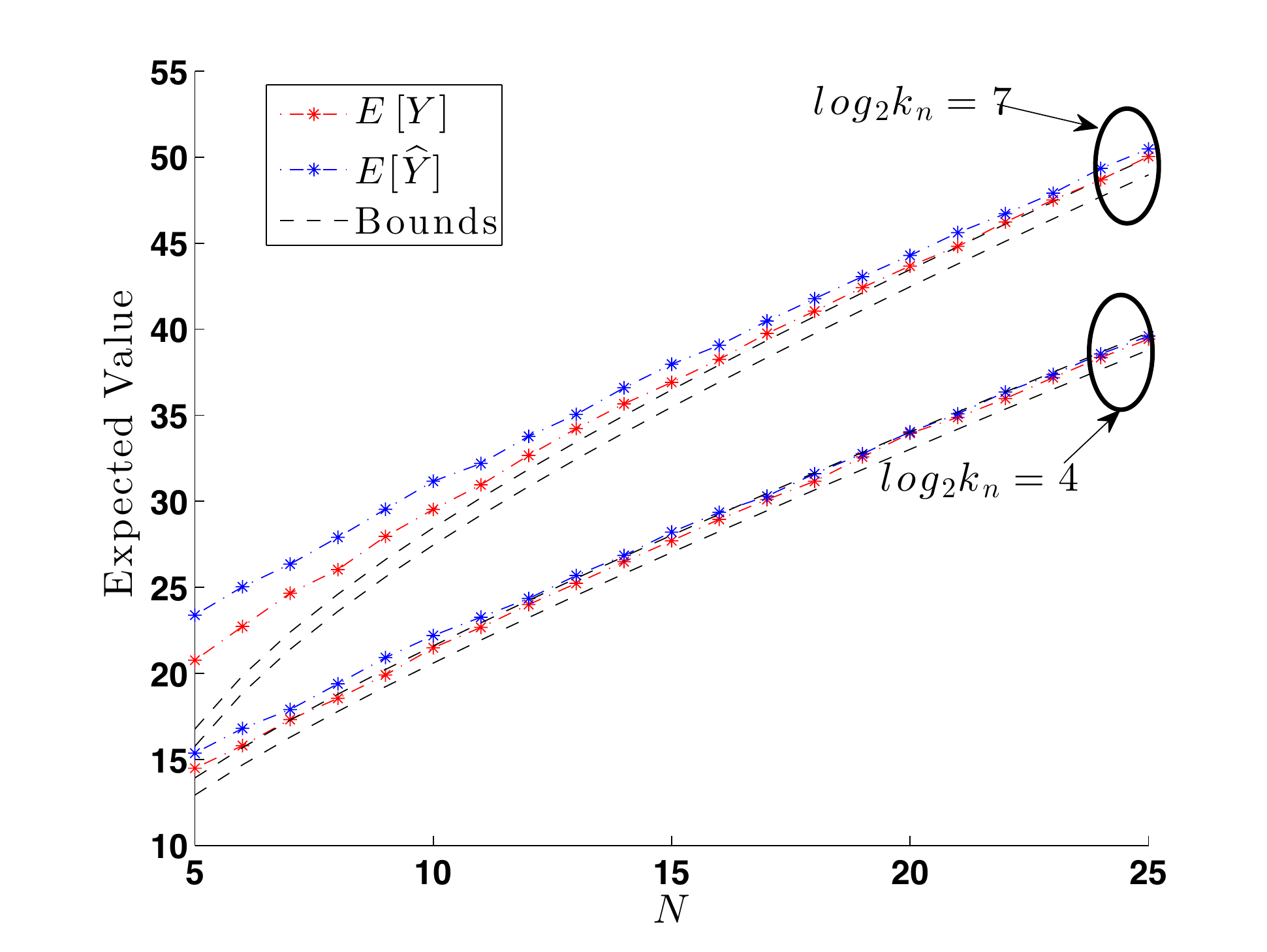}
\caption{$p=0.5$}%
\label{subfig:GraphLowerBoundp05V2}%
\end{subfigure}%
\caption{Simulation of $ \mathbb{E}\left[ Y \right]$, compare with the i.i.d.\ case $ \mathbb{E}\left[ \widehat{Y} \right]$, (averaging 500 realizations), together with their upper and lower bounds (black dashed lines) for $\log_2 k_n=4,7,10$ and $p=0.1,0.2,0.5$, given in \cref{eq:boundsOfEY}, where in \cref{subfig:GraphLowerBoundp01,subfig:GraphLowerBoundp02,subfig:GraphLowerBoundp05} we use $b_n$ suggested in \cref{ab_n}, and in  \cref{subfig:GraphLowerBoundp05}, $\widehat{b}_n$ suggested in \cref{eq:B_nVerstion2}.}
\label{fig:GraphLowerBoundp010205}
\end{figure*}

The next simulation is made for $\mathbb{E}\left[ Y \right]$, that is, a part of the lower bound as outlined in \Cref{sec:lowerBound}.
Namely, $Y$ is the maximum of $n$ random variables which are parted equally into $\frac{n}{k_n}$ subtrees with height of $\log_2 k_n$.
We checked the convergence of it to the i.i.d.\ case (denoted as $\mathbb{E}\left[ \widehat{Y} \right]$), which simply means that $\widehat{Y}=\max \left(\widehat{Y}_1,\ldots,\widehat{Y}_n \right)$, where the sequence is i.i.d.\ and follows the same marginal distribution as $\left(Y_1,\ldots,Y_n \right)$.
We also plot the analytical upper and lower bounds (which differ only by one) as in \cref{eq:boundsOfEY}, as a function of number of users $2^N$, for $N=1, \ldots,24$, and for $p=0.05,0.1,0.5$.
Results are plotted in \Cref{subfig:GraphLowerBoundp01,subfig:GraphLowerBoundp02,subfig:GraphLowerBoundp05}, respectively.
For each $p$, the figure depict the cases of $\log_2 k_n=4,7,10$.\\
The first conclusion is the convergence of $ \mathbb{E}\left[ Y \right]$ to $\mathbb{E}\left[ \widehat{Y} \right]$  depends on $k_n$.
That is, as $k_n$ is greater, so the convergence of $ \mathbb{E}\left[ Y \right]$ to $\mathbb{E}\left[ \widehat{Y} \right]$ is slower, and it can be clearly seen by the proof of \Cref{EVT_for_nonstationary_integer_valued}.
Specifically, the difference between the distributions of $Y$ and $\hat{Y}$ is reflected in \cref{ConditionD_k_nV2}, where $k_n$ is a dominant factor for the difference between them.
Note that \Cref{EVT_for_nonstationary_integer_valued} is on the distribution, not the expected valued, yet, the relationship between them is clear.
Another conclusion on the relation between $\mathbb{E}\left[ Y \right]$ and $\mathbb{E}\left[ \widehat{Y} \right]$, is that the impact of $p$ is relatively negligible for the values that were simulated, which is contrary to the proof of \Cref{ConditionDtag}, where it can be seen that $p$ may affected the convergence.
Yet, clearly $p$ has no impact on the local dependence of $\left( Y_1,\ldots,Y_n \right)$.
Next, we examine the convergence to the analytical upper and lower bounds (which are marked as black dashed lines), for both  $\mathbb{E}\left[Y \right]$ and  $\mathbb{E}\left[ \hat{Y} \right]$.
Recall that the rate of the convergence is determined according  to the choice of $b_n$, and it can be seen in the transition from \cref{Eq31} to \cref{Eq32}.
Hence, both $p$ and $k_n$ have an impact on the converge in the analytical bounds, for both  $\mathbb{E}\left[ \hat{Y} \right]$ and  $\mathbb{E}\left[ \hat{Y} \right]$.
It is clear that for small $p$, i.e., $p=0.05,0.1$ both values converge between the upper and lower bounds in the range of $N$ that was tested.
However, for $p=0.5$, we can see a relatively large error.
Yet, we should note that first, the packet loss probability $p$ is usually small, hence we should put more attention to the cases of small $p$, rather than large $p$.
Furthermore, it is clear that the distance to the analytical bounds does get smaller as $N$ increases, hence we expect it to converge for very large $N$.
Moreover, we propose that for $p=0.5$, to take $\widehat{b}_n$ such that $\widehat{b}_n$ may replace $b_n$ defined in \cref{ab_n},
\begin{align}\label{eq:B_nVerstion2}
\widehat{b}_n=\log_2 n +\left( \log_2 k_n -1 \right) \log_2 \left( \widehat{\Psi} \left(\log_2 k_n \right) \right)
 -\log_2 \left( \left( \log_2 kn -1 \right)! \right)
\end{align}
\begin{align*}
 \widehat{\Psi}  \left(x \right)= \log_2 n+ \left(x-1 \right)\log \log n -\log_2 \left( x-1 \right)!.
\end{align*}
In order to show that $\widehat{b}_n$  proposed above is an appropriate sequence, one needs to prove that the conditions for \Cref{EVT_for_nonstationary_integer_valued} hold and in particular \cref{FirstCondition} and condition $D^{'}_{k_n} \left( u_n \right)$.
We skip these proofs.
However, note that $\widehat{b}_n$ differs from $b_n$  by only  $\widehat{\Psi}$ (compare to $\Psi$  defined in \cref{eq:PsiFunction}), hence, one can see that $\widehat{b}_n-b_n \to 0$.
In \cref{subfig:GraphLowerBoundp05V2}, we plot simulations of $ \mathbb{E}\left[ Y \right]$, and the i.i.d.\ case $ \mathbb{E}\left[ \widehat{Y} \right]$, same as  \cref{subfig:GraphLowerBoundp05}, however, the upper and lower use $\widehat{b}_n$ instead of $b_n$.
We can see that the convergence of both $\log2 k_n=4$, and $\log2 k_n=7$ is significantly better.

\begin{figure}[t]
  \centering
    \includegraphics[width=0.7\textwidth]{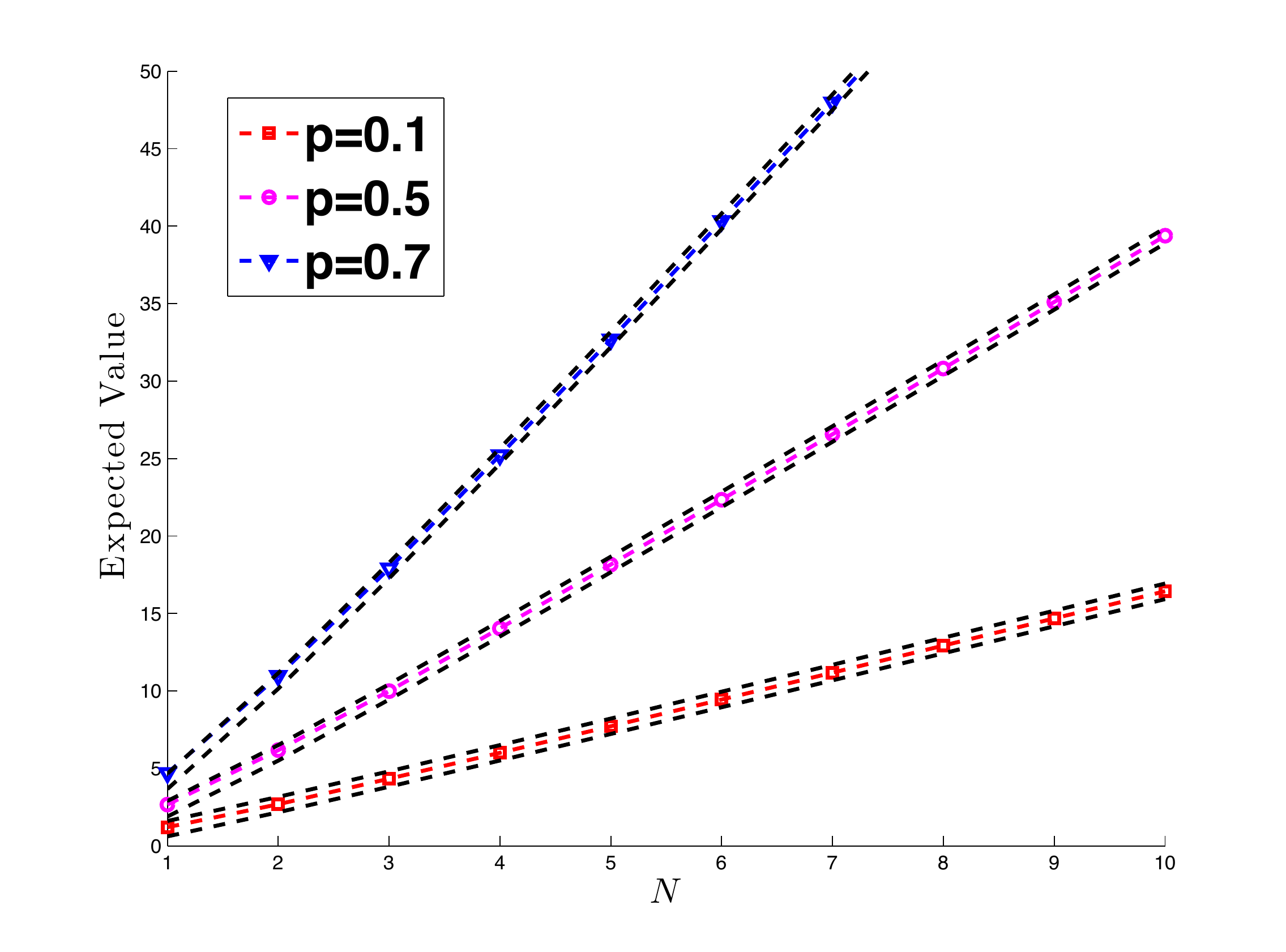}
    \caption{Simulation of the expected value of maximum of $2^N$  i.i.d.\ NB with $N$ successes and $p=0.1,0.5,0.7$ (averaging over 5000 realizations), together with their bounds (black dashed lines) given in \cref{upperLowerBoundIIDCase}}
      \label{fig:iidExpectedValue}
\end{figure}

Finally, in order to explore the convergence of the expected value of $\hat{M}_{n}$ (i.e., the upper bound for $\mathbb{E} \left[ M_n \right]$ as defined in \Cref{sec:UpperBound}), we simulated, by choosing the maximum from $n$ i.i.d.\ random variables, following NB distribution with $\log_2 n$ successes, and averaging over 5000 realizations.
The results are plotted in \Cref{fig:iidExpectedValue}, for $p=0.1,0.5,0,7$.
For each value of $p$, the figure depicts the simulation result together with the analytical upper and lower bounds, given in \cref{upperLowerBoundIIDCase}, as a function of $N$.
Note that the upper and lower bounds differ only by 1, for any valued of $N$.
We can see that \Cref{upperLowerBoundIIDCase}  holds for  even for small $N$, namely $\mathbb{E} \left[ \hat{M}_{2^N} \right]$ is confined between $\mathbb{E} \left[T_l \right]$, and $\mathbb{E} \left[T_u \right]$ for different values of $p$ and even for small $N$.
\section{Conclusion}
\label{sec:conclusion}f
In this work, we presented an asymptotic analysis of the completion time when disseminating $M$ messages through a multicast network, using EVT as a primary tool.

We characterized the limit distribution of the normalized maxima of NSIV sequences, under appropriate dependence restrictions, and specifically, we found that under appropriate conditions, the asymptotic distribution behaves as the same as if the sequence were i.i.d.
Possible extension is to study  NSIV sequences who have a non negligible local dependence, and non-identical marginals.

The result of EVT enabled us to bound the scaling law of the expected completion time.
In particular, we found that these bounds are depend on $p$, where for small $p$, the bounds are tight.
In addition, for large $M$, we found that the expected completion time scale as $\frac{M \log_2 n}{1-p}$.
Tight bounds were also found for the case where the hight of the multicast tree is small enough.
In addition, we also found an upper bound for the exact distribution of the completion time.
Possible extension is to assume heterogeneous packet losses.
Note that the distribution of $T_i$ is no longer NB, thus the main challenge will be to find an appropriate $b_n$ for the this distribution.
Another possible extension is to assume some correlation between the channels, where in this case, the main challenge is to show the dependence conditions hold.

\nomenclature{$M_n$}{Maximum of $n$ random variables.}
\nomenclature{$K$}{Number of child nodes.}
 \nomenclature{$b_n$}{Normalizing constant.}
 \nomenclature{$D \left(u_n \right)$}{Long range dependence condition.}
 \nomenclature{$D^{'} \left(u_n \right)$}{Local dependence condition.}
  \nomenclature{$\gamma$}{Euler-Mascheroni constant, $\approx 0.57721$.}
  \nomenclature{$\lbrace T_i \rbrace$}{Sequence for the arrival time.}
\nomenclature{$\mathbf{T}_n$}{Full K-ary Dissemination tree, with $n$ leaves.}
\nomenclature{$F[*]$}{CDF of discrete random variable, usually NB.}
\nomenclature{$F_c \left(* \right)$}{CDF of continuous random variable, use to bound discrete random variable, usually NB}
\nomenclature{$r$ or $r_n$}{Number of sets used for the partition of $\lbrace 1,\ldots, n \rbrace$.}
\nomenclature{$k_n$}{In \Cref{sec:EVTNonStationaty,PreliminariesSection} indicate the size of $I_l$, otherwise, the size of each subtree.}
\nomenclature{$D_{k_n}^{'}\left(u_n \right)$}{Local dependence condition. Defined in \Cref{PreliminariesSection}}
\nomenclature{$M \left(* \right)$}{The event that the maximum from the set $*$ is less then $u_n$.}
\nomenclature{$W$}{A random variable, which represents the time slots it takes to forward the message from the root to a node which is located at distance of  $\log_2 n/k_n$ hops.}
\nomenclature{$W_i$}{A random variable, which represents the time slots it takes to forward the message from the root to a node the child $i$. which is located at distance of $\log_2 n/k_n$ hops.}
\nomenclature{$Y_i^j$}{A random variable represents the number of time slots it takes to forward one packet from the root of the subtree $\mathbf{T}_{k_n}^{(i)}$ to the leaf $j$, $1\leq j \leq k_n$.}
\nomenclature{$Y^{(i)}$}{A random variable represents the completion time of the subtree $\mathbf{T}_{k_n}^{(i)}$.}
\nomenclature{$Y_{i}$}{Equivalent to $Y_{j}^{(i)}$, namely, $Y_i$ is the time slot it takes to forward one packet from the root of some dissemination tree, with height of $\log_2 k_n$ to leaf $i$, where $1 \leq i \leq n$.}
\nomenclature{$Y$}{Maximum of the sequence $\left(  Y^{(1)}, Y^{(2)},\ldots, Y^{(n/k_n)} \right)$, or equivalently, $\left(Y_1,\ldots,Y_n \right)$.}
\nomenclature{$m_n$}{Number of success of a random variable follows NB, first defined at \cref{chi_cdf}.}
\nomenclature{$B(a,b)$}{Beta function.}
\nomenclature{$B(p;a,b)$}{Incomplete beta function.}
\nomenclature{$\phi \left( * \right)$}{Defined in \cref{eq:phiValue}.}
\nomenclature{$\Psi_{\alpha} \left( * \right)$ or $\Psi\left( * \right)$}{Defined in \cref{eq:PsiFunction}, where $\Psi$ defined for $\alpha=1$.}
\nomenclature{$\epsilon \left(* \right)$}{Defined in \cref{Assumptions}.}
\nomenclature{$\epsilon$}{The probability that after $T$ time slot, the message is \emph{not} reach to all node in  $\mathbf{T}_n$.}
 \nomenclature{$\widetilde{b}_m$}{An increasing sequence, has the form $a m+d_m$, $d_m=o \left(m \right)$.}
\nomenclature{$d_m$}{An increasing sequence, where $d_m=o \left(m \right)$.}
\nomenclature{$\Gamma \left( * \right)$}{Gamma function.}
\nomenclature{$\delta$}{$\delta=\frac{a}{a-1}p$ or $\delta=\frac{\alpha_p}{\alpha_p-1}p$.}
\nomenclature{$g_p(*)$}{A function defined in \cref{eq:gPAlpha}.}
\nomenclature{$p$}{Packet loss probability.}
\nomenclature{$\bar{F}$}{$1-F$.}
\nomenclature{$\mathbb{E} \left[ * \right]$}{Expected value of $*$.}
\nomenclature{$M_n^l$}{The completion time of the lower bound.}
\nomenclature{$\hat{M}_n$}{The completion time of the upper bound, namely, $\hat{M}_n=\max \left(T_1,\ldots,T_n \right)$.}
\nomenclature{$ \lbrace \hat{T}_i \rbrace$}{i.i.d.\ sequence, follows NB, with $\log_K n$ successes.}
\nomenclature{$\alpha_p$}{The root of $g_p(\alpha)$, satisfies $\alpha_p>\frac{1}{1-p}$.}
\nomenclature{$\alpha^{*}$}{The leading multiplicative constant factor, ($\log n$), of the expected completion time, $\mathbb{E} \left[ M_n \right]$. Namely, $\mathbb{E} \left[ M_n \right]=\alpha^{*} \log n+o \left(\log n \right)$.}
\nomenclature{$M$}{Number of message to disseminate from the root of $\mathbf{T}_n$.}
\nomenclature{$h$ or $h_n$}{Height of $\mathbf{T}_n$.}
\appendices

\section{Proof of Lemma \ref{lemma:AsymtoticIndependet}}
\label{appendix:ProofAsymtoticIndependet}
Similar to the proof of  \cite[Lemma 2.1]{N10}, divide each sets $I_1,\ldots,I_{r_n}$ into subsets $I_{1,1},I_{1,2},I_{2,1},I_{2,2},\ldots,I_{1,r_n},I_{2,r_n}$ where
\begin{align}
&I_{1,1}=\lbrace 1,2,\ldots,k_n-l_n \rbrace \quad I_{1,2}=\lbrace k_n-l_n+1,\ldots,k_n \rbrace
   \\ \nonumber &
 I_{2,1}=\lbrace k_n+1,\ldots,2k_n-l_n \rbrace \quad I_{2,2}=\lbrace 2k_n-l_n+1,\ldots,2k_n \rbrace  
    \\ \nonumber &
    \vdots
    \\ \nonumber &    
 I_{r_n,1}=\lbrace (r_n-1)k_n+1,\ldots,r_n k_n-l_n \rbrace
    \\ \nonumber &   
 I_{r_n,2}=\lbrace r_n k_n-l_n+1,\ldots,n \rbrace 
\end{align}
Hence, each pair of intervals $I_{l_1,1},I_{l_2,1}$ are separated by at least $l_n$, where $l_n$ be as in the definition of $D \left(u_n \right)$. 
Obviously we have
\begin{align}\label{eq:FeeltheEnd1}
 \Pr \left( \bigcap_{i=l}^{r_n} M \left(I_{i,1} \right) \right)-\Pr \left(M_n \leq u_n \right)
 &\leq 
\sum_{l=1}^{r_n} \sum_{i \in I_{l,2}} \Pr \left(X_i >u_n \right)
    \\ \nonumber &  
\stackrel{(a)}{\leq}   (r_nl_n+k_n) \Pr \left(X_1 >u_n \right)
    \\ \nonumber &  
\stackrel{(b)}{=}   (r_nl_n+k_n) \left(\frac{K}{n} \right)
    \\ \nonumber & 
 =  K \left( \frac{r_n l_n}{n}+\frac{k_n}{n} \right) \to 0
\end{align}
as $n \to \infty$, where $(a)$, true since $X_i$ have identical distribution for all $i$, and each $ I_{l,2}$ contain $l_n$ integers, except  $I_{r_n,2}$ having $n-r_nk_n+l_n \leq k_n+l_n$, (recall that $r_n=\left[ n/k_n \right]$), and $(b)$ we use \cref{eq:limittausup}, then there exists constant $K$ such that $\Pr \left(X_1>u_n \right) \leq \frac{K}{n}$.
Next, by \cite[Lemma 2.1]{N41husler1986extreme} it follows that 
\begin{align}\label{eq:FeeltheEnd2}
\left| \Pr  \left( \bigcap_{i=l}^{r_n} M \left(I_{i,1} \right) \right)-\prod_{l=1}^{r_n}  \Pr \left( M \left(I_{i,1} \right) \right) \right| \leq r_n \alpha_{n,l_n}
\end{align}
Finally, denote $I_l= \lbrace I_{l,1} \cup I_{l,2} \rbrace$, and $M^C(*)$ be the complementary event of $M$, then
\begin{align}\label{eq:FeeltheEnd3}
\prod_{l=1}^{r_n} M \left(I_{l,1} \right)-\prod_{l=1}^{r_n} M \left(I_{l} \right)
&\stackrel{(a)}{\leq} \sum_{l=1}^{r_n} \left( \Pr \left( M \left(I_{l,1} \right) \right)- \Pr \left( M \left(I_{l} \right) \right) \right)
    \\ \nonumber &
 \stackrel{(b)}{=}\sum_{l=1}^{r_n}  \Pr \left( M \left(I_{l,1} \right),M^{C} \left(I_{l,2}  \right) \right)
    \\ \nonumber &
 \leq \sum_{l=1}^{r_n} \Pr \left(M^{C} \left(I_{l,2}  \right) \right) 
     \\ \nonumber &
= \sum_{l=1}^{r_n} \Pr \left(  \bigcup_{i \in  I_{l,2}} \left(  X_i >u_n \right)   \right) 
        \\ \nonumber &
    \stackrel{(c)}{\leq} \sum_{l=1}^{r_n-1} l_n \Pr \left(X_1 >u_n \right)+k_n \Pr \left(X_1 >u_n \right)
     \\ \nonumber &
    \leq K \left( r_n  \frac{l_n}{n}+ \frac{k_n}{n} \right) \to 0,
\end{align}
where $(a)$ using $|\prod_i x_i-\prod_i y_i|\leq |\sum_{i}x_i-y_i|$, for $x_i,y_i \in (0,1]$, $(b)$ holds since $\Pr \left(A \right)-\Pr \left(A \cap B \right)=\Pr \left(A \cap B^C \right)$, and in $(c)$ we apply union bound.
The result now follows at once by combining \cref{eq:FeeltheEnd1,eq:FeeltheEnd2,eq:FeeltheEnd3}
\section{Proof of Claim \ref{claim1}}
\label{Appendix:LimitTwiceExplanation}
Let $a_n$ be a sequence such that $a_m \to a < \infty$, as $n \to \infty$, then
\begin{align}
\lim_{m \to \infty} \left(1+ \frac{a_m}{m} \right)^m = e^a
\end{align}
\begin{proof}
We prove it by the definition of the limit of sequence.
Let $\epsilon>0$, and $\epsilon^{'}=\log \left(\frac{\epsilon}{2}e^{-a}+1 \right)$, and note that $\epsilon^{'}>0$.
Since $a_m \to a$, there exists $m_1$, where for each $m>m_1$ satisfy $|a_m-a| < \epsilon^{'}$, thus
\begin{align}
a_m <a+\log \left(\frac{\epsilon}{2}e^{-a}+1 \right).
\end{align}
Hence, for each $m>m_1$, we have
\begin{align}
\left(1+ \frac{a_m}{m} \right)^m & < 
\left(1+\frac{a+\log \left(\frac{\epsilon}{2}e^{-a}+1 \right)}{m} \right)^m
\end{align}
let $\epsilon^{''}=\frac{\epsilon}{2}$, then by the definition of the limit of the exponential function, there exist $m_2$, such that for each $m>m_2$, we have
\begin{align}
\left(1+\frac{a+\log \left(\frac{\epsilon}{2}e^{-a}+1 \right)}{m} \right)^m
 \nonumber & 
< e^{a+\log \left(\frac{\epsilon}{2}e^{-a}+1 \right)}+\epsilon^{''}
  \\ \nonumber &
= e^{a}+\frac{\epsilon}{2}+\epsilon^{''}
  \\ \nonumber &
= e^{a}+\epsilon.
\end{align}
Now set $m_0=\max \left(m_1,m_2 \right)$.
similarly, on the other side, for $\epsilon^{'}=-\log \left(1-\frac{\epsilon}{2}e^{-a} \right)$, one can verify that there exist $m_0$, such that for each $m>m_0$,
\begin{align}
\left(1+ \frac{a_m}{m} \right)^m  > e^a-\epsilon.
\end{align}
which complete the proof.
\end{proof}
\section{Proof of Lemma \ref{General}}
\label{AppendixProofOfLemmaGeneral}
By \cref{betaInComplete2} we have
\begin{multline}\label{betaPhiN}
  B(p;\phi(n)-m_n+1,m_n)=
 \\ 
 \frac{p^{\phi(n)-m_n+1}(1-p)^{m_n}}{\phi(n)-m_n+1}  \mathop{F\/}\nolimits\!\left(\phi(n)+1,1;\phi(n)-m_n+2;p\right),
\end{multline}
where, by \cref{eq:gaussianSeries}, it can be verified that
\begin{align}
&
 \mathop{F\/}\nolimits\!\left(\phi(n)+1,1;\phi(n)-m_n+2;p\right)
\\ &\nonumber
=1+ \frac{\phi \left(n \right)+1}{\phi \left(n \right)-m_m+2}p+\frac{\left(\phi \left(n \right)+1 \right)\left(\phi \left(n \right)+2 \right)}{\left(\phi \left(n \right)-m_n+2 \right)\left(\phi \left(n \right)-m_n+3 \right)}p^2+\cdots 
  \\ \nonumber &
=1+\sum_{k=1}^{\infty} \prod_{j=1}^{k} \left( \frac{\phi(n)+j}{\phi(n)-m_n+1+j} \right) p^k,
\end{align}
and
\begin{align}
\prod_{j=1}^{k} \left( \frac{\phi(n)+j}{\phi(n)-m_n+1+j} \right) &= 
\prod_{j=1}^{k} \left( 1+ \frac{m_n-1}{\phi(n)-m_n+1+j} \right)
\\ \nonumber &
 \leq \prod_{j=1}^{k} \left( 1+ \frac{m_n-1}{\phi(n)-m_n} \right)
\\ \nonumber &
= \left( 1+ \frac{m_n-1}{\phi(n)-m_n} \right)^k.
\end{align}
Denote $g(n)=p\left( 1+ \frac{m_n-1}{\phi(n)-m_n} \right)$, and note that $g(n) \to p$ as $n \to \infty$.
Then
\begin{align}\label{A6}
 \mathop{F\/}\nolimits\!\left(\phi(n)+1,1;\phi(n)-m_n+2;p\right) 
 &
 \leq 1+\sum_{k=1}^{\infty}
\left( 1+ \frac{m_n-1}{\phi(n)-m_n} \right)^k p^k
\\ \nonumber &
=\sum_{k=0}^{\infty} g(n)^k
\\ \nonumber &
\stackrel{(a)}{=}\frac{1}{1-g(n)}
\\ \nonumber &
=\frac{1}{1-p}+O \left( \frac{m_n}{\log n} \right),
\end{align}
where $(a)$ hold for large enough $n$, such that $g(n) <1$.
On the other hand,
\begin{align}\label{A7}
\mathop{F\/}\nolimits\!\left(\phi(n)+1,1;\phi(n)-m_n+2;p\right)
 &
=1+\sum_{k=1}^{\infty} \prod_{j=1}^{k} \left( \frac{\phi(n)+j}{\phi(n)-m_n+1+j} \right) p^k
\\ \nonumber &
=1+\sum_{k=1}^{\infty} \prod_{j=1}^{k} \left( 1+ \frac{m_n-1}{\phi(n)-m_n+1+j} \right) p^k
\\ \nonumber &
\geq 
1+\sum_{k=1}^{\infty} p^k
\\ \nonumber &
=\frac{1}{1-p}.
\end{align}
Hence, by \cref{betaPhiN}, for large enough $n$, we have
\begin{align}\label{189}
 B(p;\phi(n)-m_n+1,m_n)
 =\frac{p^{\phi(n)-m_n+1}(1-p)^{m_n-1}}{\phi(n)-m_n+1} \left( 1+O \left( \frac{m_n}{\log n} \right) \right).
\end{align}
In addition,
\begin{align}\label{BetaInCompleteForm}
B(\phi(n)-m_n+1,m_n) &
\stackrel{(a)}{=}
\frac{ \Gamma \left( \phi \left( n \right)-m_n+1 \right) \Gamma \left(m_n \right)}{\Gamma \left( \phi \left( n \right)+1 \right)} 
  \\ \nonumber &
\stackrel{(b)}{=}
\frac{\left(m_n-1 \right)!}{\prod_{j=0}^{m_n-1} \left(\phi(n)-j \right)},
\end{align}
where $(a)$ is due to \cref{betaComplete}, and $(b)$ results from \cref{eq:GammaRecurrenceRelation} and the fact that $m_n \in \mathbb{N}$.
Hence, we have
\begin{align}
\bar{F}_c( \phi (n)|m_n,p) 
& \stackrel{(a)}{=} \frac{B(p;\phi(n)-m_n+1,m_n)}{B(\phi(n)-m_n+1,m_n) } 
  \\ \nonumber &
\stackrel{(b)}{=}   \frac{\prod_{j=0}^{m_n-1} \left(\phi(n)-j \right)}{\left(m_n-1 \right)!} \frac{p^{\phi(n)-m_n+1}(1-p)^{m_n-1}}{\phi(n)-m_n+1}\left( 1+O \left( \frac{m_n}{\log n} \right) \right)
\\ \nonumber &
\stackrel{(c)}{=} \frac{1}{n^{\alpha}}\frac{\prod_{j=0}^{m_n-1} \left(\phi(n)-j \right) }{\left(\phi(n)-m_n+1\right)  \Psi_{\alpha} \left(m_n \right)^{m_n-1}} p^x \left( 1+O \left( \frac{m_n}{\log n} \right) \right)
\\ \nonumber &
= \frac{p^x}{n^{\alpha}}\epsilon \left(n \right) \frac{  \Psi_{\alpha} \left(m_n \right)}{\phi(n)-m_n+1}\left( 1+O \left( \frac{m_n}{\log n} \right) \right)
\\ \nonumber &
\stackrel{(d)}{=} \frac{p^x}{n^{\alpha}}\epsilon \left(n \right) \left( 1+O \left( \frac{m_n}{\log n} \right) \right),
\end{align}
where $(a)$ follow from \cref{chi_cdf}, $(b)$ from \cref{BetaInCompleteForm}, and $(c)$ is by assigning $\phi (n)$ as defined in \cref{eq:phiValue}.
Finally, $(d)$ is since $\frac{  \Psi_{\alpha} \left(m_n \right)}{\phi(n)-m_n+1}= 1+O \left( \frac{m_n}{\log n} \right)$.
\section{}
The following lemma is a technical claim, which will be required further.
\begin{lemma} \label{lemma94}
Let $f[x|m,p]$ be the PDF of NB distribution with parameters $m \in \mathbb{N}$ and $0<p<1$, as defined in \cref{eq:NBPDF}.
Then
\begin{align}
 \sum_{k=m}^{\infty}f[k|m,p]\prod_{j=0}^{m-1}\left( k-j \right) \leq \frac{(2m-1)!}{\left(m-1\right)!}\frac{1}{\left( 1-p\right)^{m}}.
\end{align}
\begin{proof}
Since
\begin{align}
\sum_{k=M}^{\infty}f[k|M,p]=\sum_{k=M}^{\infty} \binom{k-1}{M-1}p^{k-M}(1-p)^M=1.
\end{align}
Writing $k'=k-M$, we get
\begin{align}
\sum_{k'=0}^{\infty} \binom{k'+M-1}{M-1}p^{k'}(1-p)^M=1.
\end{align}
As a result, for any $M \in \mathbb{N}$
\begin{align}\label{Mysentence}
\sum_{k=0}^{\infty} p^k \prod_{j=1}^{M-1} \left( k+j \right)=  \frac{(M-1)!}{\left( 1-p\right)^M}.
\end{align}
Consequently,
\begin{align}
 \sum_{k=m}^{\infty}f[k|m,p]\prod_{j=0}^{m-1}\left( k-j \right)
 =& \sum_{k=M}^{\infty} \binom{k-1}{m-1}p^{k-m}(1-p)^m \prod_{j=0}^{m-1}\left( k-j \right)
    \\ \nonumber
 =& \sum_{k'=0}^{\infty} \binom{k'+m-1}{m-1}p^{k'}(1-p)^m \prod_{j=0}^{m-1}\left( k'+m-j \right)
  \\ \nonumber
 =& \sum_{k'=0}^{\infty} \binom{k'+m-1}{m-1}p^{k'}(1-p)^m \prod_{j=1}^{m}\left( k'+j \right)
   \\ \nonumber
 =& \frac{(1-p)^m}{\left(m-1\right)!}\sum_{k'=0}^{\infty}p^{k'}\prod_{i=1}^{m-1}\left( k'+i \right) \prod_{j=1}^{m}\left( k'+j \right)
    \\ \nonumber
    \leq &
    \frac{(1-p)^m}{\left(m-1\right)!}\sum_{k'=0}^{\infty}p^{k'}\prod_{i=1}^{2m-1}\left( k'+i \right)
         \\ \nonumber
     \stackrel{(a)}{=} &  \frac{(1-p)^m}{\left(m-1\right)!}\frac{(2m-1)!}{\left( 1-p\right)^{2m}}
      \\ \nonumber
     =&  \frac{(2m-1)!}{\left(m-1\right)! \left( 1-p\right)^{m}}
\end{align}
where $(a)$ is by using \cref{Mysentence} with $M=2m$.
\end{proof}
\end{lemma}
The following lemma is used for the proof of \Cref{JointDistributionLemma}, and in particular, bounds the Probability Generating Function of $\min \left(Z_1,Z_2 \right)$, $Z_i \sim NB \left(m,p \right)$ $i=1,2$, at point $\frac{1}{p}$.
\begin{lemma}\label{minmin}
For two i.i.d.\ random variables $Z_1,Z_2$, with $\Pr \left( Z \leq z \right)=F[z|m,p]$.
\begin{align}
\sum_{k=m}^{\infty} \left( \frac{1}{p} \right)^k \Pr \left(\min \left( Z_1,Z_2 \right)=k \right)< 2\left( \frac{4}{p}\right)^m.
\end{align}
\begin{proof}
Denote by $F_{\min}[z|m,p]=1-\left(\bar{F}[z|m,p] \right)^2$, the distribution of the minimum of two \emph{discrete} in random variables with distribution $F[z|m,p]$.
Then
\begin{align}
& \sum_{k=m}^{\infty} \left( \frac{1}{p} \right)^k \Pr \left(\min \left( Z_1,Z_2 \right)=k \right)
\\ \nonumber
= & \sum_{k=m}^{\infty} \left( \frac{1}{p} \right)^k \left(F_{\min}[k|m,p]-F_{\min}[k-1|m,p] \right)
\\ \nonumber
= & \sum_{k=m}^{\infty} \left( \frac{1}{p} \right)^k \left(\bar{F}[k-1|m,p]^2-\bar{F}[k|m,p]^2 \right)
\\ \nonumber
= & \sum_{k=m}^{\infty} \left( \frac{1}{p} \right)^k \left(\bar{F}[k-1|m,p]-\bar{F}[k|m,p] \right)
 \left(\bar{F}[k-1|m,p]+\bar{F}[k|m,p] \right)
\\ \nonumber
\leq & \sum_{k=m}^{\infty} \left( \frac{1}{p} \right)^k \left(F[k|m,p]-F[k-1|m,p] \right) \left(2\bar{F}[k-1|m,p] \right)
\\ \nonumber
= & 2\sum_{k=m}^{\infty} \left( \frac{1}{p} \right)^k f[k|m,p]  \bar{F}[k-1|m,p].
\end{align}
Note that
\begin{align}
\bar{F}[k-1|m,p]=\sum_{j=k}^\infty f[j|m,p],
\end{align}
and
\begin{align}
 f[x|m,p] &=  \binom{x-1}{m-1}p^{x-m}(1-p)^m 
 \\ \nonumber &
 =\left( \frac{1-p}{p}\right)^m  \binom{x-1}{m-1}p^{x}.
\end{align}
Thus
\begin{align}
 \sum_{k=m}^{\infty} \left( \frac{1}{p} \right)^k \Pr \left(\min \left( Z_1,Z_2 \right)=k \right)
\leq & 2\left( \frac{1-p}{p}\right)^m  \sum_{k=m}^{\infty} \binom{k-1}{m-1} \left( \sum_{j=k}^\infty f[j|m,p] \right)
 \\ \nonumber
\stackrel{(a)}{=}& 2\left( \frac{1-p}{p}\right)^m  \sum_{k=m}^{\infty}f[k|m,p]\left(  \sum_{j=m}^k  \binom{j-1}{m-1}  \right)
 \\ \nonumber
\stackrel{(b)}{= }& 2\left( \frac{1-p}{p}\right)^m  \sum_{k=m}^{\infty}f[k|m,p] \binom{k}{m}
 \\ \nonumber
 =& 2\left( \frac{1-p}{p}\right)^m  \sum_{k=m}^{\infty}f[k|m,p] \frac{\prod_{j=0}^{m-1}\left(k-j \right)}{m!}
   \\ \nonumber
 \stackrel{(c)}{<}& 2\left( \frac{1-p}{p}\right)^m  \frac{1}{m!}  \frac{(2m-1)!}{\left(m-1\right)!}\frac{1}{\left( 1-p\right)^{m}}
 \\ \nonumber
  =& 2\left( \frac{1}{p}\right)^m  \binom{2m-1}{m-1}
   \\ \nonumber
     \stackrel{(d)}{<}& 2\left( \frac{4}{p}\right)^m,
\end{align}
where $(a)$ is due to  changing the order of summation.
$(b)$ results from the identity $\sum_{j=m}^k  \binom{j-1}{m-1} = \binom{k}{m}$ \cite[pp.18]{Book1},
$(c)$ follows from \Cref{lemma94}, and finally, $(d)$ uses the bound $ \binom{2m-1}{m-1} < 4^m$, true for each $m=1,2,\ldots$
\end{proof}
\end{lemma}
\section{Proof of Lemma \ref{JointDistributionLemma}}
\label{AppendixProofJointDistibution}
By the construction of $\left( Y_1,\ldots,Y_n \right)$, the pair $\left( Y_1,Y_{2^i} \right)$ shared $\log_2 k_n -i$  common channels, and $i$ independent channels, $i=1,2,\ldots,\log_2 k_n-1$.\\
Denote the time it takes a packet to cross the  common channels by a random variable $W$, where $W$ has CDF $F \left[w|\log_2 k_n -i,p\right]$.
Also, denote the time it takes a packet to cross independent channels by the random variables $Z_1$ and $Z_2$, where $Z_1$ and $Z_2$ have CDF  $F \left[z|i,p\right]$.
Then, we can write the join distribution as
\begin{align} \label{169}
 \Pr \left(Y_1>u_n,Y_{2^i} >u_n \right)&= \Pr \left(W+Z_1>u_n,W+Z_2 >u_n \right)
\\ \nonumber
&
=\Pr \left(W+\min \left( Z_1,Z_2 \right)>u_n \right)
\\ \nonumber
&
=\sum_{k=i}^{\infty} \Pr \left(W > u_n-k\right) \Pr \left(\min \left( Z_1,Z_2 \right)=k \right).
\end{align}
The concept of the proof is to use \Cref{General} for the tail distribution $\Pr \left(W> u_n-k \right)$.
However, $k$ may be too large (since $k \in [i,\infty)$), such that $u_n-k$ would not satisfy the assumptions of \Cref{General}.\\
Thus, we will break the segment $ [i,\infty)$ into two sub-segment, $[i,v_n^i]$ and $(v_n^i,\infty)$, such that
\begin{align}
&\sum_{k=i}^{\infty} \Pr \left(W > u_n-k\right) \Pr \left(\min \left( Z_1,Z_2 \right)=k \right)
\\ \nonumber  \leq &
\sum_{k=i}^{v_n^i} \Pr \left(W > u_n-k\right) \Pr \left(\min \left( Z_1,Z_2 \right)=k \right) 
 + 
\Pr \left( \min \left( Z_1,Z_2 \right) >v_n^i \right).
\end{align}
Now, chose $v_n^i$ such that on the one hand, satisfy $\Pr \left( \min \left( Z_1,Z_2 \right) >v_n^i \right)=o \left( \frac{1}{n} \right)$, but on the other hand, enable to use \Cref{General} for the tail distibution $\Pr \left(W> u_n-k \right)$, for any $k \in [i,v_n^k]$.
Therefore,
\begin{align}
 v_n^i= \alpha \log_{\frac{1}{p}} n+ (i-1) \log_{\frac{1}{p}} \left( \frac{1-p}{p}  \Psi_{\alpha} \left(i \right) \right) - \log_{\frac{1}{p}} (i-1)!+1,
\end{align}
for some $\frac{1}{2} < \alpha < 1$.
Thus, using \Cref{General},
\begin{align} \label{172}
\Pr \left( \min \left( Z_1,Z_2 \right) >v_n^i \right) 
&= \bar{F}\left[v_n^i|i,p\right]^2
\\ \nonumber 
& \leq \bar{F}_c \left( v_n^i-1|i,p\right )^2
\\ \nonumber 
&=  \frac{1}{n^{2\alpha}} \left(1+o \left( 1 \right) \right).
\end{align}
Now, in order to derive asymptotic approximation for the tail distribution $\Pr \left(W>u_n-k \right)$, first note that the boundary level $u_n$ (which was defined in \cref{ab_n}) is related to NB with $\log_2 k_n$ successes, while $W \sim NB \left( \log_2 k_n-i,p \right)$.
Therefore, write $u_n=u_n^{\left( W \right)}+\Delta(n)$, where $u_n^{\left( W \right)}$ is the boundary level for the random variable $W$, namely,
\begin{align} 
 u_n^{(W)}=
   \log_{\frac{1}{p}} n+ \left(\log_2 k_n-i-1 \right) \log_{\frac{1}{p}} \left(  \frac{1-p}{p} \Psi \left( \log_2 k_n -i \right) \right) 
- \log_{\frac{1}{p}} (\left(\log_2 k_n-i-1 \right)!),
\end{align}
and
\begin{align}
 \Delta(n)=  x+\log_{\frac{1}{p}} \left(\frac{1-p}{p}  \Psi \left( \log_2 k_n -i \right) \right)^{i}
-\sum_{r=1}^{i} \log_{\frac{1}{p}}(\log_2 k_n -r)
 +\log_{\frac{1}{p}}  \left(\frac{\Psi \left( \log_2 k_n -i \right)}{\Psi \left( \log_2 k_n  \right)} \right)^{\log_2 k_n-1}.
\end{align}
Note that $\Delta(n) < K \log_2 k_n \log \log n $, for some constant $K$, and $k \leq v_n^i$, hence $\Delta(n)-k > \widetilde{\alpha} \log_{\frac{1}{p}} n$, for some $0< \widetilde{\alpha}<1$.
Then, by \Cref{General}, we have
\begin{align} \label{171}
\Pr \left(W > u_n-k\right) 
&
 = \Pr \left(W > u_n^W+\Delta(n) -k\right)
\\ \nonumber &
 \leq F_c \left( u_n^W+\Delta(n) -k| \log_2 k_n-i \right)
\\ \nonumber
& \leq \frac{1}{n} p^{\Delta(n) -k}
\\ \nonumber &
 =
 \frac{1}{n} p^x\frac{\prod_{r=1}^{i} (\log_2 k_n-r)}{ \left(\frac{1-p}{p}  \Psi \left( \log_2 k_n -i \right) \right)^{i} }\left(\frac{\Psi \left( \log_2 k_n  \right)}{\Psi \left( \log_2 k_n-i  \right)} \right)^{\log_2 k_n-1} p^{-k} .
\end{align}
Note that the expression  $\log_{\frac{1}{p}} \left( \frac{1-p}{p} \right)$ is positive for $p< \frac{1}{2}$, and negative for $p>\frac{1}{2}$.
Therefore,
\begin{align}
\frac{1}{\Psi \left( \log_2 k_n-i  \right)}
&
=
 \frac{1}{\log_{\frac{1}{p}} n +\left( \log_2 k_n-i-1 \right) \log_{\frac{1}{p}} \left(  \frac{1-p}{p} \right)}
\\ \nonumber 
& \leq 
\frac{1}{\log_{\frac{1}{p}} n}
\end{align}
for $p<\frac{1}{2}$, and for $p>\frac{1}{2}$, it can be verified that $\frac{1}{\Psi \left( \log_2 k_n-i  \right)} \leq \frac{1}{\Psi \left( \log_2 k_n \right)}$. \\
As a result
 \begin{align}\label{1711}
\frac{1}{\Psi \left( \log_2 k_n-i  \right)} \leq C,
\end{align}
where $C= \max \left( \frac{1}{•\log_{\frac{1}{p}} n },\frac{1}{\Psi \left( \log_2 k_n \right) }\right)$.\\
Accordantly, assign \cref{1711} into \cref{171}, would result
\begin{align}\label{171b}
 \Pr \left(W > u_n-k\right)
  \leq 
 \frac{p^x}{n}\left( C \frac{p}{1-p}  \log_2 k_n \right)^{i} \left( C \Psi \left( \log_2 k_n \right) \right)^{\log_2 k_n-1} \left( \frac{1}{p} \right)^k.
\end{align}
Now, combining  \cref{171b,172} into \cref{169}, we have
\begin{align}
&\Pr \left(Y_1>u_n,Y_{2^i} >u_n \right) 
 \\ \nonumber
 \leq &
  \frac{p^x}{n}\left( C \frac{p}{1-p}  \log_2 k_n \right)^{i} \left( C \Psi \left( \log_2 k_n \right) \right)^{\log_2 k_n-1}
 \sum_{k=i}^{v_n^i} \Pr \left(\min \left( Z_1,Z_2 \right)=k \right) \left( \frac{1}{p} \right)^k +  \frac{1}{n^{2\alpha}} \left(1+o \left( 1 \right) \right)
\\ \nonumber
 \leq
&  \frac{p^x}{n}\left( C \frac{p}{1-p}  \log_2 k_n \right)^{i} \left( C \Psi \left( \log_2 k_n \right) \right)^{\log_2 k_n-1}
 \sum_{k=i}^{\infty} \Pr \left(\min \left( Z_1,Z_2 \right)=k \right) \left( \frac{1}{p} \right)^k +  \frac{1}{n^{2\alpha}} \left(1+o \left( 1 \right) \right)
 \\ \nonumber
\leq & 2 \frac{p^x}{n} \left( C \Psi \left( \log_2 k_n \right) \right)^{\log_2 k_n-1} \left( C \frac{4}{1-p}  \log_2 k_n \right)^{i} 
+ \frac{1}{n^{2\alpha}} \left(1+o \left( 1 \right) \right)
\end{align}
Where the last inequality derived from \Cref{minmin}.
\section{Proof of Lemma \ref{CLoseFormExpressionCDFNegativeBinomial}}
\label{ProofCLoseFormExpressionCDFNegativeBinomial}
In this proof, we will show that the tail distribution at $\widetilde{b}_m$ can be represented as the PDF at $\widetilde{b}_m+1$ multiple some  hypergeometric function.
\begin{align}
 \Pr \left(X \geq \widetilde{b}_m \right)
 &= 
\sum_{k=\widetilde{b}_m+1}^{\infty} \Pr \left( X=k \right)
\\ \nonumber
&=\sum_{k=\widetilde{b}_m+1}^{\infty} \binom{k-1}{m-1}(1-p)^m p^{k-m}
\\ \nonumber
&
=\binom{\widetilde{b}_m}{m-1}(1-p)^m p^{\widetilde{b}_m+1-m} \mathop{F\/}\nolimits\!\left(1,\widetilde{b}_m+1;\widetilde{b}_m-m+2;p\right)
\\ \nonumber
&
=\Pr \left( X=\widetilde{b}_m+1 \right)  \mathop{F\/}\nolimits\!\left(1,\widetilde{b}_m+1;\widetilde{b}_m-m+2;p\right),
\end{align}
where $\mathop{F\/}\nolimits\!\left(1,\widetilde{b}_m+1;\widetilde{b}_m-m+2;p\right)$, the hypergeometric function is defined by the Gauss series (\cref{eq:gaussianSeries}).
We have
\begin{align}
&\mathop{F\/}\nolimits\!\left(1,\widetilde{b}_m+1;\widetilde{b}_m-m+2;p\right)
\\ &\nonumber
=1+ \frac{\widetilde{b}_m+1}{\widetilde{b}_m-m+2}p+\frac{\left(\widetilde{b}_m+1 \right)\left(\widetilde{b}_m+2 \right)}{\left(\widetilde{b}_m-m+2 \right)\left(\widetilde{b}_m-m+3 \right)}p^2+\cdots
\\ &\nonumber
=1+\sum_{k=1}^{\infty} p^k \prod_{j=1}^{k} \left( \frac{\widetilde{b}_m+j}{\widetilde{b}_m-m+1+j} \right) 
\\ & \nonumber
 =1+\sum_{k=1}^{\infty} p^k \prod_{j=1}^{k} \left( \frac{a m + d_m+j}{(a-1) m + d_m+1+j} \right) 
\\ & \nonumber
=1+\sum_{k=1}^{\infty} p^k \prod_{j=1}^{k} \left( 1+ \frac{m}{(a-1) m + d_m+1+j}  -\frac{1}{(a-1) m + d_m+1+j}  \right)
\\ & \nonumber
\stackrel{(*)}{\leq } 
1+\sum_{k=1}^{\infty} p^k \prod_{j=1}^{k} \left(  1+\frac{m}{(a-1) m + d_m} \right)
\\ & \nonumber
=1+\sum_{k=1}^{\infty} p^k \left(  1+\frac{m}{(a-1) m + d_m} \right)^k
\\ & \nonumber
=1+\sum_{k=1}^{\infty} \left(p \frac{am+d_m}{(a-1)m+d_m} \right)^k
\\ & \nonumber
\stackrel{(*)}{=}  \frac{1}{1-p \frac{am+d_m}{(a-1)m+d_m} } 
\\ & \nonumber
= \frac{1}{1-p \frac{am}{(a-1)m}+O \left(\frac{d_m}{m} \right) } 
\\ & \nonumber
= \frac{1}{1-\frac{a}{a-1}p}+O \left(\frac{d_m}{m} \right),
\end{align}
where $(*)$ holds for large enough $m$.
On the other hand, define an increasing sequence $l_m$, such that $l_m=o (m)$.
Then
\begin{align}
&\mathop{F\/}\nolimits\!\left(1,\widetilde{b}_m+1;\widetilde{b}_m-m+2;p\right)
 \\ & \nonumber
= 1+\sum_{k=1}^{\infty} p^k \prod_{j=1}^{k} \left( \frac{a m + d_m+j}{(a-1) m + d_m+1+j} \right) 
\\ & \nonumber
 \geq 1+\sum_{k=1}^{l_m} p^k \prod_{j=1}^{k} \left( \frac{a m + d_m+j}{(a-1) m + d_m+1+j} \right) 
\\ & \nonumber
=1+\sum_{k=1}^{l_m} p^k \prod_{j=1}^{k} \left( 1+ \frac{m-1}{(a-1) m + d_m+1+j} \right)
\\ & \nonumber
\geq 1+\sum_{k=1}^{l_m} p^k \prod_{j=1}^{k} \left( 1+ \frac{m-1}{(a-1) m + d_m+1+l_m} \right)
\\ & \nonumber
= 1+\sum_{k=1}^{l_m} p^k \left( 1+ \frac{m-1}{(a-1) m + d_m+1+l_m} \right)^k
\\ & \nonumber
=\sum_{k=0}^{l_m} \left( p  \frac{a m +d_m +l_m }{(a-1)m+d_m+l_m+1}  \right)^k,
\\ & \nonumber
 \text{set $l_m=-d_m$ (recall that $d_m<0$), then}
\\ & \nonumber
=\sum_{k=0}^{l_m} \left( \frac{a}{a-1}p+O \left(\frac{1}{m} \right)  \right)^k
\\ & \nonumber
=\frac{1}{1-\frac{a}{a-1}p+O\left(\frac{1}{m} \right)}-\frac{\left(\frac{a}{a-1}p+O \left(\frac{1}{m} \right) \right)^{-d_m+1}}{1-\frac{a}{a-1}p+O\left(\frac{1}{m} \right)}
\\ & \nonumber
=\frac{1}{1-\frac{a}{a-1}p}+O \left(\frac{1}{m} \right)+O \left( \left( \frac{a}{a-1}p \right)^{-d_m} \right).
\end{align}
\section{Proof of Lemma \ref{lemma259}}
\label{ProofLemma259}
Remember that by \cref{Fc},
\begin{align}
\bar{F}_c \left(\widetilde{b}_m \right)= C_m \frac{\Gamma(\widetilde{b}_m+1)}{\Gamma(m) \Gamma(\widetilde{b}_m-m+2)}(1-p)^m p^{\widetilde{b}_m-m+1}.
\end{align}
Using Stirling's approximation \cite[pp.257]{N46abramowitz1972handbook},
\begin{align}\label{eq:StirlingApprox}
\Gamma(x) = \sqrt{\frac{2 \pi }{x}} \left( \frac{x}{e} \right)^x \left(1+O\left( \frac{1}{x} \right) \right),
\end{align}
it can be verified that
\begin{align}\label{264}
\frac{\Gamma(\widetilde{b}_m+1)}{\Gamma(m) \Gamma(\widetilde{b}_m-m+2)}
 & 
=
\frac{ \left( \frac{\widetilde{b}_m+1}{e} \right)^{\widetilde{b}_m+1} \sqrt{\frac{2 \pi}{ \widetilde{b}_m+1}}}{ \left( \frac{m}{e} \right)^m \sqrt{\frac{2 \pi} {m}} \left( \frac{\widetilde{b}_m-m+2}{e} \right)^{\widetilde{b}_m-m+2} \sqrt{\frac{2 \pi} {\widetilde{b}_m-m+2}}}
 \left(1+O \left( \frac{1}{m}\right) \right)
\\ \nonumber
 &
= e\frac{(\widetilde{b}_m+1)^{\widetilde{b}_m+1}}{m^m (\widetilde{b}_m-m+2)^{\widetilde{b}_m-m+2}}\frac{\sqrt{m}\sqrt{\widetilde{b}_m-m+2}}{\sqrt{2 \pi}\sqrt{\widetilde{b}_m+1}}
\left(1+O \left( \frac{1}{m}\right) \right)
\end{align}
Assign $\widetilde{b}_m=am+d_m$, we have,
\begin{align}\label{265}
 & e\frac{(\widetilde{b}_m+1)^{\widetilde{b}_m+1}}{m^m (\widetilde{b}_m-m+2)^{\widetilde{b}_m-m+2}}
 \\ & \nonumber
 =\frac{\left(am+d_m+1 \right)^{am+d_m+1}}{\left((a-1)m+d_m+2 \right)^{(a-1)m+d_m+2}}
 \\ \nonumber &
 =e \left( \frac{a^a}{\left(a-1 \right)^{a-1}} \right)^m
\frac{\left( 1+\frac{d_m}{am} \right)^{am}}{\left( 1+\frac{d_m+1}{(a-1)m} \right)^{(a-1)m}}
 \frac{ \left( am+d_m+1 \right)^{d_m+1}}{\left( (a-1)m+d_m+2 \right)^{d_m+2}}.
\end{align}
Note that
\begin{align}\label{eq:Technique}
\left( 1+\frac{d_m}{am} \right)^{am} &= \exp \left( a m \log \left( 1+\frac{d_m}{a m} \right) \right)
 \\ & \nonumber
=
\exp \left( a m \left( \frac{d_m}{a m} +O \left( \frac{d^2_m}{m^2} \right) \right) \right)
 \\ & \nonumber
=
\exp \left( d_m+O\left(\frac{d^2_m}{m} \right) \right)
 \\ & \nonumber
=
e^{d_m}e^{O\left(\frac{d^2_m}{m} \right)}
 \\ & \nonumber
=
e^{d_m} \left(1+  O \left( \frac{d_m^2}{m} \right) \right).
\end{align}
Similarly,
\begin{align}\label{eq:148}
\left( 1+\frac{d_m+1}{(a-1)m} \right)^{(a-1)m}=e^{d_m+1}
\left(1+ O \left( \frac{d_m^2}{m}  \right) \right).
\end{align}
In addition, 
\begin{align}\label{eq:149}
\frac{ \left( am+d_m +1\right)^{d_m+1}}{\left( (a-1)m+d_m+2 \right)^{d_m+2}}
&
= \left(\frac{a}{a-1}+O \left(\frac{d_m}{m} \right) \right)^{d_m+1}\frac{1}{(a-1)m+d_m+2}
 \\ & \nonumber
= \left(\frac{a}{a-1}\right)^{d_m+1}\frac{1}{(a-1)m+d_m+2}\left(1+ O \left(\frac{d_m}{m} \right) \right)^{d_m+1}.
\end{align}
Using the same technique as in \cref{eq:Technique}, it can be verified that
\begin{align}\label{eq:150}
\left(1+ O \left(\frac{d_m}{m} \right) \right)^{d_m+1}&=e^{O \left(\frac{d_m^2}{m} \right)}
 \\ & \nonumber
=\left(1+ O \left( \frac{d_m^2}{m}  \right) \right).
\end{align}
Substituting \cref{eq:150,eq:149,eq:148,eq:Technique} into \cref{265}, we have
\begin{align}\label{eq:151b}
 & e\frac{(\widetilde{b}_m+1)^{\widetilde{b}_m+1}}{m^m (\widetilde{b}_m-m+2)^{\widetilde{b}_m-m+2}}
  \\ & \nonumber
=  \left( \frac{a^a}{\left(a-1 \right)^{a-1}} \right)^m
  \left(\frac{a}{a-1}\right)^{d_m+1}\frac{1}{(a-1)m+d_m+2}
  \left(1+ O \left( \frac{d_m^2}{m}  \right) \right) 
  \\ & \nonumber
=  \left( \frac{a^a}{\left(a-1 \right)^{a-1}} \right)^m
  \left(\frac{a}{a-1}\right)^{d_m+1}\frac{1}{(a-1)m}
  \left(1+ O \left( \frac{d_m^2}{m}  \right) \right)   .
\end{align}
In addition, 
\begin{align}\label{eq:151}
\sqrt{\frac{ \widetilde{b}_m-m+2}{(\widetilde{b}_m+1)}} = \sqrt{\frac{a-1}{a}}+O\left(\frac{d_m}{m} \right).
\end{align}
Now, combining \cref{eq:151,eq:151b} into  \cref{264}
\begin{multline}
\frac{\Gamma(\widetilde{b}_m+1)}{\Gamma(m) \Gamma(\widetilde{b}_m-m+2)} 
 = \left( \frac{a^a}{\left(a-1 \right)^{a-1}} \right)^m \left( \frac{a}{a-1} \right)^{d_m}\frac{1}{(a-1)\sqrt{m}}\frac{1}{\sqrt{2 \pi}}\sqrt{\frac{a}{a-1}}
 \\ 
   \left(1+ O \left( \frac{d_m^2}{m}  \right) \right),
\end{multline}
and therefore
\begin{align}
 \bar{F}_c \left(\widetilde{b}_m \right)
 &= C_m\frac{\Gamma(\widetilde{b}_m+1)}{\Gamma(m) \Gamma(\widetilde{b}_m-m+2)}(1-p)^m p^{\widetilde{b}_m-m+1}
 \\ \nonumber &
= C_m \left( \frac{a^a}{\left(a-1 \right)^{a-1}} \right)^m \left( \frac{a}{a-1} \right)^{d_m} \frac{1}{(a-1)\sqrt{m}}\frac{1}{\sqrt{2 \pi}}\sqrt{\frac{a}{a-1}}
 \\ \nonumber &
(1-p)^m p^{\widetilde{b}_m+1-m}  \left(1+ O \left( \frac{d_m^2}{m}  \right) \right)
  \\ \nonumber &
= C_m \left(\frac{a^a}{\left(a-1 \right)^{a-1}}p^{a-1}(1-p) \right)^m  \delta^{d_m}  \frac{1}{\sqrt{m}}\frac{1}{(a-1)} \sqrt{\frac{\delta p}{2 \pi}}
  \\ \nonumber &
 \left(1+ O \left( \frac{d_m^2}{m}  \right) \right).
\end{align}
\section{Proof of \Cref{lemma:OneSolRoot}}
\label{Appendix:ExaclyOneRoot}
First note that $g_p \left(\alpha \right)$ is a continuous function of $\alpha$ satisfying $\lim_{\alpha \to \infty} g_p(\alpha) =-1$.
In addition, note that
\begin{align}
\frac{\partial}{\partial \alpha} g_p\left(\alpha \right)=\left(g_p(\alpha)+1 \right) \log \left(\frac{\alpha}{\alpha-1}p \right).
\end{align}
For  $\alpha=\frac{1}{1-p}$, we have $\log \left(\frac{\alpha}{\alpha-1}p \right)=0$, and one might check that this point is indeed a maximum value, with $g_p \left( \frac{1}{1-p} \right) =1$.
Furthermore, for each $\alpha>\alpha_{\max}$, $\log \left(\frac{\alpha}{\alpha-1}p \right)<0$, and $\left(g_p(\alpha)+1 \right)>0$, hence, we can conclude that $g_p(\alpha)$ has exactly one stationary point (maximum), at $\alpha_{\max}=\frac{1}{1-p}$.\\
Now, by the Intermediate Value Theorem, there exist exactly one $\alpha_p \in \left(x_{\max},\infty \right)$, such that $g\left(\alpha_p,p \right)=0$.

\begin{remark}
Note that, $\lim_{\alpha \to 1^+} g_p(\alpha) =1-2p$, which implies that for $1/2<p<1$, there is another root to $g$, at some point within the interval $\left(1, \frac{1}{1-p} \right)$.
\end{remark}
\section{Proof of Lemma \ref{Lemma:MmessageSolution}}
\label{appendix:ProofTheorem6}
Similar to the proof in \Cref{lemma:OneSolRoot}, we show by the Intermediate Value Theorem, that $g_p(\alpha|M)>0$, for $\alpha=\frac{1}{1-p}$, while, for $\alpha= \frac{1}{1-p}+\epsilon_M$, we have $g_p(\alpha|M)< 0$.
Hence, the Intermediate Value Theorem guarantees that the root $\alpha_{p,M}$ exist at some point within this interval.

On the one hand, it can be verified that for $\alpha=\frac{1}{1-p}$, we obtain $g_p(\alpha|M)=2^{\frac{1}{M}}-1>0$, for each $M=1,2,\ldots$
On the other hand, using Taylor approximation for $g_p(\alpha|M)$, at $\alpha=\frac{1}{1-p}$, we have
\begin{align}
g_p(\alpha|M)= 2^{\frac{1}{M}}-1-\frac{(1-p)^2}{2p}2^{\frac{1}{M}} \left(\alpha-\frac{1}{1-p} \right)\\\ +O \left( \alpha-\frac{1}{1-p} \right)^3.
\end{align}
Assign $\alpha=\frac{1}{1-p}+\frac{1}{M^s}$, for $s>0$, we have
\begin{align}
g_p(\alpha|M)& = 2^{\frac{1}{M}}-1-\frac{(1-p)^2}{2p}2^{\frac{1}{M}} \frac{1}{M^{2s}} +O \left( \frac{1}{M^{3s}} \right)
\\ \nonumber &
\stackrel{(*)}{=} \frac{\log 2}{M}-\frac{(1-p)^2}{2p}\frac{1}{M^{2s}}+O \left(\frac{1}{M^2} \right)
+O \left(\frac{1}{M^{1+2s}} \right),
\end{align}
where in $(*)$, we use again Taylor approximation, as $M \to \infty$.
Now, clearly see that if $s<\frac{1}{2}$, we have 
\begin{align}
g_p(\alpha|M)= -\frac{(1-p)^2}{2p} \frac{1}{M^{2s}}+O \left(\frac{1}{M} \right) <0,
\end{align}
for sufficiently large $M$, which complete the proof.
\section{Proof of asymptotic distribution for normalized $Y$, for general $K$-ary tree}
\label{Appendix:ExtentionKChildLowerBound}
In this chaper, we show the existence of  asymptotic distribution for normalized $Y$, i.e. $Y-b_n$, which base on the structure of \Cref{sec:lowerBound}.
First, similar to \Cref{ConditionA}, it can be easily seen, by \Cref{General} that for the $b_n$ suggested in \cref{eq:b_nLowerBoundGeneralK} satisfy
\begin{align}\label{eq:tautau}
\lim_{n \to \infty} n \bar{F}_c \left(u_n |\log_{K}k_n,p \right)=p^x,
\end{align}
for $u_n=b_n+x$, and $m_n=\log_K k_n$.\\
Next, recall that the correlation between two random variable, is reflected by the number of shared channels, thus we can restate \Cref{JointDistributionLemma} for $K$ child as:
for $i=1,2,\ldots,\log_K k_n-1$, the joint distribution
\begin{align}\label{eq:JointDistGen}
 \Pr \left(Y_1>u_n,Y_{K^i} >u_n \right) 
\leq  \frac{2}{n} \left( C \Psi \left( \log_K k_n \right) \right)^{\log_K k_n-1} \left( C \frac{4}{1-p}  \log_K k_n \right)^{i} +\frac{1}{n^{2\alpha}}.
\end{align}
For \Cref{ConditionDtag} we show the complete proof:
\begin{lemma}\label{lemma:conditionDKun}
The sequence  $\left( Y_1,\ldots,Y_n \right)$ satisfy condition $D^{'}_{k_n} \left( u_n \right)$
\begin{proof}
Let 
\begin{align}
\alpha_n= n \sum_{i=2}^{k_n/K} \Pr \left( X_1>u_n,X_i>u_n \right) .
\end{align}
Then by \cref{ConditionD_k_nV2}, condition $D^{'}_{k_n} \left( u_n \right)$ satisfy if $\alpha_n \to 0$, as $n \to \infty$.\\
As mentioned before we can write the joint distribution $ \Pr \left(Y_1,Y_i \right)$ as $ \Pr \left(W+Z_1,W+Z_2 \right)$, where $W$ is random variable, represent the shared branches of the tree (dependent channels), of $Y_1,Y_i$ and $Z_1,Z_2$ as the independent branches (channels) of $Y_1,Y_i$ respectively.\\
For the general case, one might check that $Y_1$ have $(K-1)K^{i-1}$ random variables who shared $\log_K k_n-i$ channels
Then we can write  $\alpha_{n}$ as
\begin{align}
 \alpha_{n}= & n  \sum_{i=2}^{k_n/K}  \Pr \left(Y_1>u_n,Y_i >u_n \right)
 \\ \nonumber =&
n \sum_{i=1}^{\log_K k_n-1}  (K-1)K^{i-1} \Pr \left(Y_1>u_n,Y_{K^i}>u_n \right)
  \\ \nonumber \stackrel{(a)}{\leq }  &
n \sum_{i=1}^{\log_K k_n-1}  K^i 
\left( \frac{2}{n} \left( C \Psi \left( \log_K k_n \right) \right)^{\log_K k_n-1} \left( C \frac{4}{1-p}  \log_K k_n \right)^{i} +\frac{1}{n^{2\alpha}} \right) 
  \\ \nonumber = &
  2 \left( C \Psi \left( \log_K k_n \right) \right)^{\log_K k_n-1}
   \sum_{i=1}^{\log_K k_n-1}\left( C \frac{4K}{1-p}  \log_K k_n \right)^{i}+\frac{1}{2}\frac{1}{n^{2 \alpha-1}} \sum_{i=1}^{\log_K k_n-1} K^i
\end{align}
where in $(a)$, we used \cref{eq:JointDistGen}.
Now, let $g(n)=  C \frac{4K}{1-p}  \log_K k_n$, and note that $g(n) \to 0$ as $n \to \infty$, then for large enough $n$, we have
\begin{align}\nonumber
 \alpha_{n} \leq  &     \left( C \Psi \left( \log_K k_n \right) \right)^{\log_K k_n-1}
   \sum_{i=1}^{\log_K k_n-1}\left( g(n) \right)^{i}
   +\frac{1}{n^{2 \alpha-1}} \frac{k_n-K}{K-1}
   \\ \nonumber = &
 \left( C \Psi \left( \log_K k_n \right) \right)^{\log_K k_n-1}
   \frac{g(n)}{1-g(n)} \left(1-g(n) \right)^{\log_K k_n}
   +\frac{k_n}{n ^{2 \alpha-1}}  
    \\ \nonumber \leq &
  \left( C \Psi \left( \log_K k_n \right) \right)^{\log_K k_n-1}    \frac{g(n)}{1-g(n)}+\frac{k_n}{n ^{2 \alpha-1}}     \to 0
 \end{align}
 as $n \to \infty$. since similar to \Cref{ConditionDtag},  $  \left( C \Psi \left( \log_K k_n \right) \right)^{\log_K k_n-1} \to 1$, $    \frac{g(n)}{1-g(n)} \to 0$, and also by the choose of $k_n$, we obtain $\frac{k_n}{n^{2 \alpha-1}} \to 0$\\
 The lemma was proved.
 \end{proof}
Hence by \cref{eq:tautau}, and  \Cref{lemma:conditionDKun},  it is possible to apply \Cref{EVT_for_nonstationary_integer_valued}, i.e., for sufficiently large $n$,
\begin{align}
e^{-p^{x-1}} \leq \Pr \left(Y \leq u_n \right) \leq e^{-p^x},
\end{align}
\end{lemma}
\section{}
\label{Appendix:NonExsistenceConditionDun}
Let $\left( T_1,\ldots,T_n \right)$ be the completion time for the $n$ leaves, on multicast tree $\mathbf{T}_n$ with height $h$, $h=1,2,\ldots$.
Hence, each node have exactly $\sqrt[h]{n}$ child nodes (i.e. $K_n$).
Thus, $\mathbf{T}_n$ consider as $K_n$ dependent, namely the sequence $T_1,\ldots,T_n$ can be parted into $K_n$ sets, where each one of them contain $n/K_n$ elements.
Therefore condition $D_{k_n}^{'}\left(u_n \right)$, for $k_n=\frac{n}{K_n}$ can be written as
\begin{align}\label{eq:ConditionDtagV3}
D_{k_n}^{'}\left(u_n \right): \limsup n \sum_{i=2}^{k_n} \Pr \left( X_1>u_n,X_i>u_n \right) =0,
\end{align}
where
\begin{align}
n \sum_{i=2}^{k_n} \Pr \left( T_1>u_n,T_i>u_n \right) 
= n \binom{K_n}{2} \sum_{i=1}^{h-1} K_n^{i-2} \Pr \left( T_1>u_n,T_{K_n^i}>u_n \right).
\end{align}
Clearly, the sum go to zero iff each element of the sum go to zero, and in particular for $i=h-1$.
Namely,
\begin{align}
\binom{K_n}{2} \frac{n^2}{K_n^3}\Pr \left( T_1>u_n,T_{K_n}^{h-1}>u_n \right),
\end{align} 
where the pair $\left(   T_1 ,T_{K_n}^{h-1} \right) $ shared one channel in common and $h-1$ independent channels for each one, and therefore the number of pairs is $\binom{K_n}{2} \frac{n^2}{K_n^4}$.
Write the mutual channels as $W$ where $W$ follows geometric random variable, and $Z_1,Z_2$ be the independent channels of $\left(   T_1 ,T_{K_n}^{h-1} \right) $ , respectively.
As a result, We have
\begin{align}
\binom{K_n}{2} \frac{n^2}{K_n^3}\Pr \left( T_1>u_n,T_{K_n}^{h-1}>u_n \right)
& =
\frac{K_n-1}{2}\frac{n^2}{K_n^2} \Pr \left( T_1>u_n,T_{K_n}^{h-1}>u_n  \right)
\\ \nonumber 
& =\frac{K_n-1}{2}\frac{n^2}{K_n^2}  \Pr \left(W+Z_1>u_n, W+Z_2>u_n \right)
\\ \nonumber 
& \geq \frac{K_n-1}{2}\frac{n^2}{K_n^2}  \Pr \left(W>u_n \right)
\\ \nonumber 
& = \frac{K_n-1}{2}\frac{n^2}{K_n^2}  p^{u_n}
\\ \nonumber 
& = \frac{n^{\frac{1}{h}}-1}{2}n^{2-\frac{2}{h}}\frac{1}{n}\frac{(h-1)!}{\left( \frac{1-p}{p} \Psi \left(h \right)  \right)^{h-1}}
\\ \nonumber 
& = \frac{n^{\frac{1}{h}}-1}{2}n^{1-\frac{2}{h}}\frac{(h-1)!}{\left( \frac{1-p}{p} \Psi \left(h \right) \right)^{h-1}}
\\ \nonumber 
& =  o \left(\frac{n^{1-\frac{1}{h}}}{ \left( \log n \right)^{h-1}} \right) \to \infty
\end{align} 
for each $h=1,2,\ldots$.
\section{}
\label{Appendix:ExtensionIIDKnChildren}
Let 
\begin{align}
\hat{M}_n=\max \left(T_1,\ldots,T_n \right),
\end{align}
where $\left(\hat{T}_1,\ldots,\hat{T}_n \right)$ is an i.i.d.\ sequence follow NB distribution with $h_n$ successes, where
\begin{align}
h_n=\frac{\log n}{\log K_n},
\end{align}
$h_n=1,2,\ldots$, and let $u_n=b_n+x$, where
\begin{align}
b_n= \alpha_{p,K_n} \log n+ \log_{\delta} \sqrt{\log n \log K_n}+\beta
\end{align}
where $\alpha_{p,K_n},\delta,\beta$ defined in \Cref{sec:ExtentionsK_nH_n}, then
\begin{align}
&\limsup_{n \to \infty} \Pr \left( \hat{M}_n \leq u_n \right) \leq e^{-\delta^x}
 \\ \nonumber &
\liminf_{n \to \infty} \Pr \left(  \hat{M}_n \leq u_n \right) \geq e^{-\delta^{x-1}}
\end{align}
\begin{proof}
The constellation of the proofs will be similar to \Cref{sec:UpperBound}, hence, we write only sketch of the proof and necessary differences will be mention.\\
First  let present $b_n$ as $b_n=a \log n+c_n$, where $c_n$ satisfy $\frac{c_n^2}{\log n} \to 0$, 
\begin{align}
\Pr \left(T_i > b_n \right)  = C_n \Pr \left( X=b_n+1 \right)
\end{align}
 where $C_n = \frac{1}{1-\delta}+o(1)$, and $\delta=\frac{a}{a-\frac{1}{K_n}}p$.
The prove is based on \Cref{CLoseFormExpressionCDFNegativeBinomial}, and in particular we show that hypergeometric function $\mathop{F\/}\nolimits\!\left(1,b_n+1;b_n-h_n+2;p\right)$ is convergence to $\frac{1}{1-\delta}$ as $n \to \infty$.
\begin{align}
\mathop{F\/}\nolimits\!\left(1,b_n+1;b_n-h_n+2;p\right)
 &
=1+\sum_{k=1}^{\infty} p^k \prod_{j=1}^{k} \left( \frac{b_n+j}{b_n-h_n+1+j} \right) 
\\ & \nonumber
 =1+\sum_{k=1}^{\infty} p^k \prod_{j=1}^{k} \left( \frac{a m + c_n+j}{(a-\frac{1}{K_n}) n + c_n+1+j} \right) 
\end{align}
Hence the only change from the proof in \Cref{CLoseFormExpressionCDFNegativeBinomial} is that we write $a-\frac{1}{K_n} $ instead of $a-1$,
therefore we can define $\bar{F}_c$.
\begin{align}\label{eq:cdf3}
\bar{F}_c \left(b_n \right)= & C_n \frac{\Gamma(b_n+1)}{\Gamma(h_n) \Gamma(b_n-h_n+2)}
(1-p)^{h_n} p^{b_n-h_n+1}.
\end{align}
Next, we show that $\bar{F}_c \left(b_n \right)=\frac{1}{n} \left(1+o \left(1 \right) \right)$.\\
Using Stirling's approximation, it can be verified that
\begin{align}\label{eq:139}
\frac{\Gamma(b_n+1)}{\Gamma(h_n ) \Gamma(b_n-h_n+2)}
 & 
 =
\frac{ \left( \frac{b_n+1}{e} \right)^{b_n+1} \sqrt{\frac{2 \pi}{ b_n+1}}}{ \left( \frac{h_n}{e} \right)^{h_n} \sqrt{\frac{2 \pi} {h_n}} \left( \frac{b_n-h_n+2}{e} \right)^{b_n-h_n+2} \sqrt{\frac{2 \pi} {b_n-h_n+2}}}
 \left(1+O \left( \frac{1}{h_n} \right) \right)
\\ \nonumber
 &= e\frac{(b_n+1)^{b_n+1}}{h_n^{h_n} (b_n-h_n+2)^{b_n-h_n+2}}
 \frac{\sqrt{h_n}\sqrt{b_n-h_n+2}}{\sqrt{2 \pi}\sqrt{b_n+1}}
 \left(1+O \left( \frac{1}{h_n} \right) \right).
\end{align}
Assign $b_n=a \log n+c_n$, (where $c_n=O \left(\log \log n \right)$) we obtain
\begin{align}\label{eq:140}
 & e\frac{(b_n+1)^{b_n+1}}{h_n^{h_n} (b_n-h_n+2)^{b_n-h_n+2}}
   \\ \nonumber &
  =e \frac{\left( a \log n +c_n+1 \right)^{a \log n+c_n+1}}{\left( \left( a-\frac{1}{\log K_n} \right) \log n+c_n+2 \right)^{ \left( a-\frac{1}{\log K_n} \right) \log n+c_n+2}}
  \\ \nonumber &
 =e \left( \log K_n^{\frac{1}{\log K_n}} \frac{a^a}{\left( a-\frac{1}{\log K_n} \right)^{ a-\frac{1}{\log K_n}}} \right)^{\log n} 
\frac{\left( 1+\frac{c_n+1}{a\log n} \right)^{a \log n}}{\left( 1+\frac{c_n+2}{\left( a-\frac{1}{\log K_n} \right) \log n} \right)^{\left( a-\frac{1}{\log K_n} \right)\log n}} 
  \\ \nonumber &
\frac{\left( a \log n +c_n+1 \right)^{c_n+1}}{\left( \left( a-\frac{1}{\log K_n} \right) \log n+c_n+2 \right)^{ c_n+2}}.
\end{align}
Based on the technique in Appendix \ref{ProofLemma259}, \cref{eq:Technique}, it can be verified that
\begin{align}\label{eq:141}
e \frac{\left( 1+\frac{c_n+1}{a \log n} \right)^{a \log n}}{\left( 1+\frac{c_n+2}{\left( a-\frac{1}{\log K_n} \right)  \log n} \right)^{\left( a-\frac{1}{\log K_n} \right) \log n}} = 1+O \left(\frac{c_n^2}{\log n} \right),
\end{align}
and
\begin{multline}\label{eq:142}
\frac{\left( a \log n +c_n+1 \right)^{c_n+1}}{\left( \left( a-\frac{1}{\log K_n} \right) \log n+c_n+2 \right)^{ c_n+2}} 
= \left(\frac{a}{a-\frac{1}{\log K_n}} \right)^{c_n+1}\frac{1}{(a-\frac{1}{\log K_n}) \log n+c_n+2}
\\
\left( 1+O \left(\frac{c_n^2}{\log n} \right) \right).
\end{multline}
In addition
\begin{align}\label{eq:143}
\sqrt{\frac{ b_n-h_n+2}{b_n+1}} = \sqrt{\frac{a-\frac{1}{\log K_n}}{a}} \left(1+O \left(\frac{c_n}{\log n} \right) \right).
\end{align}
Hence by  \cref{eq:139,eq:140,eq:141,eq:142,eq:143} 
\begin{align}
& \frac{\Gamma(b_n+1)}{\Gamma(h_n  
) \Gamma(b_h-h+2)} 
 \\ \nonumber
& =  \left( \log K_n^{\frac{1}{\log K_n}} \frac{a^a}{\left( a-\frac{1}{\log K_n} \right)^{ a-\frac{1}{\log K_n}}} \right)^{\log n}\left(\frac{a}{a-\frac{1}{\log K_n}} \right)^{c_n+1}
 \\ \nonumber
&
\frac{1}{(a-\frac{1}{\log K_n}) \log n+c_n+2}\sqrt{\frac{a-\frac{1}{\log K_n}}{a}}
\sqrt{\frac{h}{2 \pi}} \left(1+o(1) \right).
\end{align}
And therefore
\begin{align}
&\bar{F}_c \left(b_n \right) = C\frac{\Gamma(b_n+1)}{\Gamma(h_n) \Gamma(b_n-h_n+2)}(1-p)^{h_n} p^{b_n-h_n+1}
 \\ \nonumber &
= C \left( \log K_n^{\frac{1}{\log K_n}} \frac{a^a}{\left( a-\frac{1}{\log K_n} \right)^{ a-\frac{1}{\log K_n}}} \right)^{\log n}\left(\frac{a}{a-\frac{1}{\log K_n}} \right)^{c_n+1}
 \\ \nonumber &
\frac{1}{(a-\frac{1}{\log K_n}) \log n+c_n+2}
\sqrt{\frac{a-\frac{1}{\log K_n}}{a}}
\sqrt{\frac{h_n}{2 \pi}} (1-p)^{h_n} p^{b_n-h_n+1}
  \\ \nonumber &
 \left(1+o(1) \right)
  \\ \nonumber &
= C \left( \log K_n^{\frac{1}{\log K_n}} \frac{a^a}{\left( a-\frac{1}{\log K_n} \right)^{ a-\frac{1}{\log K_n}}} p^{a-\frac{1}{\log K_n}}\left(1-p \right)^{\frac{1}{\log K_n}} \right)^{\log n}
  \\ \nonumber &
\delta^{c_n} \frac{\sqrt{h_n}}{(a-\frac{1}{\log K_n}) \log n+c_n+2} \sqrt{\frac{\delta p}{2 \pi}}\left(1+o(1) \right)
    \\ \nonumber &
    \text{ writing $h_n=\frac{\log n}{\log K_n}$, and by \cref{eq:GeneralizedalphaP}}
        \\ \nonumber &
   = \frac{1}{n} C \sqrt{\frac{\delta p}{2 \pi}} \frac{\sqrt{\frac{\log n}{\log K_n}}}{\left(a-\frac{1}{\log K_n } \right)\log n+o\left(\log n \right)} \delta^{c_n}\left(1+o(1) \right)
\end{align}
Now set $c_n=\log_{\delta} \sqrt{\log n \log K_n}+\log_{\delta} \left(  \frac{a-\frac{1}{\log K_n }}{C}\sqrt{\frac{2 \pi}{\delta p}} \right) $, we obtain
\begin{align} \nonumber
&=\frac{1}{n}{\left( a-\frac{1}{\log K_n } \right) \log n }{\left(a-\frac{1}{\log K_n } \right)\log n+o\left(\log n \right)} 
\left(1+o \left( 1 \right) \right)
  \\ \nonumber &
= \frac{1}{n} \left(1+o \left( 1 \right) \right)
\end{align}
and therefore for $u_n=b_n+x$, for $b_n$ 
\begin{align}
&\limsup_{n \to \infty} \Pr \left( \hat{M}_n \leq u_n \right) \leq e^{-\delta^x}
 \\ \nonumber &
\liminf_{n \to \infty} \Pr \left(  \hat{M}_n \leq u_n \right) \geq e^{-\delta^{x-1}}
\end{align}
which imply $\mathbb{E} \left[\hat{M}_n \right] \leq b_n -\frac{\gamma}{\log \delta}+1$
\end{proof}

\bibliographystyle{IEEEtran}
\bibliography{bibliugrafy}{}
\end{document}